%% file: cas-sc-template.tex
\documentclass[a4paper,fleqn]{cas-sc}

\usepackage{preamble}

\input{acro.tex}

\begin{document}
	\let\WriteBookmarks\relax
	\def\floatpagepagefraction{1}
	\def\textpagefraction{.001}
	\shorttitle{Assessment of alternative covariance functions for joint input-state estimation via Gaussian Process latent force models in structural dynamics}
	\shortauthors{Silvia Vettori et~al.}
	
	\title [mode = title]{Assessment of alternative covariance functions for joint input-state estimation via Gaussian Process latent force models in structural dynamics}                      
	
	

	\author[1,2]{S. Vettori}[type=editor,
	auid=000,bioid=1,
	prefix=,
	role=,
	orcid= 0000-0001-6097-4278]
	\cormark[1]
	\ead{silvia.vettori@siemens.com}
	
	
	\address[1]{Siemens Digital Industries Software, Interleuvenlaan 68, 3001
		Leuven, Belgium}
	
	\author[1]{E. {Di Lorenzo}}[%
	role=,
	suffix=,
	]
	\ead{emilio.dilorenzo@siemens.com}
	
	\author[1]{B. Peeters}[%
	role=,
	suffix=,
	]
	\ead{bart.peeters@siemens.com}
	

	\address[2]{Institute of Structural Engineering, ETH Zürich, Stefano-Franscini-Platz 5, CH-8093 Zürich, Switzerland}

	\author%
	[2]
	{E. Chatzi}
	\ead{chatzi@ibk.baug.ethz.ch}

	\cortext[cor1]{Corresponding author}
	
	\begin{abstract}
		Digital technologies can be used to gather accurate information about the behavior of structural components for improving systems design, as well as for enabling advanced \ac{SHM} strategies. New avenues for achieving automated and continuous structural assessment are opened up via development of virtualization approaches delivering so-called \acp{DT}, i.e., digital mirrored representations of physical systems relying on fusion of simulation models and real-time monitoring data. In this framework, the main motivation of the work presented in this paper stems from the existing challenges in the implementation and deployment of a real-time predictive framework for virtualization of dynamic systems. Kalman-based filters are usually employed in this context to address the task of joint input-state prediction in structural dynamics. A \ac{GPLFM} approach is exploited in this work to construct flexible data-driven a priori models for the unknown inputs, which are then coupled with a combined deterministic-stochastic state-space model of the structural component under study for Kalman-based input-state estimation. The use of \ac{GP} regression for this task overcomes the limitations of the conventionally adopted random-walk model, thus limiting the necessity of offline user-dependent calibration of this type of data assimilation methods. This paper proposes the use of alternative covariance functions for \ac{GP} regression in structural dynamics. A theoretical analysis of the \acp{GPLFM} linked to the investigated covariance functions is offered. The outcome of this study provides insights into the applicability of each covariance type for \ac{GP}-based input-state estimation. The proposed framework is validated via an illustrative simulated example, namely a 3 \ac{DOF} system subjected to an array of different loading scenarios. Additionally, the performance of the method is experimentally assessed on the task of joint input-state estimation during testing of a 3D-printed scaled \ac{WT} blade.
	\end{abstract}

	
	\maketitle
	
	\section{Introduction}
	Design, identification, verification and monitoring comprise primary tasks  for effectuating an efficient life-cycle management of structural components. Several challenges render the execution of these tasks non-trivial, relating to system complexity, inaccessibility and further deployment limitations, time and cost constraints, as well as external sources of uncertainty. The process of virtualization offers a strategy for narrowing these uncertainties, by harnessing the fusion of physics-based models with data to build a virtual replica of the physical system; a \acf{DT} \cite{AIAA_DT}. This further allows for generating knowledge beyond the sparse physical sensing locations, by permitting to predict response in "virtual sensor" locations. The main challenge behind the definition of \acp{DT} lies in the implementation of reliable methodologies for combining model-based and data-based information to create a predictive tool that can evolve over time. To this end, data assimilation techniques are widely adopted in the structural dynamics context for implementation of \acf{VS} strategies, i.e., for inferring \ac{QoI} such as system responses and/or unknown loads in a dynamic environment \cite{maes2016joint, maes2016verification, maes2016dynamic, cumbo2019kalman, naets2015stable,azam2017experimental, lourens2012joint,tchemodanova2021strain, tamarozzi2016noise, TARPO2020105280, avitabile2012prediction, iliopoulos2016modal}. The working principle of these strategies relies on use of data for reducing the uncertainties arising from the limitations of mechanistic modeling, which is in turn exploited to expand the available data and gain further knowledge of the system dynamic behavior. Despite \ac{VS} methods nowadays being recognized as powerful tools for twinning of dynamic systems, several limitations are still to be addressed for improvement of the accuracy and efficiency of the underlying data assimilation process. 
	
	Kalman-type filters have been widely exploited in the recent years for constructing virtual sensors \cite{dertimanis2019input, papadimitriou2011fatigue, TATSIS2021107223, tatsis2019response, tatsis2020adaptive, risaliti2019multibody} in linear or even nonlinear contexts \cite{chatzi2009unscented,mariani2007unscented, tatsis2018state}. Several adaptations of the standard \ac{KF} algorithm have been constructed to simultaneously address response prediction and inverse load identification of linear dynamic systems in a stochastic setting. To this end, a random-walk equation is commonly adopted as prior assumption on the unknown input dynamics. The resulting equation is then used in combination with the mechanistic stochastic state-space model of the structure, for simultaneously estimating its responses at unmeasured locations and the loads it is subjected to. For instance, within the so-called \ac{AKF} algorithm, the random-walk assumption allows to include the unknown input term within a new augmented state vector. The \ac{AKF} and its applicability to structural dynamics have been proposed in \cite{lourens2012augmented}. In \cite{naets2015stable}, the un-observability of the augmented system matrix appearing when only acceleration measurements are considered has been proved and tackled by means of dummy displacement measurements. Alternatively, the numerical issues due to un-observability exhibited by the \ac{AKF} can be overcome by adopting the so-called \ac{DKF} for joint input-state estimation. This Kalman-based recursive estimator, presented in \cite{azam2015dual} and tested in \cite{azam2017experimental}, recasts the input and state prediction equations into two sequential stages. Several additional algorithms have been proposed for joint input-state estimation with a time-delayed scheme \cite{hsieh2009optimal, hsieh2017unbiased,maes2018smoothing}. It has been proven in \cite{maes2018smoothing} that these approaches, classified as smoothing schemes, allow to significantly reduce the estimation uncertainty due to measurement noise when measurements collocated with the estimated forces are not available. The challenges derived from instantaneous system inversion for load identification within input-state estimation algorithms have been investigated in \cite{maes2015design}, where the requirements in terms of sensors quantity and type are investigated. In this sense, an \ac{OSP} strategy has been developed in \cite{cumbo2021advanced,mazzanti2020improved} for input-state estimation using the \ac{AKF}, while a more extended framework based on information theory is proposed in \cite{s21103400,ercan2022optimal,ercan2023information}. Although the random-walk model and its adaptations are widely adopted as prior input information in generic scenarios when applying joint input-state estimation schemes, this is often not representative of the actual loading conditions. Indeed, the random-walk model can be treated as a special first order autoregressive process with unit root, i.e., a non-stationary process \cite{dertimanis2019input}, thus resulting in a too strict simplification for loads which deviate from the popular ambient noise excitation scheme. This leads to challenging tuning efforts of the estimators and difficulties in providing reliable loading predictions in generic circumstances. To this end, a more generic model for the input, i.e., a first order stationary autoregressive process, has been used in \cite{dertimanis2019input} to address joint input-state-parameter estimation via a combination of the \ac{DKF} and the \ac{UKF} \cite{julier1995new, julier1997new, safarinejadian2015kalman, chatzi2015online}. \ac{ML} approaches such as \acf{GP} regression \cite{williams2006gaussian} have been also exploited to construct more comprehensive models for the unknown input. A state-space representation for \ac{GP} models has been proposed in \cite{hartikainen2010kalman, hartikainen2012sequential}, where a recursive regression solution has been implemented using a combination of Kalman filtering and smoothing. The sequential \ac{GP} regression solution has been used to construct \acfp{GPLFM} with applications in several domains \cite{sarkka2018gaussian,pmid:24051729,alvarez2009latent}. In \cite{nayek2019gaussian}, \acp{GPLFM} have been introduced as flexible alternatives for unknown input modeling in the framework of joint input-state estimation of structures. With this purpose, a \ac{GPLFM} with a Matérn kernel is used in combination with the deterministic-stochastic state-space model of the system for joint input-state prediction via Kalman filtering and smoothing. Besides its higher flexibility, it has been demonstrated in \cite{nayek2019gaussian} that this method overcomes the un-observability issues arising from the use of the \ac{AKF} with acceleration-only data sets. The \ac{GPLFM} has been validated for in-situ measurements in \cite{zou2022virtual}, further exploited for input-state-parameter estimation in \cite{rogers2020application} and adapted for identification of mechanical systems with discontinuous nonlinearity in \cite{marino2023switching}.  Additionally, \ac{GP} regression has been implemented in both time and space dimensions for recursive distributed load prediction in \cite{tatsis2020spatiotemporal}.
	
	A critical review on the necessary tools for building a state-space representation of a \ac{GP} is presented in this paper. Specifically, a detailed analysis of conventional and non-conventional \ac{GP} covariance functions in structural dynamics is offered, along with the derivation of their \ac{SDE} representations. The latter has been analyzed by means of an analogy with the harmonic oscillators theory, which has served to identify the inner dynamic features of each investigated covariance function. This work proposes the use of bespoke covariance functions according to the analyzed experimental case study in order to maximize the joint input-state estimation performance. To this end, the theoretical study has been complemented with a simulated example concerning a 3 \acf{DOF} system subjected to several loading conditions. The outcome is also experimentally validated on the case study of a 3D-printed scaled \acf{WT} blade undergoing dynamic testing in clamped-free conditions.
	
	The paper is organized as follows. \Autoref{sec:theory} offers the formulation for the \ac{MOR} technique adopted in this work, along with a description of the combined deterministic-stochastic state-space equations for a linear structural system. \Autoref{sec:theory} continues by proposing the theoretical background on \ac{GP} regression and providing a thorough analysis of both conventional and non-conventional kernels (covariance functions) in the structural dynamics context. \Autoref{sec:regr_statespace} reports on the state-space formulation for recursive \ac{GP} regression and investigates its features as a function of the adopted kernel by means of an analogy with harmonic oscillators. The construction of \acp{LFM} for joint input-state estimation is then presented in \autoref{sec:LFM}, where some considerations regarding the practical implementation of the method are highlighted. Following the reported notions, a \ac{GPLFM} framework is built and tested in \autoref{sec:LFM} for a 3 \acp{DOF} system. An experimental case study is provided in \autoref{sec:3D_blade}, where the novel bespoke kernel selection for \ac{GP}-based input-state estimation is assessed during testing of a 3D-printed scaled \ac{WT} blade. Finally, the work is concluded in \autoref{sec:concl}.	
	
	\section{Theoretical background}
	\label{sec:theory}
	\subsection{Bayesian dynamic model of a structural system}
	\label{subsec:SSM}
	
	The \ac{EoM} of a linear structural system can be formulated as a second order vector differential equation of the form:
	\begin{equation}
		\label{eq:eqmotion}
		\mathbf{M \ddot{z}}(t)+\mathbf{D\dot{z}}(t)+\mathbf{Kz}(t)=\mathbf{S}_i\mathbf{u}(t)
	\end{equation}
	where $\mathbf{z}(t) \in \mathbb{R}^{n_{dof}}$ is the vector of displacements, usually corresponding to the \ac{FE} model \acp{DOF}, $\mathbf{M} \in \mathbb{R}^{n_{dof}\times n_{dof}}$, $\mathbf{D} \in \mathbb{R}^{n_{dof}\times n_{dof}}$ and $\mathbf{K} \in \mathbb{R}^{n_{dof}\times n_{dof}}$ denote the mass, damping and stiffness matrices respectively;  $\mathbf{u}(t)\in\mathbb{R}^{n_i}$ (with $n_i$ representing the number of loads) is the input vector and $\mathbf{S}_{i}\in\mathbb{R}^{n_{dof}\times n_{i}}$ is the Boolean input shape matrix that selects the \acp{DOF} where loads (inputs) are applied. A \ac{MOR} technique \cite{craig1985review} can be applied to reduce the model size by formulating the dynamic behavior of a structure as a superposition of modal contributions: 
	\begin{equation}
		\label{eq:redcoord}
		\mathbf{z} (t)\: \approx \:  \mathbf{\Psi p}(t)
	\end{equation}
	where $\mathbf{\Psi}\in\mathbb{R}^{n_{dof}\times{n_r}}$ is the reduction basis and  $\mathbf{p}\in\mathbb{R}^{n_r}$ is the vector of the generalized coordinates of the system, with $n_r$ being the dimension of the reduced coordinates. Inserting the reduction basis into \autoref{eq:eqmotion} and premultiplying each term by $\mathbf{\Psi}^T$, the resulting equation is of the form: 
	
	\begin{equation}
		\label{eq:redeqmotion}
		\mathbf{M}_r \mathbf{\ddot{p}}(t)+\mathbf{D}_r\mathbf{\dot{p}}(t)+\mathbf{K}_r \mathbf{p}(t)=\mathbf{S}_r \mathbf{u}(t)
	\end{equation}
	where the mass, damping, stiffness and input shape matrices of the reduced system are respectively $\mathbf{M}_r = \mathbf{\Psi}^{T}\mathbf{M\Psi}$, $\mathbf{D}_r=\mathbf{\Psi}^{T}\mathbf{D\Psi}$, $\mathbf{K}_r = \mathbf{\Psi}^{T}\mathbf{K\Psi}$ and $\mathbf{S}_r=\mathbf{\Psi}^{T}\mathbf{S}_i$. 
	
	The reduction basis adopted in this work is expressed as:
	\begin{equation}
		\label{eq:redbasis}
		\mathbf{\Psi} = \begin{bmatrix} \mathbf{\Psi}_n & \mathbf{\Psi}_{\alpha} \end{bmatrix}  
	\end{equation}
	where $\mathbf{\Psi}_n\in\mathbb{R}^{n_{dof}\times n_{k}}$ is the matrix of the numerical normal modes to be included in the \ac{ROM}, i.e., the eigenmodes of the entire structure in the frequency range of interest, and $\mathbf{\Psi}_{\alpha}\in\mathbb{R}^{n_{dof}\times n_{\alpha}}$ is the residual attachment modes matrix \cite{craig1985review,VETTORI2023109654}, with $n_k + n_{\alpha} = n_r$.
	
	In the structural dynamics context, it is common practice to transform \autoref{eq:eqmotion} into a state-space model \cite{VETTORI2023109654}. In this work, the use of a combined deterministic-stochastic state-space model is introduced for consistency with the Bayesian approach hereby adopted for the task of input-state estimation. This type of models assume the evolution of states over time to follow a stochastic process with Gaussian errors modeling uncertainties on both the transition and observation equations. The discrete-time version of a structural combined deterministic-stochastic state-space model is of the type:
	\begin{equation} 
		\label{eq:BDM}
		\begin{cases}
			\mathbf{x}_{k} = \mathbf{A}_{d}\mathbf{x}_{k-1}+\mathbf{B}_d\mathbf{u}_{k-1}+\mathbf{w}_{k-1} \\ \mathbf{y}_k = \mathbf{C} \mathbf{x}_k+\mathbf{G}\mathbf{u}_k+\mathbf{v}_k.
		\end{cases}	
	\end{equation}
	where the state vector $\mathbf{x}_k = \begin{bmatrix} {\mathbf{p}_k}^T & {\mathbf{\dot{p}}_k}^T \end{bmatrix}^T \in \mathbb{R}^{2n_{r}}$ is described by a Gaussian distribution with mean $\mathbf{\hat{x}}_k \in \mathbb{R}^{2n_r}$ and covariance matrix $\mathbf{P}_k \in \mathbb{R}^{2n_r \times 2n_r}$. Stationary zero-mean uncorrelated white noises $\mathbf{w}_k$ and $\mathbf{v}_k$ have been introduced to respectively take into account model uncertainties and measurement noise. The covariance matrices associated to $\mathbf{v}_k$ and $\mathbf{w}_k$ are respectively denoted as $\mathbf{R} \in \mathbb{R}^{n_o \times n_o}$ and $\mathbf{Q}\in \mathbb{R}^{2n_{r}\times 2n_r}$ such that:
	\begin{equation}\label{eq:covariance_mat_all}
		\mathbb{E} \begin{Bmatrix}
			\begin{pmatrix}
				\mathbf{w}_k \\ \mathbf{v}_k
			\end{pmatrix}
			&
			\begin{pmatrix}
				{\mathbf{{w}}_{l}}^{T} & {\mathbf{{v}}_{l}}^{T}
			\end{pmatrix}
		\end{Bmatrix} = \begin{bmatrix}
			\mathbf{Q} & \mathbf{0} \\ \mathbf{0}^T & \mathbf{R}
		\end{bmatrix} \delta_{kl}
	\end{equation}
	where $\delta_{kl}$ is the Kronecker delta function and the autocovariance terms $\mathbf{Q}$ and $\mathbf{R}$ represent the covariance matrices of $\mathbf{w}_k$ and $\mathbf{v}_k$, respectively. In \autoref{eq:covariance_mat_all}, the mutual correlation of these processes has been discarded for simplicity, i.e., the matrix off-diagonal terms are null. Additionally, further simplification is introduced in \autoref{eq:BDM} via the assumption of uncorrelated process noise sources, i.e. diagonal $\mathbf{Q}$, and uncorrelated measurement noise sources, i.e., diagonal $\mathbf{R}$. 
	
	\subsection{Gaussian Process regression}
	Regression is a well-known approach for generating models via a data-driven paradigm; given labeled training data sampled independently, regression algorithms learn about the latent data input-output relationships. The simplest approach consists in assuming a linear relationship between outputs and inputs of the underlying function \cite{galton1886regression, kenney1962linear, zhang2004solving, lai1979strong, hoerl1970ridge}. However, this assumption is not always suitable for real-life problems, for which often outputs cannot be computed as a linear combination of inputs. A method to add flexibility to regression consists in assuming distributions over functions modeled via stochastic processes, i.e., collections of random variables \cite{edition2002probability, gikhman2004theory}. In particular, a \ac{GP} is a stochastic process for which any finite number of variables have a joint Gaussian distribution \cite{williams2006gaussian}. Given an independent variable $t \in \mathbb{R}^d$ and a dependent variable $f(t): \mathbb{R}^d \rightarrow \mathbb{R}$, a \ac{GP} is defined as:
	\begin{equation}\label{eq:GP}
		f(t)\sim \mathcal{GP}(\mu(t),k(t,t',\theta))
	\end{equation}
	where $\mu(t)= \mathbb{E} [\ f(t) ]\ $ and $ k(t,t',\theta)=\mathbb{E} [\ (f(t)-\mu(t))(f(t')-\mu(t')) ]\ $ are respectively its mean and covariance function, while $\theta$ denotes the hyperparameters of the covariance function (see \autoref{subsec:cov_functs} for examples). Since this work will treat signals, which evolve in time domain, it is convenient to introduce the definition of a one-dimensional zero-mean \ac{GP}:
	\begin{equation} \label{eq:GPtime}
		f (t) \sim \mathcal{GP}(0, k(t,t', \theta))
	\end{equation}
	where $t \in \mathbb{R}$. The zero mean assumption allows to simplify the notation without loss of generality, since a realization of a \ac{GP} with a non-zero mean function can be retrieved from a function drawn from a zero-mean \ac{GP} with the same covariance by adding the mean function. 
	
	Given a training data set of $n$ observations $D = \lbrace \lbrace  \tilde{t}_1,\tilde{y}_1 \rbrace , \lbrace \tilde{t}_2,\tilde{y}_2 \rbrace, \dots,  \linebreak[0] \lbrace \tilde{t}_n,\tilde{y}_n \rbrace \rbrace  $ and a \ac{GP} mean (conventionally 0) and covariance prior, \ac{GP} regression consists in obtaining the latent function $f(t)$ at unobserved test points such that:
	\begin{equation}\label{eq:GP_eps}
		y_k = f(t_k)+\varepsilon_k
	\end{equation} 
	where $\varepsilon_k \sim \mathcal{N}(0,\sigma_n^2)$ is white Gaussian noise corrupting the data. 
	The \ac{GP} posterior, i.e., the distribution of the latent function at the unobserved locations given the $n$ data points available in $D$, is a Gaussian distribution with mean and covariance of the following closed-form: 
	\begin{equation} 
		\label{eq:GPposterior}
		\begin{aligned}
			&\bm{\hat{\mu}} = \mathbf{K}^T_{\mathbf{{t,t}}}\left(\mathbf{K_{\tilde{t},\tilde{t}}}+\sigma_n^2\mathbf{I}\right)^{-1}\mathbf{\tilde{y}};\\ 
			&\mathbf{\hat{K}} = \mathbf{K_{t,t}} - \mathbf{K}^T_{\mathbf{{t,\tilde{t}}}} \left(\mathbf{K_{\tilde{t},\tilde{t}}}+\sigma_n^2\mathbf{I}\right)^{-1} \mathbf{K_{\tilde{t},t}}
		\end{aligned} 
	\end{equation}
	where $\tilde{\mathbf{t}} = [ \tilde{t}_1, \dots, \tilde{t}_n ]^T$ is the vector of observed time points, $\tilde{\mathbf{y}} = [ \tilde{y}_1, \dots, \tilde{y}_n ]^T$ is the vector of observations, $\mathbf{K_{\tilde{t},\tilde{t}}}$ is the covariance matrix between the observations in $D$, $\mathbf{K_{t,t}}$ is the covariance matrix between the unobserved test points and  $\mathbf{K_{\tilde{t},t}}$ is the covariance matrix between the observations and the unobserved test points. \Autoref{eq:GPposterior} is obtained following Bayes' rule by conditioning the joint Gaussian distribution on the observations $\mathbf{\tilde{y}}$ in $D$ and exploiting the properties of multivariate Gaussian distributions \cite{williams2006gaussian}. The posterior computation features a computational cost that scales with $O(n^3)$ because of the term $\left(\mathbf{K_{\tilde{t},\tilde{t}}}+\sigma_n^2\mathbf{I}\right)^{-1}$.
	
	An important step in the use of \acp{GP} for prediction consists in the selection of the prior model to be employed for regression given a set of available measurements. Model selection can be viewed as the set of a priori decisions taken to configure the regression model, e.g., the choice of the covariance form and the selection of its hyperparameters $\theta$, so defined since they are parameters of a non-parametric model. Indeed, a \ac{GP} is a function that does not depend on fixed parameters since its parameters increase as soon as new data is available \cite{williams2006gaussian}. While the covariance function choice is essentially performed by the user, which can rely on any available prior knowledge regarding the nature of the analyzed data, the selection of hyperparameters, whose nature depends on the chosen type of function (see \autoref{subsec:cov_functs}), needs to be performed via tailored algorithms based on a training data set. In this work, the hyperparameter selection is succeeded by maximizing the marginal likelihood \cite{williams2006gaussian, nayek2019gaussian}, i.e., maximizing the probability of the data given the model, w.r.t the hyperparameters. Following this method, values of the covariance hyperparameters are chosen such that the likelihood of the data is maximized when marginalizing, i.e., averaging, over all the possible \ac{GP} realizations for the chosen covariance function. This method is particularly useful for \acp{GP}, since the marginal likelihood computed exploiting the properties of multivariate Gaussian distributions features a simple and analytically tractable expression \cite{williams2006gaussian}. The resulting optimization problem is reported in \autoref{eq:marginal_opt_GP}, where maximization of the marginal likelihood has been converted into minimization of the negative logarithmic marginal likelihood for practicality:
	\begin{equation}\label{eq:marginal_opt_GP}
		\operatorname*{argmin}_{\bm{\theta}} \: -log \quad p(\mathbf{y}|\bm{\theta}) = \operatorname*{argmin}_{\bm{\theta}} \quad \frac{n}{2}log 2\pi+\frac{1}{2} log|\mathbf{K}+\sigma_n^{2}\mathbf{I}|+\frac{1}{2} \mathbf{y}^{T}( \mathbf{K}+\sigma_n^2 \mathbf{I})^{-1}\mathbf{y}\text{.}
	\end{equation}
	
	For solving the optimization problem in a \ac{GP} regression setting, gradient-based optimization is widely exploited since derivatives of \autoref{eq:marginal_opt_GP} can be easily computed. This class of methods can easily face entrapment in local minima when the adopted cost function is non-convex. As a result, different initial guess on the hyperparameters may be tested to avoid deterioration of the training outcome. With regards to computational complexity, the computation of derivatives implies a flop count of $O(n^2)$ per hyperparameter (if $n$ is the number of data points, then the dimension of the covariance matrix is $n \times n$). Therefore, the most substantial contribution is provided by the inversion of the positive definite covariance matrix, which scales cubically with $n$.

	\subsection{Covariance functions for regression via \acp{GP}}
	\label{subsec:cov_functs}
	The covariance function is a key tool for regression via \acp{GP} since it embeds any prior knowledge about the sought-after latent function. Validity of covariance functions within the context of \ac{GP}-theory is subject to certain specifications. An arbitrary time-domain function $k$ of input pairs $t$ and $t'$ is a valid covariance (kernel) function if it satisfies the symmetry and positive semi-definiteness properties \cite{williams2006gaussian}. Valid covariance functions can be manipulated to construct more complex versions by exploiting composition rules, e.g., sum, product, scaling rules \cite{williams2006gaussian}.
	
	\subsubsection{Conventional covariance functions in structural dynamics}           
\begin{figure} [!ht]
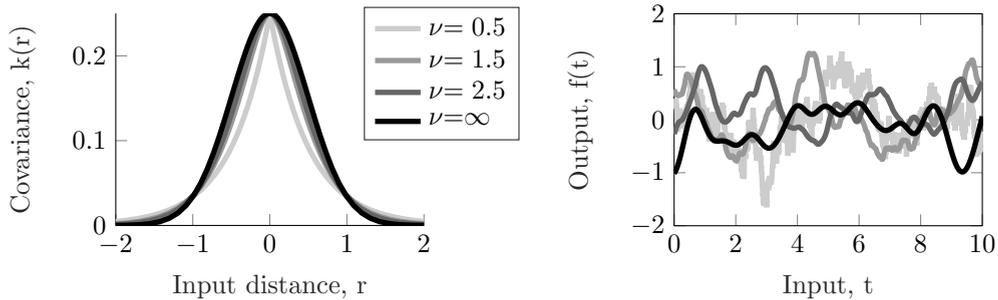

		\hspace{-4.4em}
		\begin{subfigure}{0.3\textwidth}
			\includestandalone{matern_nu}
		\end{subfigure}
		\hspace{6.4em}
		\begin{subfigure}{0.3\textwidth}
			\includestandalone{GP_nu}
		\end{subfigure}
		\caption{Matérn covariance functions with $\sigma = l = 0.5$ as a function of the input distance $r=t-t'$, for different values of $\nu$ (left). Realizations drawn from \acsp{GP} with Matérn covariance functions ($\sigma = l = 0.5$) for different values of $\nu$ (right). }
		\label{fig:matern_nu}
	\end{figure}

	Isotropic covariance functions are commonly adopted for regression via \acp{GP} in the structural dynamics context. The most common types of covariance functions are reported below for time-domain one-dimensional \acp{GP}.
	\paragraph{Squared exponential covariance function:}\mbox{} 
	
	\begin{equation}
		k(\tau)=\sigma^2 exp\left( - \frac{\tau^2}{2 l^2} \right), \quad \tau = |t-t'|
	\end{equation}
	where $l$ is the characteristic length-scale controlling the bandwidth of the resulting process and $\sigma$ is the signal variance controlling the magnitude of the process. This covariance function is infinitely differentiable, which implies infinite mean square differentiability of the \ac{GP} with a squared exponential covariance function. The latter has therefore a very smooth shape.
	\paragraph{Exponential covariance function:}\mbox{} 
	
	\begin{equation}
		k(\tau)=\sigma^2 exp\left( - \frac{\tau}{2 l} \right), \quad \tau = |t-t'|
	\end{equation}
	where $l$ is the characteristic length-scale and $\sigma$ is the magnitude-scale. This type of functions generate mean square continuous but not mean square differentiable \acp{GP}, which thus have a non-smooth shape.
	\paragraph{Matérn covariance function:}\mbox{} 
	
	\begin{equation}
		k(\tau; \nu, \sigma, l) = \sigma^2 \frac{2^{1-\nu}}{\Gamma(\nu)} \left(\frac{\sqrt{2\nu}}{l} \tau \right)^{\nu} \text{K}_{\nu} \left(\frac{\sqrt{2\nu}}{l} \tau \right), \quad \tau = |t-t'|
	\end{equation}
	where $l$ is the length-scale hyperparameter, $\sigma$ is the magnitude-scale hyperparameter, $\nu$ is the smoothness parameter, $\Gamma(\nu)$ is the Gamma function and $K_{\nu}$ is a modified Bessel function of second kind \cite{abramowitz1988handbook}. For half-integer values of $\nu$, i.e., non-negative integer values of $p=\nu-\frac{1}{2}$, the covariance function can be written as the product of a decaying exponential and a polynomial of order $p$ \cite{abramowitz1988handbook}:
	\begin{equation}\label{eq:matern}
		k_{\nu = p+1/2} = \sigma^2 exp ( \frac{- \sqrt{2\nu} \tau}{l}) \frac{\Gamma(p+1)}{\Gamma(2p+1)} \sum_{i=0}^{p} \frac{(p+i)!}{i!(p-1)!} (\frac{\sqrt{8\nu}\tau}{l})^{p-i}.
	\end{equation}
	The smoothness parameter $\nu$ determines the decaying rate of the covariance function, its differentiability and the mean square differentiability of the corresponding \ac{GP}. For $p=0$ ($\nu = 0.5$),  the Matérn covariance function corresponds to an exponential covariance function, i.e., it is continuous but not differentiable. For $\nu \rightarrow \infty$, the function converges to a squared exponential covariance function, i.e., it is infinitely differentiable. More generally, Matérn covariance functions generate \acp{GP} that are $\nu-1$ times mean square differentiable \cite{williams2006gaussian}. 
	These concepts are visually demonstrated by \autoref{fig:matern_nu}, which displays Matérn covariance functions with assigned $l$ and $\sigma$ for different $\nu$ values (left) and corresponding random \ac{GP} realizations (right). \Autoref{fig:matern_nu} reports the square exponential and the exponential covariance functions as limit cases for the Matérn class, respectively achieved for $\nu = 0.5$ and $\nu \rightarrow \infty$. Additionally, \autoref{fig:matern_nu} provides a visible insight regarding the functions differentiability scaling with $\nu$.

	\subsubsection{Alternative covariance functions in structural dynamics}
	The previously presented covariance functions are the most commonly employed for \ac{GP} regression in the structural dynamics context. In this section, several alternative covariance functions are extensively described.
	\paragraph{Periodic covariance functions}\mbox{} 
	
	Periodic stationary covariance functions can be derived via the so-called warping method, i.e., introducing a periodic mapping $u(t)$ of the input $t$ and using a stationary covariance function in the $u$ space \cite{solin2014explicit}. A typical choice for the periodic function is $u={\left(sin(t),cos(t)\right)}^T$, for which the following stationary property holds:
	\begin{equation}
		\label{eq:periodic_stat}
		||u(t)-u(t^{\prime } )||^2 =(sin(t)-sin(t^{\prime } ))^2 +(cos(t)-cos(t^{\prime } ))^2 +4sin^2 \left(\frac{t-t^{\prime } }{2}\right).
	\end{equation}

\begin{figure} [!ht]
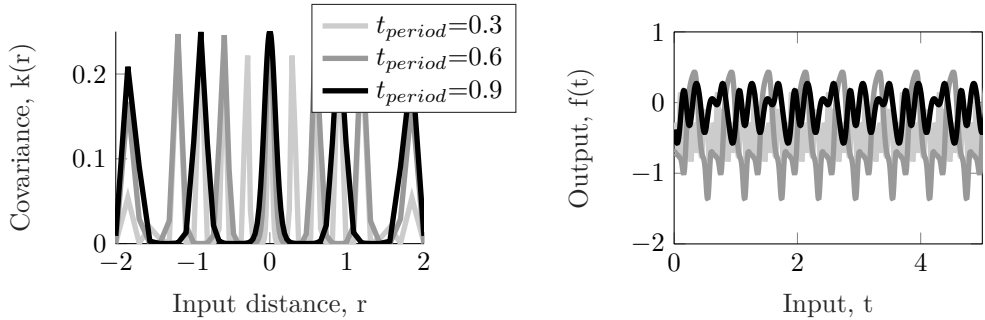

		\hspace{-4.4em}
		\begin{subfigure}{0.3\textwidth}
			\includestandalone{periodic_period}
		\end{subfigure}
		\hspace{6.4em}
		\begin{subfigure}{0.3\textwidth}
			\includestandalone{GP_period}
		\end{subfigure}
		\caption{Periodic covariance functions with $\sigma = l = 0.5$ as a function of the input distance $r=t-t'$, for different values of $t_{period}$ (left). Realizations drawn from \acsp{GP} with periodic covariance functions ($\sigma = l = 0.5$) for different values of $t_{period}$ (right). }
		\label{fig:periodic_cov}
	\end{figure}

	A commonly adopted periodic covariance function is the so-called canonical periodic covariance function, which is derived as a squared exponential covariance in the previously introduced $u$ space: 
	\begin{equation}
		\label{eq:periodic_can}
		k(\tau ;\sigma ,l,\omega_0 )=\sigma^2 exp\left(-\frac{2sin^2 \left(\frac{\omega_0 \tau }{2}\right)}{l^2}\right)
	\end{equation}
	where $\tau =|t-t^{\prime } |$, $l$ is the length-scale hyperparameter, $\sigma$ the magnitude-scale hyperparameter and  $\omega_{0\;}$ the frequency. It is common practice to refer to the period length parameter, i.e., $t_{period} =2\pi /\omega_0$, rather than the frequency.
	\Autoref{fig:periodic_cov} illustrates a representation of the canonical periodic covariance function (left) and corresponding \ac{GP} realization at several values of $t_{period}$.

	\paragraph{Quasiperiodic covariance functions}\mbox{} 
		
\begin{figure} [H]
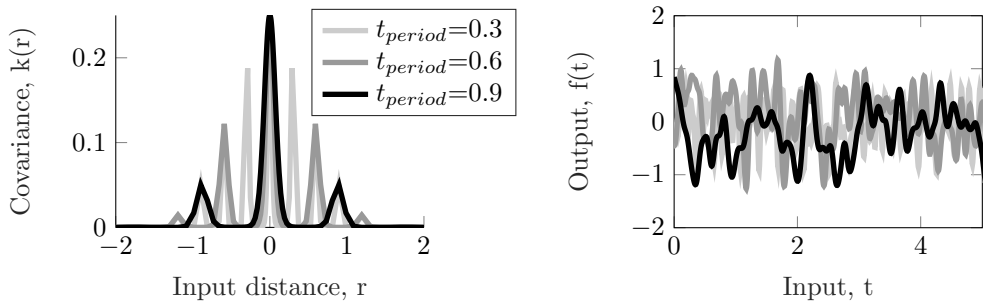

		\hspace{-4.4em}
		\begin{subfigure}{0.3\textwidth}
			\includestandalone{quasiperiodic_period}
		\end{subfigure}
		\hspace{6.4em}
		\begin{subfigure}{0.3\textwidth}
			\includestandalone{GP_quasiperiod}
		\end{subfigure}
		\caption{Quasiperiodic covariance functions with $\sigma_{per} = l_{per} = \sigma_{se} = l_{se} = 0.5$ as a function of the input distance $r=t-t'$, for different values of $t_{period}$ (left). Realizations drawn from \acsp{GP} with periodic covariance functions ($\sigma_{per} = l_{per} = \sigma_{se} = l_{se} = 0.5$) for different values of $t_{period}$ (right). }
		\label{fig:quasiperiodic_cov}
	\end{figure}

	In practical applications, using a periodic covariance function may result in a too strict assumption. To account for variability effects, a quasiperiodic covariance function can be adopted. This type of covariance function is obtained from the product of a periodic covariance function and a covariance function with long characteristic length scale, i.e., Matérn covariance functions. This feature allows for the introduction of a slow decay effect, i.e., damping effect, which is controlled by the smoothness parameter of the Matérn covariance function as described in \autoref{fig:matern_nu}. \Autoref{fig:quasiperiodic_cov} (left) shows a representation of a quasiperiodic covariance function obtained from the product of a canonical periodic function and a squared exponential covariance function (Matérn with $\nu \to \infty \;$) for several values of $t_{period}$. The relative \ac{GP} realizations are reported in \Autoref{fig:quasiperiodic_cov} (right). The visual information provided by \autoref{fig:quasiperiodic_cov} proves that the quasiperiodic covariance function is a damped version (with damping dependent on the smoothness of the Matérn covariance function) of the periodic covariance function. 
	\paragraph{Constant covariance functions}\mbox{} 
	
\begin{figure} [!ht]
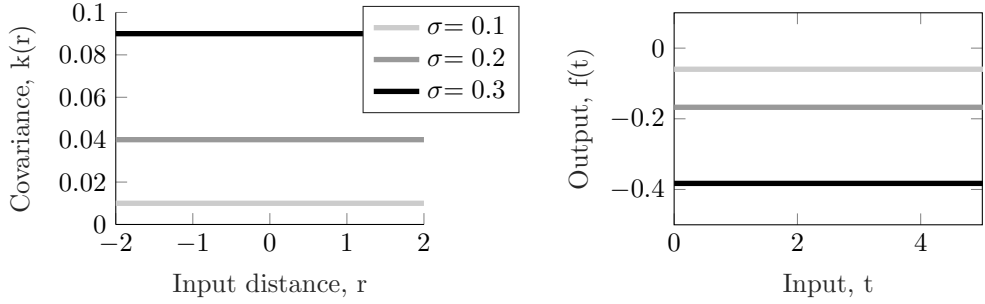

		\hspace{-4.4em}
		\begin{subfigure}{0.3\textwidth}
			\includestandalone{const_sigma}
		\end{subfigure}
		\hspace{6.4em}
		\begin{subfigure}{0.3\textwidth}
			\includestandalone{GPconst_sigma}
		\end{subfigure}
		\caption{Constant covariance functions as a function of the input distance $r=t-t'$, for different values of $\sigma$ (left). Realizations drawn from \acsp{GP} with constant covariance functions for different values of $\sigma$ (right). }
		\label{fig:constant_cov}
	\end{figure}

	Constant covariance functions are conventionally adopted to model biases in \acp{GP}. A constant covariance function is defined as:
	\begin{equation}
		k(t,t^{\prime})=\sigma^2
	\end{equation}
	where $\sigma \;$ is the magnitude-scale hyperparameter. \Autoref{fig:constant_cov} depicts constant covariance functions and relative \ac{GP} realizations for several values of the magnitude scale hyperparameter.
	\vspace{0.7em}
	\paragraph{Biased quasiperiodic covariance functions}\mbox{} 
	
\begin{figure} [!ht]
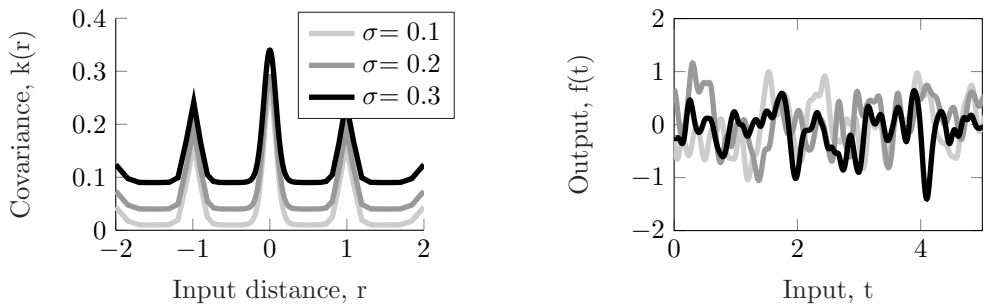

		\hspace{-4.4em}
		\begin{subfigure}{0.3\textwidth}
			\includestandalone{biasedquasper_sigma}
		\end{subfigure}
		\hspace{6.4em}
		\begin{subfigure}{0.3\textwidth}
			\includestandalone{GPbiasedquasper_sigma}
		\end{subfigure}
		\caption{Biased quasiperiodic covariance functions as a function of the input distance $r=t-t'$, for different values of $\sigma_{constant}$ (left). Realizations drawn from \acsp{GP} with biased quasiperiodic covariance functions for different values of $\sigma_{constant}$ (right). }
		\label{fig:biasedquasper_cov}
	\end{figure}

	A biased quasiperiodic covariance function is obtained from a constant and a quasiperiodic covariance function as:
	\begin{equation}
		k = k_{constant} + k_{quasiperiodic}.
	\end{equation}
	The sum rule is thus exploited to build a more complex covariance function in which a constant offset is added on top of the dynamics, which are fully modeled via the quasiperiodic function. This operation is commonly adopted in regression problems where data are distributed around a bias, which would not be detected without the introduction of a static term via the constant covariance function. A visualization of the biased quasiperiodic covariance function for several values of the constant covariance magnitude-scale is reported in \autoref{fig:biasedquasper_cov} (left), while \autoref{fig:biasedquasper_cov} (right) displays the respective \ac{GP} realizations.
		\vspace{0.7em}
	\paragraph{Linear covariance functions}\mbox{} 

\begin{figure} [!ht]
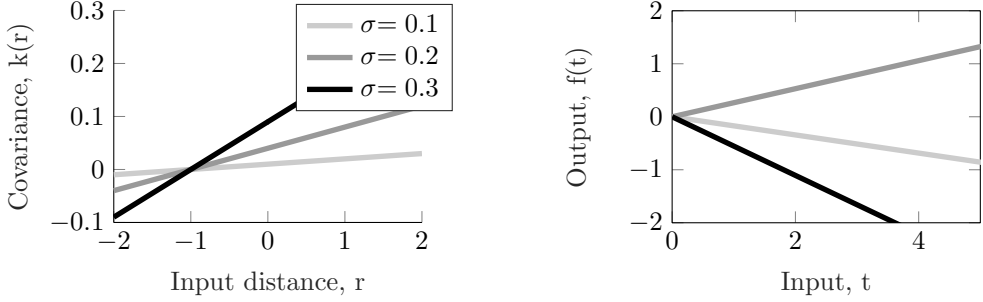

		\hspace{-4.4em}
		\begin{subfigure}{0.3\textwidth}
			\includestandalone{linear_sigma}
		\end{subfigure}
		\hspace{6.4em}
		\begin{subfigure}{0.3\textwidth}
			\includestandalone{GPlinear_sigma}
		\end{subfigure}
		\caption{Linear covariance functions as a function of the input distance $r=t-t'$, for different values of $\sigma$ (left). Realizations drawn from \acsp{GP} with linear covariance functions for different values of $\sigma$ (right). }
		\label{fig:linear_cov}
	\end{figure}

	Linear covariance functions can be adopted for regression via \acp{GP} to model linear trends in data. A linear covariance function is defined as:
	\begin{equation}
		k(t,t^{\prime } )=\sigma^2 tt^{\prime }
	\end{equation}
	where $\sigma$ is the magnitude-scale hyperparameter. \Autoref{fig:linear_cov} illustrates linear covariance functions and relative \ac{GP} realizations for several values of the magnitude-scale hyperparameter.

	\paragraph{Wiener covariance functions}\mbox{} 
	
\begin{figure} [!ht]
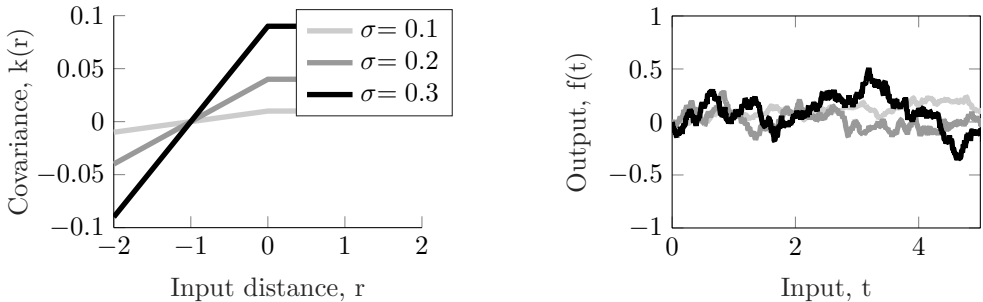

		\hspace{-4.4em}
		\begin{subfigure}{0.3\textwidth}
			\includestandalone{wiener_sigma}
		\end{subfigure}
		\hspace{6.4em}
		\begin{subfigure}{0.3\textwidth}
			\includestandalone{GPwiener_sigma}
		\end{subfigure}
		\caption{Wiener covariance functions as a function of the input distance $r=t-t'$, for different values of $\sigma$ (left). Realizations drawn from \acsp{GP} with Wiener covariance functions for different values of $\sigma$ (right). }
		\label{fig:wiener_cov}
	\end{figure}

	The Wiener process, i.e., a pure Brownian motion, is a widely adopted non-stationary process. Its covariance function is defined as:
	\begin{equation}
		\label{eq:wiener_cov}
		k(t,t^{\prime } )=\sigma^2 min(t,t^{\prime } )
	\end{equation}
	valid for $t,t^{\prime } \ge 0$, with $\sigma$ being the magnitude-scale hyperparameter. A representation of the Wiener covariance function and relative \ac{GP} realizations for several values of $\sigma$ is provided in \autoref{fig:wiener_cov}. \Autoref{fig:wiener_cov} (right) proves that a covariance function such as \autoref{eq:wiener_cov} generates \ac{GP} samples that are in the form of random-walk models. In this sense, a \ac{GP} with a Wiener covariance function is an alternative to the direct random-walk expression adopted for input-state estimation within Kalman-type filters such as the \ac{AKF}.
	
	\section{Gaussian Process regression in state-space form} 
	\label{sec:regr_statespace}
	The main drawback of the direct solution for regression via \acp{GP} (see \autoref{eq:GPposterior}) consists in the computational complexity scaling as $O(n^3)$, where $n$ in this work corresponds to the number of time instances in the observations used for regression. In actual monitoring measurements, the collected data is relatively big (and highly sampled), resulting in a large amount of time required for \ac{GP} regression. To remedy this problem, a sequential inference scheme, whose computational complexity scales linearly with $n$, can be implemented using Kalman filtering and smoothing \cite{hartikainen2010kalman}. In view of adopting a sequential approach, the \ac{GP} needs to be formulated as a dynamical system rather than adopting the conventional kernel formalism. This consists in expressing a \ac{GP} as a solution of the following $m^{th}$ order \ac{LTI} \ac{SDE}:
	\begin{equation}\label{eq:GP_LTI_SDE}
		\frac{d^m f(t)}{dt^m }+a_{m-1} \frac{d^{m-1} f(t)}{dt^{m-1} }+\dots +a_1 \frac{df(t)}{dt}+a_0 f(t)=w(t) 
	\end{equation}
	where $w$ is a zero-mean continuous-time white noise process. Every solution $f(t)$ of \autoref{eq:GP_LTI_SDE} is a sample function of a \ac{GP} defined by a certain covariance function $k$. Collecting the derivative terms in \autoref{eq:GP_LTI_SDE} builds the so-called companion form, i.e., a linear state-space model to be employed in the recursive inference scheme:
	\begin{equation} 
		\label{eq:GPSF}
		\begin{aligned}
			\begin{cases}
				\dot{\mathbf{z}}(t) & = \mathbf{F} \mathbf{z}(t) + \mathbf{L} w(t) \\
				f(t)                & = \mathbf{H}\mathbf{z}(t)
			\end{cases}
		\end{aligned}
	\end{equation}
	where the input state $\mathbf{z} \in \mathbb{R}^m$ and the matrices $\mathbf{F} \in \mathbb{R}^{m \times m}$, $\mathbf{L} \in \mathbb{R}^{m \times 1}$ and $\mathbf{H} \in \mathbb{R}^{1 \times m}$ are given by:
	
	\begin{equation}\label{eq:GPSFmat}
		\resizebox*{1\textwidth}{!}{$
			\mathbf{z} (t) = \left[\begin{array} {c}
				f(t)                           \\
				\frac{d f(t)}{d t}             \\
				\vdots                               \\
				\frac{d^{m-2} f(t)}{d t^{m-2}} \\
				\frac{d^{m-1} f(t)}{d t^{m-1}}
			\end{array}\right], \quad
			\mathbf{F}=\left[\begin{array}{ccccc}
				0      & 1      & 0      & \ldots & 0        \\
				0      & 0      & 1      & \ldots & 0        \\
				\vdots & \vdots & \vdots & \ddots & \vdots   \\
				0      & 0      & 0      & \ldots & 1        \\
				-a_{0} & -a_{1} & \ldots & \ldots & -a_{m-1}
			\end{array}\right], \quad \mathbf{L}=\left[\begin{array}{c}
				0      \\
				0      \\
				\vdots \\
				0      \\
				1
			\end{array}\right], \quad \mathbf{H}=\left[\begin{array}{lllll}
				1 & 0 & \ldots & 0 & 0
			\end{array}\right].
			$}
	\end{equation}
	
	To be employed for \ac{GP} sequential inference, the first equation in \autoref{eq:GPSF} must be transformed in a discrete-time model of the form: 
	\begin{equation}\label{eq:GP_discr}
		\mathbf{z}_k = \mathbf{F}_{d_{k-1}}\mathbf{z}_{k-1}+\mathbf{q}_{k-1}, \quad \mathbf{q}_{k-1}\sim\mathcal{N}(\mathbf{0},\mathbf{Q}_{k-1})
	\end{equation}
	where the state transition and process noise covariance matrices can be calculated analytically as:
	\begin{equation} \label{eq:GP_discr_mat}
		\begin{aligned}
			&\mathbf{F}_{d_{k-1}} = exp(\mathbf{F}\Delta t)\\
			&\mathbf{Q}_{k-1} = \int_{0}^{\Delta t} exp(\mathbf{F}(\Delta t - \tau))\mathbf{L} q_c \mathbf{L}^{T} exp(\mathbf{F}^{T}(\Delta t - \tau)) d\tau
		\end{aligned}
	\end{equation}
	where $\Delta t = t_{k}- t_{k-1}$ \cite{sarkka2006recursive}. The second equation in \autoref{eq:GPSF} instead represents the intrinsically discrete measurement model, on top of which the noise term $\varepsilon_k$ (introduced in \autoref{eq:GP_eps}) must be added to account for noisy observations:
	\begin{equation}\label{eq:GP_output_eq}
		f_k = \mathbf{H}\mathbf{z}_{k}+\varepsilon_k, \quad \varepsilon_k \sim \mathcal{N}(0,\sigma_n^2).
	\end{equation}
	
	Given the model postulated by \multiref{eq:GP_discr}{eq:GP_output_eq}, \ac{GP} sequential inference consists in estimating the state vector $\mathbf{z}_k$ at any $k$ given the available measurements for $k = 0:\Delta t:T$. This operation can be performed in two sequential steps: the filtered posterior distribution of $\mathbf{z}$ is computed via a \ac{KF}, then a \ac{RTS} smoother is employed to evaluate the smoothing distribution of $\mathbf{z}$ \cite{hartikainen2010kalman}. Both the \ac{KF} and the \ac{RTS} smoother algorithms scale with  $O(m^3n)$ in computational complexity, hence allowing for substantial computational time reduction with respect to the \ac{GP} regression naive implementation. The same benefit applies for covariance hyperparameters optimization via the minimization of the negative log-likelihood since the marginal likelihood in \autoref{eq:marginal_opt_GP} can be sequentially evaluated as a by-product of the \ac{KF} update \cite{sarkka2013bayesian}.
	
	The \ac{SDE} representation of the \ac{GP} in \autoref{eq:GP_LTI_SDE} is determined by the \ac{GP} kernel (covariance function). According to the adopted class of covariance function, either an exact \ac{LTI} state-space \ac{SDE} representation or an approximated version can be retrieved. A collection of state-space \ac{SDE} representations for the previously presented covariance functions is reported in the following. 
	
	\paragraph{Matèrn class}\mbox{} 
	
	For \acp{GP} with a covariance function of the Matérn type, an exact state-space representation can be extracted following the rational spectrum approach. The latter derives from the assumption of a rational spectral density for this class of covariance functions. In order to introduce the concept of spectral density, it is convenient to define covariance functions for zero-mean complex-valued \acp{GP}: $k(\mathbf{t},\mathbf{t'})=\mathbb{E}\left[f(\mathbf{t})f^*(\mathbf{t'})\right]$ with $\mathbf{t,t'}\in\mathbb{C}^d$. Bochner's theorem \cite{williams2006gaussian,gikhman2004theory, stein1999interpolation} states that a complex-valued function $k$ defined on $\mathbb{R}^d$ is the covariance function of a weakly stationary square continuous complex-valued \ac{GP} in $\mathbb{R}^d$ if and only if it can be represented as:
    \begin{equation}\label{eq:spec_density_or}
		k( \bm{\tau} ) = \int_{ \mathbb{R}^d}^{} e^{2\pi i \mathbf{f} \cdot \bm{\tau}} \mu  (d\mathbf{f}) = \int_{ \mathbb{R}^d}^{} e^{2\pi i \mathbf{f} \cdot \bm{\tau}}  dF (\mathbf{f})
	\end{equation}
	where $\bm{\tau} = \mathbf{t}-\mathbf{t}'$, $\mu$ is a positive finite measure and $F(\mathbf{f})$ is defined such that $\mu  (d\mathbf{f})= dF (\mathbf{f})$. If $\mu$ is absolutely continuous, then $F$ is differentiable almost everywhere and $\mu  (d\mathbf{f})= S(\mathbf{f}) d\mathbf{f}$, where $S(\mathbf{f})$ is defined as the spectral density or power spectrum corresponding to $k$. Introducing $S(\mathbf{f})$ into \autoref{eq:spec_density_or} and exploiting Wiener-Khintchine theorem \cite{chatfield2003analysis}, $k$ and $S(\mathbf{f})$ can be defined as Fourier duals: 	
	\begin{equation}\label{eq:spec_density}
		k( \bm{\tau} ) = \int S(\mathbf{f}) e^{2\pi i \mathbf{f} \cdot \bm{\tau}} d\mathbf{f}, \quad S(\mathbf{f}) = \int k(\bm{\tau}) e ^{-2\pi i \mathbf{f} \cdot \bm{\tau}} d \bm{\tau}.
	\end{equation}
	The relationships in \autoref{eq:spec_density} can be expressed for a one-dimensional temporal problem as:
	\begin{equation}\label{eq:spec_density_1D}
		k( \tau ) = \int S(f) e^{2\pi i f \cdot \tau} df, \quad S(f) = \int k(\tau) e ^{-2\pi i f \cdot \tau} d \tau, \quad f = \frac{\omega}{2\pi}.
	\end{equation}
	
\begin{figure} [!ht]
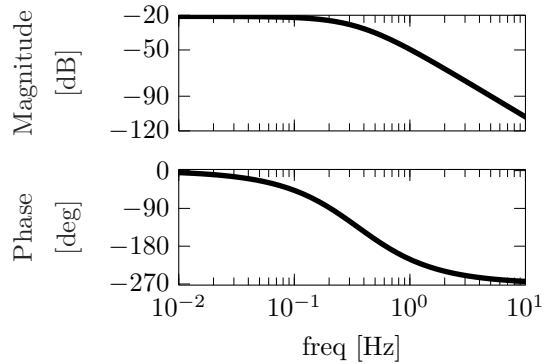

		\centering
		\includestandalone{matern_tf}
		\caption{\acs{GP} with Matérn covariance function: \acs{TF} of the equivalent \acs{SDE} system}
		\label{fig:matern_tf}
	\end{figure}

	 Following \autoref{eq:spec_density_1D}, this paragraph will make use of the \ac{FT} to link time and frequency domains and compute the spectral density for a Matérn covariance function. In particular, when $k$ falls in the Matérn class of functions, its \ac{FT} yields:
	\begin{equation}
		S(\omega )=\sigma^2 \frac{2\pi^{1/2} \Gamma (\nu +1/2)}{\Gamma (\nu )}\lambda^{2\nu } (\lambda^2 +\omega^2 )^{-(\nu +1/2)}, \quad \lambda =\frac{\sqrt{2\nu }}{l}=\frac{\sqrt{2(p+1)}}{l}.
	\end{equation}
	Under the assumption of $\nu$ being a half-integer, the spectral density can be reformulated as:
	\begin{equation}
		S(\omega )=\sigma^2 \frac{2\pi^{1/2} \Gamma (p+1)}{\Gamma (p+1/2)}\lambda^{2p+1} (\lambda^2 +\omega^2 )^{-(p+1)}. 
	\end{equation}
	After grouping all constant terms in a single constant $q_c$, the spectral density can be factorized in a rational fraction form:
	\begin{equation}\label{eq:spec_density_rational}
		S(\omega )=q_c (\lambda^2 +\omega^2 )^{-(p+1)} =q_c (\lambda +i\omega )^{-(p+1)} (\lambda -i\omega )^{-(p+1)} =q_c G(i\omega )G(-i\omega ). 
	\end{equation}
	\Autoref{eq:spec_density_rational} can be interpreted as the spectral density of the output of a system with \ac{TF} $G\left(i\omega \right)={\left(\lambda +i\omega \right)}^{-\left(p+1\right)}$, excited by an input $w\left(t\right)$ (white noise with spectral density $q_c$):
	\begin{equation}\label{eq:spec_density_system}
		F(\omega )=G(i\omega )W(\omega ). 
	\end{equation}
	In \autoref{eq:spec_density_system}, $F\left(\omega \right)$ (\ac{FT} of $f\left(t\right)$) is the system output with spectral density $S\left(\omega \right)$ and $W\left(\omega \right)$ is the \ac{FT} of the exciting term $w\left(t\right)$, i.e., $W=\sqrt{\left(q_c \right)}$. The function $f\left(t\right)$ is the \ac{GP} with Matérn covariance function $k$, i.e., with desired spectral density $S(\omega )$.
	For the Matérn class of covariance functions, the \ac{TF} $G\left(i\omega \right)$ has a rational form of the type $G(i\omega )=((i\omega )^m +a_{m-1} (i\omega )^{m-1} +...+a_0)^{-1}$. Hence, \autoref{eq:spec_density_system} can be reformulated as:
	\begin{equation}\label{eq:spec_density_final}
		F(\omega )[(i\omega )^m +a_{m-1} (i\omega )^{m-1} +...+a_0 ]=W(\omega).
	\end{equation}
	The inverse \ac{FT} of \autoref{eq:spec_density_final} provides the \ac{LTI} \ac{SDE} in \autoref{eq:GP_LTI_SDE}, which thus models the evolution in time of sample functions drawn from a \ac{GP} with covariance $k$ expressed by \autoref{eq:matern} \cite{hartikainen2010kalman}. The associated \ac{LTI} \ac{SDE} can be easily translated into the linear state-space model postulated by \multiref{eq:GPSF}{eq:GPSFmat}, where the involved coefficients can be determined by analyzing the form of the Matérn covariance functions \ac{TF}:
	\begin{equation}\label{eq:matern_TF}
		G(i\omega )=\frac{1}{(\lambda +i\omega )^{p+1} }. 
	\end{equation}
	For example, \autoref{eq:matern_TF} for $\nu =\frac{5}{2}\left(p=2\right)$ has the form:
	\begin{equation}
		\label{eq:G_matern}
		G(i\omega )=\frac{1}{(\lambda +i\omega )^3}
	\end{equation}
	where the denominator can be rewritten as $\lambda^3 +(i\omega )^3 +3\lambda^2 (i\omega )+3\lambda (i\omega )^2$. Hence, the coefficients $a_m ,\ldotp \ldotp \ldotp ,a_0 \;\textrm{with}\;m=3$ are: $a_3 =\;1,a_2 =3\lambda ,a_1 =3\lambda^{2\;} ,a_0 =\lambda^3$ and the resulting \ac{SDE} is: 
	\begin{equation}\label{eq:GP_LTI_SDE3}
		\frac{d^3 f(t)}{dt^3 }+3\lambda \frac{d^2 f(t)}{dt^2 }+3\lambda^2 \frac{df(t)}{dt}+\lambda^3 f(t)=w(t).
	\end{equation}
	The roots of this \ac{SDE} are real and coincident with multiplicity equal to 3: $\omega_1 =\omega_2 =\omega_3 =-\lambda$. These correspond to the poles of the system with \ac{TF} postulated by \autoref{eq:G_matern}, whose bode plot is displayed in \autoref{fig:matern_tf}. The system poles can be computed by setting the denominator of \autoref{eq:G_matern} equal to zero and solving the so-called characteristic equation. This result can be extended to all orders as roots of \acp{SDE} associated to Matérn kernels are always equal to $-\lambda$ with multiplicity $p+1$. Therefore, the system with \ac{TF} in \autoref{eq:matern_TF} is ``critically damped'' for any integer $p$, i.e., the time waveform is an exponentially decaying function with a decaying rate driven by the value of $\lambda =\frac{\sqrt{2\nu }}{l}=\frac{\sqrt{2(p+1)}}{l}$.
	This result is confirmed by the \ac{TF} shape visible in \autoref{fig:matern_tf} for $p=2$. The higher $\lambda \;$ (and $\nu$) is, the faster will be the decay. A mathematical proof to this is provided by the homogeneous solution for \autoref{eq:GP_LTI_SDE3}, which could only correspond to: 
	\begin{equation}\label{eq:GP_LTI_SDEsol}
		g\left(t\right)=\tilde{g} \;e^{\left\lbrace -\lambda t\right\rbrace}
	\end{equation}
	where $\tilde{g}$ can be determined by substituting this expression into the unforced \ac{SDE}, yielding (for $p=2$):
	\begin{equation}\label{eq:GP_LTI_SDEhom}
		\frac{d^3 \tilde{g} }{dt^3 }e^{-\lambda t} =0.
	\end{equation}
	Since the exponential function never equals 0, the only possible solution for \autoref{eq:GP_LTI_SDEhom} is given by $\frac{d^3 \tilde{g} }{dt^3 }=0$. This expression can be double integrated to get $\tilde{g} =c_1 +c_2 t+c_3 t^2$. Therefore, the explicit form of \autoref{eq:GP_LTI_SDEhom} is given by: 
	\begin{equation}
		g=(c_1 +c_2 t+c_3 t^2 )e^{-\lambda t}
	\end{equation}
	where constants can be determined by applying the initial conditions on $g,\frac{d\;g}{d\;t},\frac{d^2 g}{d\;t^{2\;} }$. This homogeneous solution should be convoluted with the forcing term $w\left(t\right)$ to get the forced response of the system with \ac{TF} $G\left(i\omega \right)$: $f\left(t\right)=g\left(t\right)*w\left(t\right)$ \cite{tenenbaum1985ordinary}. The same result could also be obtained by exploiting partial fraction expansion from control theory to manipulate the \ac{TF} $G\left(\omega\right)$. 
	\paragraph{Periodic class}\mbox{} 
	
\begin{figure} [!ht]
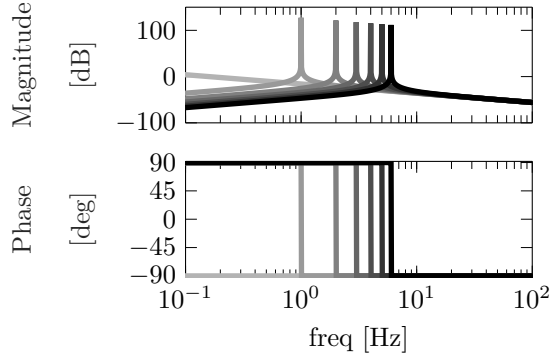

		\centering
		\includestandalone{periodic_tf}
		\caption{\acs{GP} with periodic covariance function: \acsp{TF} of the equivalent \acs{SDE} system}
		\label{fig:periodic_tf}
	\end{figure}

	A periodic and symmetric covariance function can be expanded into a convergent Fourier series of the form:
	\begin{equation}
		\label{eq:periodic_fourier}
		k(\tau )=\sum_{j=0}^{\infty } q_j^2 cos(j\omega_0 \tau ).
	\end{equation}
	The spectral density corresponding to \autoref{eq:periodic_fourier} consists of delta peaks at the harmonic frequencies defined by $\omega_0$, i.e., the angular frequency defining the periodicity of the function:
	\begin{equation}
		S_p (\omega )=\sum_{j=0}^{\infty } q_j^2 \pi \left\lbrack \delta (\omega -j\omega_0 )+\delta (\omega +j\omega_0 )\right\rbrack. 
	\end{equation}	
	This spectral density does not exhibit a rational form, thus implying that the procedure adopted for Matérn covariance functions can not be used for building the state-space representation of a \ac{GP} with a periodic covariance function. However, each $j^{th}$ term in \autoref{eq:periodic_fourier} can be seen as the covariance function of the sum of statistically independent resonators $\sum_{j=0}^{\infty } x_j (t)$ such that:
	\begin{equation}
		f_j (t)={\left(x_j (t),y_j (t)\right)}^T 
	\end{equation}		
	with the initial condition $f_j (0)\sim \mathcal{N}\left(0,q_j^2 I\right)$ and the following differential equations defining the harmonic oscillator:
	\begin{equation}
		\label{eq:periodic_diff}
		\begin{aligned}
			&\frac{dx_j (t)}{dt}=-j\omega_0 y_j (t), \quad \frac{dy_j (t)}{dt}=j\omega_0 x_j (t).
		\end{aligned}
	\end{equation}
	Solving \autoref{eq:periodic_diff} yields:
	\begin{equation}
		\label{eq:periodic_sol}
		x_j (t)=x_j (0)cos(\omega_0 jt)-y_j (0)sin(\omega_0 jt)
	\end{equation}
	with associated covariance $\mathbb{E}\left\lbrack x_j (t)x_j (t+\tau )\right\rbrack =q_j^2 cos(j\omega_0 \tau )$. As a result, these processes are deterministic with initial state drawn from a Gaussian distribution.
	The corresponding state-space formulation features block diagonal matrices composed of $J$ matrices as follows:
	\begin{equation}
		\begin{aligned}
			&F_j =\left\lbrack \begin{array}{cc}
				0 & -\omega_{0\;} j\\
				\omega_{0\;} j & 0
			\end{array}\right\rbrack, \quad 
			L_j =I_2, \quad	P_{\infty ,j} =q_j^2 I_2, \quad Q_c =0, \quad H_j =\left\lbrack \begin{array}{cc}
				1 & 0
			\end{array}\right\rbrack.
		\end{aligned}		
	\end{equation}
	In order to retrieve the form in \autoref{eq:periodic_fourier} for the canonical covariance function in \autoref{eq:periodic_can}, the relation $2\sin^{2\;} \left(\frac{\tau }{2}\right)=1-\cos \left(\tau \right)$ must be substituted in the canonical covariance function expression. After expanding the exponential function via a Taylor series, the following expression is obtained:
	\begin{equation}
		k(\tau )=exp(-l^2 )\sum_{j=0}^{\infty } \frac{1}{j!}cos^j (\tau ).
	\end{equation}
	This series can be truncated at an order $J\;$ and the powers of cosines can be written as summation of cosine terms with multiplied angles. This leads to the following expression for the canonical periodic covariance function:
	\begin{equation}
		\label{eq:periodic_can_cov}
		k(\tau )=\sum_{j=0}^J {\hat{q} }_{j,J}^2 cos(j\tau )
	\end{equation}
	with coefficients ${\hat{q} }_{j,J}^2 =\frac{2}{exp(l^{-2} )}\sum_{i=0}^{\lfloor \frac{J-j}{2}\rfloor } \frac{(2l^2 )^{-j-2i} }{(j+i)!i!}$, where $j=1,\ldotp \ldotp \ldotp J$. An approximation of this expression can be retrieved taking the limit $J\to \infty \;$ \cite{solin2014explicit}.
	The eigenvalues associated to the dynamical system corresponding to a \ac{GP} with covariance function described by \autoref{eq:periodic_can_cov} are complex conjugates with zero real part (each pair corresponds to the eigenvalues associated to the $j^{\textrm{th}} \;$ component, with the first two eigenvalues being zero). Referring to the single $j^{\textrm{th}\;}$ component, a differential equation with imaginary conjugate poles corresponds to an undamped system with frequency specified by the magnitude of the imaginary part. Solving the eigenvalue solution for the reconstructed state-space model yields eigenvalues that are multiples of $\omega_0 =2\pi /t_{period}$. The solution of the $j^{\textrm{th}\;}$ differential equation is therefore: $x_j (t)=x_j (0)cos(\omega_0 jt)-y_j (0)sin(\omega_0 jt)$, as anticipated in \autoref{eq:periodic_sol}. It follows that the resulting \ac{GP} will be a summation of $J\;$ harmonic components (harmonics of $\omega_{0\;}$). The latters are visible in \autoref{fig:periodic_tf}, which shows the bode plot of the \ac{TF} for the dynamical system associated to a \ac{GP} with canonical periodic covariance function with $\omega_0 = 1$ and truncation order $J=6$. \Autoref{fig:periodic_tf} proves that the \ac{TF} features a $0 Hz$ component and $J$ undamped harmonics (one for each $j^{\textrm{th}} \;$ component).
	
	\paragraph{Quasiperiodic class}\mbox{} 
	
	Following the kernel product rule, a quasiperiodic covariance function can be constructed as a product of a periodic and a Matérn covariance function. As a result, the Kronecker product can be exploited as follows to build the related state-space model:
	\begin{equation}
		\label{eq:quasiperiodic_ss}
		\begin{aligned}
			&F_j =F^q \otimes I_2 +I_q \otimes F_j^p, \quad
			&&L_j =L^q \otimes L_j^p,\quad
			Q_{c,j} =Q_c^q \otimes q_j^2 I_2\\
			&P_{\infty ,j} =P_{\infty }^q \otimes P_{\infty ,J}^P,\quad
			&&H_j =H^q \otimes H_j^P\\
		\end{aligned}
	\end{equation}
	where the $p$ and $q$ notations are respectively used for matrices associated with the periodic and the Matérn covariance function.
	
\begin{figure} [!ht]
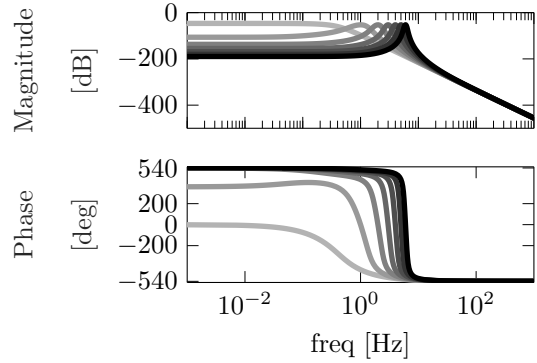

		\centering
		\includestandalone{quasiperiodic_tf}
		\caption{\acs{GP} with quasiperiodic covariance function: \acsp{TF} of the equivalent \acs{SDE} system}
		\label{fig:quasiperiodic_tf}
	\end{figure}

	The eigenvalue solution for the system in \autoref{eq:quasiperiodic_ss} provides, for each $j^{\textrm{th}\;} \;$ component, complex conjugate pairs of eigenvalues. The dynamical system representing a quasiperiodic covariance function thus corresponds to an underdamped system with solution of the type $x_j (t)=e^{\alpha t} (x_j (0)cos(\omega_0 jt)-y_j (0)sin(\omega_0 jt))$. This is confirmed by the bode plot of the \ac{TF} reported in \autoref{fig:quasiperiodic_tf}, which features a $0 Hz$ component and $J$ damped harmonics (one for each $j^{\textrm{th}} \;$ component).
	\paragraph{Constant class}\mbox{} 
	
\begin{figure} [!ht]
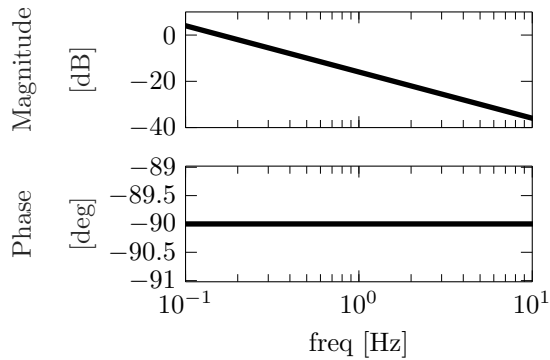

		\centering
		\includestandalone{constant_tf}
		\caption{\acs{GP} with constant, linear or Wiener covariance function: \acs{TF} of the equivalent \acs{SDE} system}
		\label{fig:const_wiener_tf}
	\end{figure}

	A constant covariance function is degenerate and the corresponding state-space model representation is given by the matrices:
	\begin{equation}
		\begin{aligned}
			&F=0, \quad L=1, \quad Q_c =0,\quad H=1,\quad P_0 =\sigma^2.
		\end{aligned}
	\end{equation}
	The bode plot for the \ac{TF} associated with this model is reported in \autoref{fig:const_wiener_tf}.
	
	\paragraph{Linear class}\mbox{} 
	
	A linear covariance function is degenerate and the corresponding state-space model is defined by the matrices: 
	\begin{equation}
		\begin{aligned}
			&F=\left\lbrack \begin{array}{cc}
				0 & 1\\
				0 & 0
			\end{array}\right\rbrack, \quad
			L=\left\lbrack \begin{array}{c}
				0\\
				1
			\end{array}\right\rbrack, \quad
			Q_c = 0, \quad
			H=\left\lbrack \begin{array}{cc}
				1 & 0
			\end{array}\right\rbrack, \quad
			P_0 =\sigma^2 \left\lbrack \begin{array}{cc}
				t_0^2  & t_0 \\
				t_0  & 1
			\end{array}\right\rbrack. 
		\end{aligned}
	\end{equation}
	The bode plot for the \ac{TF} associated with this model coincides with the bode plot related to the constant covariance function (see \autoref{fig:const_wiener_tf}).  
	
	\paragraph{Wiener class}\mbox{} 
	
	The state-space model representation of a Wiener process is defined by the matrices: 
	\begin{equation}
		\begin{aligned}
			&F=0, \quad L=1, \quad Q_c =\sigma^2,\quad H=1, \quad P_0 =0 \ldotp
		\end{aligned}
	\end{equation}
	The bode plot for the \ac{TF} associated with this model coincides with the bode plot related to the constant covariance function (see \autoref{fig:const_wiener_tf}).
	\section{Latent force modeling for Bayesian input-state estimation}
	\label{sec:LFM}
	
	The most common Kalman-based estimators, e.g., the \ac{AKF} or the \ac{DKF}, rely on a prior assumption regarding the dynamic model of the unknown input, which is constructed via a random-walk model. Although the random-walk is a widely adopted solution, it is sometimes not representative of the actual loading conditions a structure is subjected to. This results in the need of a user-dependent dedicated offline tuning procedure for the unknown input covariance, which strongly controls the force prediction accuracy. Indeed, for loads featuring complex dynamics, thus strongly deviating from the assumption of Brownian motion to which the random-walk belongs, the prediction accuracy may not reach satisfactory values. For this reason, this chapter proposes a prior transition equation for the unknown input shaped by a \ac{LFM} derived from regression via \acp{GP} using training data.
	
	\subsection{Latent Force Models}
	A more flexible and comprehensive alternative to the conventional random-walk for establishing a prior model for the unkwown input in a Kalman filtering setting consists in employing \acp{LFM} \cite{alvarez2009latent,pmid:24051729}, i.e., hybrid schemes that incorporate data-driven paradigms in a relatively simple mechanistic model. The key idea behind \acp{LFM} construction is to make use of a mechanistic model of the system which is augmented via data-driven techniques in order to provide enough flexibility to allow for the resulting model to fit the actual system even when mechanistic assumptions are not precisely met. According to this approach, any model such as the one in \autoref{eq:BDM}, in which the system is forced by latent functions can be referred as \ac{LFM} if the components of the forcing vector $\mathbf{u}$ are modeled as zero-mean independent stochastic processes. Specifically, in a \ac{GPLFM} framework, the $j$th component  $f^{j}$ of the forcing vector $\mathbf{u}$ of the system in \autoref{eq:BDM} is modeled as a zero-mean time-domain one-dimensional \ac{GP}.
	
	The approach proposed in \cite{alvarez2009latent,pmid:24051729} stems from the assumption that the structural system state vector $\mathbf{x}$ can be modeled as a multi-dimensional \ac{GP}:
	\begin{equation}\label{eq:states_GP}
		\mathbf{x}(t)\sim\mathcal{GP}(\mathbf{0},\mathbf{K}_{\mathbf{xx}}(t,t')). 
	\end{equation}
	In absence of the independence assumption of the forcing terms, correlated terms would be admitted within the forcing vector, which would thus need to be modeled by correlated \acp{GP}, i.e., with non-zero off-diagonal terms in the covariance matrix . \ac{GP} regression for full covariance matrices is addressed in literature via use of instantaneous mixing \cite{journel1976mining,goovaerts1997geostatistics,evgeniou2004regularized} or convolution \cite{ver1998constructing, boyle2004dependent,alvarez2011computationally} of a series of independent processes to construct correlated processes. These approaches require more advanced calculations and, consequently, additional computational effort. The independent forcing terms hypothesis is thus adopted in this work to simplify the construction of \acp{LFM}. This assumption does not introduce a too strict simplification within the context of Kalman-based input-state estimation algorithms, where prior information regarding the location and the direction of the unknown forcing terms is assumed to be available. Despite this simplification, the resulting \ac{LFM} turns into a \ac{GP} regression problem that still requires complex numerical integration for computing the covariance matrices, which may not always have a closed-form. A workaround to this problem consists in adopting the temporal state-space \ac{GP} formulation to model the latent forcing terms, which are then used to augment the system mechanistic model resulting in a practical joint state-space form. 
	\vspace{.6em}
	\subsection{Latent Force Models for joint input-state estimation}
	\label{subsec:LFMinputstate}
	The \ac{GPLFM} approach has been originally adopted for input-state estimation in a Kalman filtering framework in \cite{nayek2019gaussian} and further exploited in \cite{rogers2020application} for offline input-state-parameter estimation. An in-situ validation of the method has been provided in \cite{zou2022virtual} and an extension to identification of mechanical systems with discontinuous nonlinearity has been proposed in \cite{marino2023switching} via the use of switching \acp{GP}. The basic principle consists in adopting a \ac{GP} state-space representation derived to model the (unknown) latent forces acting on the system under the assumption of conventional covariance functions, i.e., squared exponential, exponential or Matérn type. The corresponding state-space model is then used in combination with the system state-space (mechanistic) model by augmenting the state vector $\mathbf{x}$ with the input state vector $\mathbf{z}$ defined in \autoref{eq:GPSFmat}, yielding the final augmented model:
	\begin{equation} \label{eq:GPLFMSSMstate}
		{\left[\begin{array}{c}
				\dot{\mathbf{x}}(t)   \\
				\dot{\mathbf{z}}^{(1)}(t) \\
				\dot{\mathbf{z}}^{(2)}(t) \\
				\vdots                    \\
				\dot{\mathbf{z}}^{\left(n_{i}\right)}(t)
			\end{array}\right]=\left[\begin{array}{ccccc}
				\mathbf{A} & \mathbf{b}_{1} \mathbf{H}^{(1)} & \mathbf{b}_{2} \mathbf{H}^{(2)} & \ldots & \mathbf{b}_{n_{i}} \mathbf{H}^{\left(n_{i}\right)} \\
				0              & \mathbf{F}^{(1)}                    & 0                                   & \ldots & 0                                                                        \\
				0              & 0                                   & \mathbf{F}^{(2)}                    & \ldots & 0                                                                        \\
				\vdots         & \vdots                              & \vdots                              & \ddots & \vdots                                                                   \\
				0              & 0                                   & 0                                   & \ldots & \mathbf{F}^{\left(n_{i}\right)}
			\end{array}\right]\left[\begin{array}{c}
				\mathbf{x}(t)   \\
				\mathbf{z}^{(1)}(t) \\
				\mathbf{z}^{(2)}(t) \\
				\vdots              \\
				\mathbf{z}^{\left(n_{i}\right)}(t)
			\end{array}\right]+\left[\begin{array}{c}
				\mathbf{w}(t)                   \\
				\widetilde{\mathbf{w}}^{(1)}(t) \\
				\widetilde{\mathbf{w}}^{(2)}(t) \\
				\vdots                              \\
				\widetilde{\mathbf{w}}^{\left(n_{i}\right)}(t)
			\end{array}\right]} \\
	\end{equation}
	\begin{equation} \label{eq:GPLFMSSMoutput}
		\qquad \mathbf{y}(t) \quad =\left[\begin{array}{lllll}
			\mathbf{C} & \mathbf{g}_{1} \mathbf{H}^{(1)} & \mathbf{g}_{2} \mathbf{H}^{(2)} & \ldots & \mathbf{g}_{n_{i}} \mathbf{H}^{\left(n_{i}\right)}
		\end{array}\right]\left[\begin{array}{c}
			\mathbf{x}(t)   \\
			\mathbf{z}^{(1)}(t) \\
			\mathbf{z}^{(2)}(t) \\
			\vdots              \\
			\mathbf{z}^{\left(n_{i}\right)}(t)
		\end{array}\right]+\mathbf{v}(t)
	\end{equation}
	where $\mathbf{b}_1$, $\mathbf{b}_2$, $\cdots$, $\mathbf{b}_{n_i}$ are the columns of the structural system $\mathbf{B}$ matrix and $\mathbf{g}_1$, $\mathbf{g}_2$, $\cdots$, $\mathbf{g}_{n_i}$ are the columns of the structural system $\mathbf{G}$ matrix. $\widetilde{\mathbf{w}}^{(j)}(t) \in \mathbb{R}^m$ is a vector-valued \ac{GP} given by $\widetilde{\mathbf{w}}^{(j)}(t) = \mathbf{L}^{(j)} w^{(j)}(t)$, with spectral density $\mathbf{Q}^{(j)}_c \in \mathbb{R}^{m \times m}$ expressed as $\mathbf{Q}^{(j)}_c = \mathbf{L}^{(j)} q_c (\mathbf{L}^{(j)})^T$. In shorthand notation, the discrete-time version of the augmented state-space model shown in \multiref{eq:GPLFMSSMstate}{eq:GPLFMSSMoutput}, computed via an exponential time discretization scheme, can be expressed as follows:
	\begin{equation}
		\begin{cases}\label{eq:GPLFMSSM_discr}
			\mathbf{x}^a_{k} & =\mathbf{A}^a_d \mathbf{x}_{k-1}^{a}+\mathbf{w}^a_{k-1} \\
			\mathbf{y}_k       & =\mathbf{C}^a \mathbf{x}_{k}^{a}+\mathbf{v}_k
		\end{cases}
	\end{equation}
	where $\mathbf{A}^a_d = e^{\left(\mathbf{A}^a \, \Delta t\right)}$ with $\mathbf{A}^a \in \mathbb{R}^{n_{aug} \times n_{aug}}$ being the state matrix in \autoref{eq:GPLFMSSMstate}, while the output matrix  $\mathbf{C}^a \in \mathbb{R}^{n_o \times n_{aug}}$, where $n_{aug} = 2n_r + m \times n_i$ if the \ac{LTI} \ac{SDE} in \autoref{eq:GP_LTI_SDE} is of the same order $m$ for all the latent forces. $\mathbf{w}^a_{k-1}$ is a discretized version of the augmented process noise vector $\mathbf{w}^a(t) \in \mathbb{R}^{n_{aug}}$, which is associated to the covariance matrix $\mathbf{Q}^a = blkdiag \left[ \mathbf{Q},\mathbf{Q}_c\right]$. In the last expression, $\mathbf{Q}$ is the discretized version of the process noise covariance matrix associated with the state vector $\mathbf{x}$, while $\mathbf{Q}_c = blkdiag \left[\mathbf{Q}_c^1, ..., \mathbf{Q}_c^{n_i} \right]$. As a result of the independence assumption made on the latent inputs, the error covariance matrix $\mathbf{P}_k^a$ associated with the augmented state vector $\mathbf{x}^a_k$ has a block diagonal form of the type:
	\begin{equation}
		\mathbf{P}_k^a = blkdiag \left[ \mathbf{P}_k, \mathbf{P}^1_k,...,\mathbf{P}^{(j)}_k,...\mathbf{P}^{n_i}_k \right]
	\end{equation}
	where $\mathbf{P}_k$ is the covariance matrix for the non-augmented state vector $\mathbf{x}_k$ and $\mathbf{P}_k^{(j)}$ is the covariance matrix for the $j$-th latent force.
	
	Joint recursive inference of states and inputs can be performed by employing the augmented system in \autoref{eq:GPLFMSSM_discr} in a sequential scheme made up of a \ac{KF} and a \ac{RTS} smoother. Once the estimated augmented state vector $\mathbf{\hat{x}}_{k}^{a}$ has been obtained, it can be used for predicting the vector $\mathbf{\hat{y}}^e_k$ of the $n_e$ unmeasured responses using the following formula:
	\begin{equation}
		\hat{\mathbf{y}}_k^e = \mathbf{C}_e^a\mathbf{\hat{x}}_{k}^{a}
		\label{eq:resp_est_AKF}
	\end{equation}
	where $\mathbf{C}_e^a$ is the augmented output matrix computed at the \acp{DOF} where the response has to be estimated.  It is worth noting that the GPLFM can be considered as a generalization of the augmented state-space model used in the well-known \ac{AKF}, which normally features marginally stable transmission zeros when no diplacement-level measurement is adopted. To the contrary, it has been found that the \ac{GPLFM} always admits stable inversion since it has no marginally stable transmission zeros regardless of the adopted type of observations \cite{nayek2019gaussian}. This property adds further flexibility to the use of the \ac{GPLFM} for joint input-state estimation in real-life problem for any type of data set. 
	
	Although the validity of using \acp{GPLFM} as reported so far has been demonstrated in \cite{nayek2019gaussian,rogers2020application,zou2022virtual,marino2023switching}, the following considerations must be made.
	\begin{itemize}
		\item The \ac{GPLFM} approach has been proposed in literature for specific and rather conventional classes of covariance functions, i.e., Matérn, squared exponential, exponential. In \autoref{sec:regr_statespace}, the equivalent state-space representation of a \ac{GP} featuring such covariance functions has been found to always have real and coincident roots with multiplicity dependent on the adopted Matérn smoothness parameter. The achieved state-space representation hence corresponds to a ``critically damped'' dynamic system whose response, i.e., the model imposed for the unknown input in the joint input-state estimation framework, has a time waveform dominated by the term $w(t)$ in \autoref{eq:GP_LTI_SDE3} and modulated by an exponentially decaying trend. As a result, conventional covariance functions represent a generic and easy-to-derive modeling choice which, however, does not account for specific content that the input may feature, e.g. harmonics, biases, sudden changes. For this reason, additional covariance functions have been introduced in this work and the associated state-space models have been derived in \autoref{sec:regr_statespace}. Among the proposed functions, periodic covariance functions allow for a dynamic model including undamped harmonics, thus representing the most suitable choice for regression of sinusoidal signals. However, real-life signals too often deviate from pure sinusoidals as they are noise-contaminated, affected by disturbances or rather obtained as compositions of signals of different nature, i.e., random multisines. To take these aspects into account, quasiperiodic covariance functions can be employed for regression. Additionally, linear and constant covariance functions can be adopted in combination with other covariance functions (to obtain, e.g. the biased quasiperiodic covariance function) to respectively model linear trends and biases in data. Finally, the Wiener covariance function can be exploited to construct an alternative and more flexible state-space model for the unknown input satisfying the random-walk assumption, thus finding its best application in regression of ambient noise signals. The hereby reported considerations will be extensively proved in \autoref{subsec:3DOFsexample}, where the mentioned covariance functions will be employed for joint input-state estimation of a 3 \acp{DOF} system subject to inputs of different nature. Additionally, an experimental validation will be provided in \autoref{sec:3D_blade}.
		\item A sequence of a \ac{KF} and a \ac{RTS} smoother is proposed in literature as a solution for recursive regression via \acp{GP}. Including smoothing in the algorithm allows to obtain estimations at the current instant of time on the basis of past and future observations. The combination of a \ac{KF} and a \ac{RTS} smoother thus allows to exactly transfer \ac{GP} regression from its batch to its recursive form. Nevertheless, this operation is in contrast with the more practical need of deploying real-time estimators for operational and experimental data since it foresees a certain lag to be taken into account before acquiring estimations at a given instant of time. Additionally, \autoref{subsec:3DOFsexample} will show that including the smoothing step in the algorithm renders the resulting predictions more prone to accumulation errors. This work therefore proposes to exclude the \ac{RTS} from the joint input-state estimation algorithm based on the \ac{GPLFM} framework, thus running a forward \ac{KF} only. 
		\item The \ac{GP}-based approach adopted in literature for joint input-state estimation foresees the use of the entire set of measurements used for the joint input-state estimation for selecting the covariance function hyperparameters. This implies that a substantially large set of observations needs to be acquired prior to the operational employment of the estimator. Additionally, the use of multiple signals for Bayesian model selection drastically increases the required computational time. To remedy these problems, the use of a single pre-recorded measurement will be proposed in this paper for both the simulated and the experimental case studies. It is worth noting that the most convenient choice for the pre-recorded signal corresponds to an acceleration signal at the unknown input location. Indeed, a collocated acceleration features a direct link to the unknown input via the feedthrough matrix. In absence of a collocated measurement, any other acceleration on the structure may be used to select the covariance function hyperparameters. Acceleration signals are instead not suitable in presence of biased inputs, e.g. a static load, as they do not provide static information. In this situations, a displacement-level signal should be adopted.
	\end{itemize}
	\subsection{Joint input-state estimation of a 3 \acsp{DOF} system via \acp{GPLFM}}
	\label{subsec:3DOFsexample}
\begin{figure} [H]
		\centering
		\input{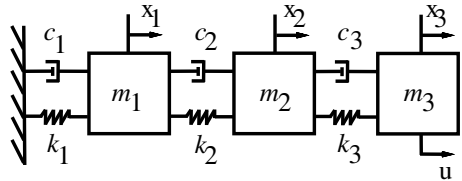}
		\caption{3 \acsp{DOF} system}
		\label{fig:3DOF}
	\end{figure}

\begin{figure} [!ht]
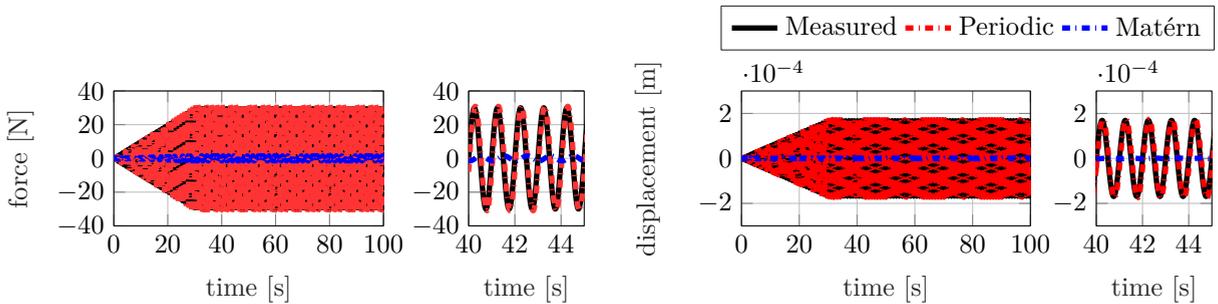

		\hspace{-9em}
		\begin{subfigure}{0.3\textwidth}
			\includestandalone{3DOF_sine_input}
		\end{subfigure}
		\hspace{9em}
		\begin{subfigure}{0.3\textwidth}
			\includestandalone{3DOF_sine_resp}
		\end{subfigure}
		\caption{3 \acsp{DOF} system: time and detailed time histories of the sine input and the resulting $1^{st}$ \acs{DOF} displacement. ``Measured'' signals are shown by a solid black line, while those estimated (via filtering only) by making use of a periodic and a Matérn covariance function ($\nu=1.5$) are respectively denoted via a dashed red and blue line.}
		\label{fig:3DOF_sine}
	\end{figure}

\begin{figure} [!ht]
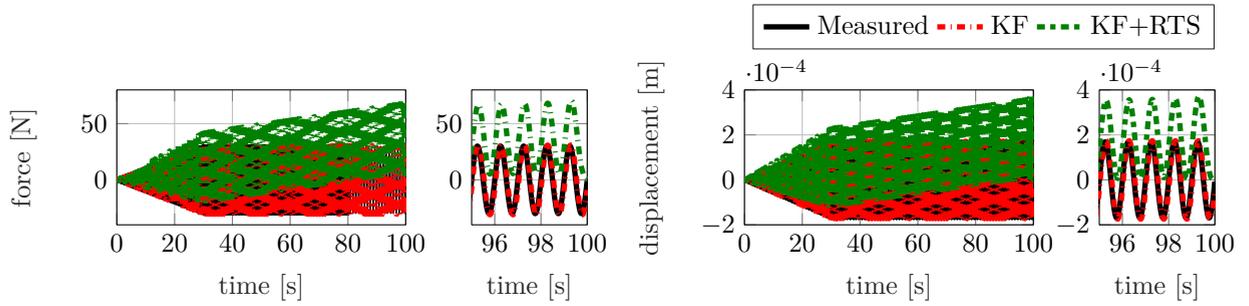

		\hspace{-9em}
		\begin{subfigure}{0.3\textwidth}
			\includestandalone{3DOF_sine_input_smooth}
		\end{subfigure}
		\hspace{9em}
		\begin{subfigure}{0.3\textwidth}
			\includestandalone{3DOF_sine_resp_smooth}
		\end{subfigure}
		\caption{3 \acsp{DOF} system: time and detailed time histories of the sine input and the resulting $1^{st}$ \acs{DOF} displacement. ``Measured'' signals are shown by a solid black line, while those estimated by filtering only and a sequence of filtering and smoothing (periodic  covariance function) are respectively denoted via a dashed red and green line.}
		\label{fig:3DOF_sine_smooth}
	\end{figure}

\begin{figure} [!ht]
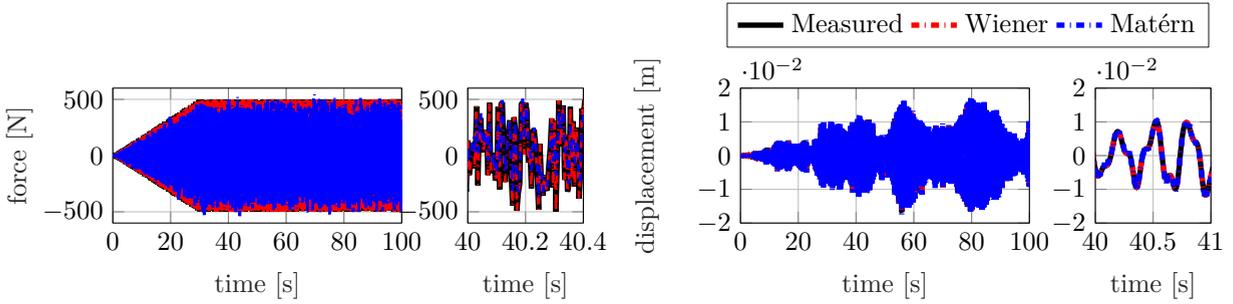

		\hspace{-9em}
		\begin{subfigure}{0.3\textwidth}
			\includestandalone{3DOF_random_input}
		\end{subfigure}
		\hspace{9em}
		\begin{subfigure}{0.3\textwidth}
			\includestandalone{3DOF_random_resp}
		\end{subfigure}
		\caption{3 \acsp{DOF} system: time and detailed time histories of the random input and the resulting $1^{st}$ \acs{DOF} displacement. ``Measured'' signals are shown by a solid black line, while those estimated (via filtering only) by making use of a Wiener and a Matérn covariance function ($\nu=1.5$) are respectively denoted via a dashed red and blue line.}
		\label{fig:3DOF_random}
	\end{figure}

\begin{figure} [!ht]
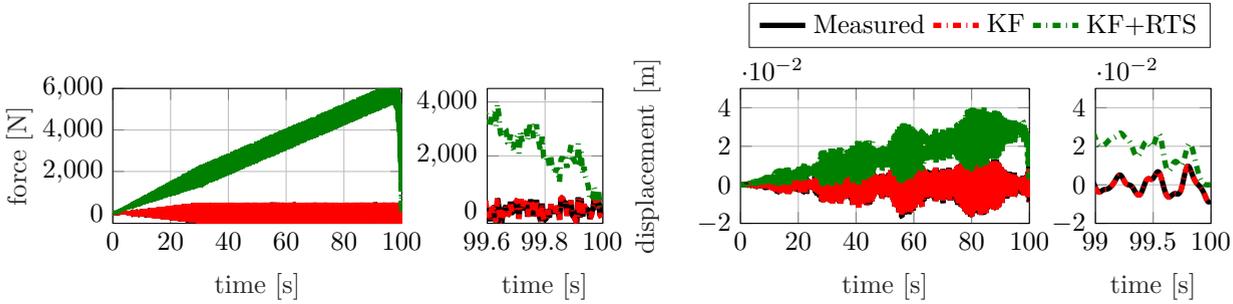

		\hspace{-9em}
		\begin{subfigure}{0.3\textwidth}
			\includestandalone{3DOF_random_input_smooth}
		\end{subfigure}
		\hspace{9em}
		\begin{subfigure}{0.3\textwidth}
			\includestandalone{3DOF_random_resp_smooth}
		\end{subfigure}
		\caption{3 \acsp{DOF} system: time and detailed time histories of the random input and the resulting $1^{st}$ \acs{DOF} displacement. ``Measured'' signals are shown by a solid black line, while those estimated by filtering only and a sequence of filtering and smoothing (Wiener  covariance function) are respectively denoted via a dashed red and green line.}
		\label{fig:3DOF_random_smooth}
	\end{figure}

	This subsection makes use of the \ac{GPLFM} approach for input-state estimation of the 3 \acp{DOF} system in \autoref{fig:3DOF}, excited by a force applied at the $3^{rd}$ mass. The system masses $m_1$, $m_2$ and $m_3$ are assumed to be $100$, $80$ and $80$ kg, respectively. The springs stiffness values have been selected as follows: $k_1=2\times10^{5}$, $k_2=1.5\times10^{5}$, $k_3=1.5\times10^{5}$ N/m. Damping has been defined assuming a proportional behavior such that $\mathbf{C} =  2\times 10^{-2}\mathbf{M} + 3\times 10^{-4} \mathbf{K}$. The system natural frequencies and damping ratios are provided in \autoref{tab:3DOF_frequencies}.
	\vspace{-1em}
	\vspace{-1em}
	\begin{table} [H]
		\centering
		\caption{Natural frequencies and damping ratios of the 3 \acsp{DOF} system}
		\label{tab:3DOF_frequencies}
		\begin{tabular}[t]{@{}cccc@{}}
			\toprule
			Modes                  & 1    & 2    & 3     \\ \midrule
			Natural frequency [Hz] & 3.26 & 8.52 & 12.16 \\ \midrule
			Damping ratio [\%]  & 0.36 & 0.82 & 1.16 \\ \bottomrule
		\end{tabular}
	\end{table}
	The proposed simulated example is hereby exploited to implement and prove the comments listed in the previous section. To explore the spectrum of covariance functions proposed in this work, the input-state prediction problem has been implemented for the 3 \acp{DOF} system under the assumption of several types of inputs acting on the $3^{rd}$ mass, i.e., random, sine, random multisine, impulse, step. Every loading scenario has been addressed adopting the corresponding most suitable covariance function and the achieved input and response predictions have been compared with the results obtained via a conventional Matérn covariance function. Moreover, the hereby applied estimator is constructed using a \ac{KF} only in order to guarantee real-time applicability. To validate this choice, predictions are compared to the ones achieved via sequential \ac{KF} and \ac{RTS}, for every loading case. The state process noise covariance matrix $\mathbf{Q}$ and the measurement noise covariance matrix $\mathbf{R}$, necessary for the implementation of the adopted filtering and smoothing algorithms, have been set by trial and error and kept unchanged across all the loading scenarios for consistency. The trial and error process conducted to select these matrices has been based on differentiating the covariance order of magnitude according to the associated quantity, e.g. displacements and velocities for $\mathbf{Q}$ and displacements and accelerations for $\mathbf{R}$, to ensure a realistic quantification of the modeled errors. A single collocated  ``measurement'', i.e., a response simulated at the $3^{rd}$ mass and contaminated by noise, is adopted as observation for input-state estimation. When the input has a step profile, the $3^{rd}$ mass displacement is used as measurement to ensure that a static information is acquired \cite{maes2015design}. Otherwise, a collocated acceleration measurement is preferred in order to guarantee a feedthrough term with the unknown input, and thus direct input-output coupling \cite{maes2015design}. The same observed response time signal is also used in \autoref{eq:marginal_opt_GP} for determining the covariance function hyperparamenters prior to online estimation. For implementing the marginal likelihood optimization step, the initial hyperparameters values are chosen so as to avoid the method to incur in erroneous local minima.
	\paragraph{Sine load}\mbox{} \\This paragraph reports on the input-state estimation results for the analyzed system excited by a sinusoidal input (1 Hz) applied at the $3^{rd}$ mass. In this loading scenario, the use of a periodic covariance function (with period selected according to the sinusoidal excitation frequency) for constructing the \ac{GPLFM} is strongly preferable to the conventional Matérn function since it allows to account for harmonic features within the adopted \ac{GP} rather than relying on the ``critically damped'' Matérn model. \Autoref{tab:3DOF_sine} summarizes the selected initialization values of the parameters necessary for the algorithm. \Autoref{fig:3DOF_sine} compares the actual input and $1^{st}$ mass displacement against the predictions achieved using the \ac{GPLFM} approach with filtering only. Both the predictions obtained via the proposed periodic covariance function and a Matérn covariance function ($\nu=1.5$) are displayed in \autoref{fig:3DOF_sine}.  A visual inspection of the results confirms that the periodic covariance function is more suitable for estimating signals featuring pure harmonic components. This is specifically applicable for simulated systems such as the one hereby analyzed, where external noise sources are limited and damping follows a precise model. To the contrary, when dealing with physical systems, selecting a periodic covariance function may correspond to a too strict assumption on the content of the analyzed signals. Indeed, noise can perturb the signal periodicity and damping may not follow the model used when defining the system state-space equations. To mitigate these effects, a quasiperiodic covariance function is to be preferred when treating real-life problems. 
	\vspace{-1em}
	\begin{table} [!ht]
		\centering
		\caption{3 \acsp{DOF} system, sine load: estimators initialization values}
		\label{tab:3DOF_sine}
		\renewcommand{\arraystretch}{2.0}
		\resizebox*{1\textwidth}{!}{
			\begin{threeparttable}			
				\begin{tabular}[H]{ccccccc}
					\toprule
					\begin{tabular}[c]{@{}c@{}} Initial state \\[-.4cm] mean\end{tabular}  & \begin{tabular}[c]{@{}c@{}} Initial error cov. \\[-.4cm] matrix\end{tabular}    & \begin{tabular}[c]{@{}c@{}}  Initial hyperparameters\\[-.4cm] periodic\end{tabular}     & \begin{tabular}[c]{@{}c@{}} Initial hyperparameters\\[-.4cm] Matérn\end{tabular}    & \begin{tabular}[c]{@{}c@{}} Process noise \\[-.4cm] cov. matrix ($\mathbf{Q}$)\end{tabular} & \begin{tabular}[c]{@{}c@{}}  Measurement noise \\[-.4cm] cov. matrix ($\mathbf{R}$) \end{tabular}                                                                                                                                                                            
					\\ \midrule
					$\mathbf{\hat{x}}^a_{0|0} = \mathbf{0}$     & $\mathbf{\hat{P}}^x_{0|0} = \mathbf{0}$     & \begin{tabular}[c]{@{}c@{}} $\sigma^2 =1 \times 10^{-1}$ \\[-.3cm] $l = 5 \times 10^{-1} $ \\[-.3cm] $t_{period}=1$ \end{tabular}                                                                                                                                                                                            & \begin{tabular}[c]{@{}c@{}} $\sigma^2 =5 $ \\[-.3cm] $l = 1 \times 10^{-2} $ \end{tabular} & \begin{tabular}[c]{@{}c@{}} $\mathbf{Q}_{\text{displ}} = 10^{-20} \times \mathbf{\mathbf{I}}_{\text{displ}}$   \\[-.3cm]     $\mathbf{Q}_{\text{vel}} = 10^{-10} \times \mathbf{\mathbf{I}}_{\text{vel}} $    \end{tabular}                                              & \begin{tabular}[c]{@{}c@{}} $\mathbf{R}_{\text{displ}} = 10^{-15} \times \mathbf{I}_{\text{displ}}$ \\[-.3cm] $\mathbf{R}_{\text{acc}} = 10^{-12} \times \mathbf{I}_{\text{acc}}$ \end{tabular} \\
					\bottomrule
				\end{tabular}
			\end{threeparttable}
		}
	\end{table}
	\\A time-domain comparison between the results obtained via filtering only and the ones achieved via sequential \ac{KF} and \ac{RTS} smoother is offered in \autoref{fig:3DOF_sine_smooth}. By analyzing the displayed signals, it can be observed that combining filtering and smoothing generates accumulation errors which result in low frequency components affecting the estimated time histories.
	\\A quantification of the estimation errors obtained by the analyzed estimators is reported in \autoref{tab:RMSE_3DOF_sine} for both the unknown force and the estimated responses by means of the \ac{NRMSE}, i.e., the \ac{RMSE} between the estimated and the corresponding ``measured'' signal, normalized with respect to the ``measured" signal \ac{RMS} value. \Autoref{tab:RMSE_3DOF_sine} demonstrates that the employment of a periodic covariance function within a \ac{GPLFM} solved with filtering only allows for the optimal performance of both input and response estimation in a sinusoidal loading scenario.
	\vspace{-.5em}
	\begin{table} [!ht]
		\centering
		\caption{3 \acsp{DOF} system, sine load: \acs{NRMSE} values between ``measured'' and estimated signals (responses and force).}
		\label{tab:RMSE_3DOF_sine}
		\resizebox*{1\textwidth}{!}{
			\begin{threeparttable}
				\begin{tabular}[H]{c|ccc|ccc|ccc|c|c}		
					& 	\multicolumn{3}{c|}{$\acs{NRMSE}_{displ} $} & \multicolumn{3}{c|}{$\acs{NRMSE}_{vel}$} & \multicolumn{3}{c|}{$\acs{NRMSE}_{acc} $} & $\acs{NRMSE} $ & \acs{NRMSE}\\
					& \acs{DOF}1 & \acs{DOF}2 & \acs{DOF}3 & \acs{DOF}1 & \acs{DOF}2 & \acs{DOF}3 & \acs{DOF}1 & \acs{DOF}2 & \acs{DOF}3 & mean resp. & force\\
					\hline
					Periodic (\acs{KF}) &0.245&0.245&0.245&0.247&0.247&0.247&0.248&0.248&0.003&0.220&0.261\\
					Matérn (\acs{KF}) & 1.001&1.001&1.000&0.999&1.001&0.994&0.963&1.042&0.982&0.998&1.001\\
					Periodic (\acs{KF}+\acs{RTS})  &1.048&1.059&1.076&0.257&0.257&0.257&0.253&0.251&0.003&0.496&1.177\\
				\end{tabular}
			\end{threeparttable}
		}
	\end{table}
	\paragraph{Random load}\mbox{} \\
	This paragraph reports on the input-state estimation results for the analyzed system excited by a pure random (white noise) input applied at the $3^{rd}$ mass. In this loading scenario, the use of a Wiener covariance 
	function for constructing the \ac{GPLFM} is proposed as an alternative to the conventional Matérn function. As mentioned in \autoref{subsec:cov_functs}, a \ac{GP} with a Wiener covariance function is a random-walk model. Implementing a \ac{GPLFM} approach for input-state prediction with a Wiener covariance function thus represents a non-conventional method for adopting a random-walk model for the unknown input. Unlike traditional Kalman-based estimators, where the random-walk assumption implies the need of a challenging tuning exercise, the \ac{GP}-based formulation offers a flexible way for employing the random-walk model for input-state prediction. Indeed, in this framework the tuning effort is substituted with the preliminary training phase aimed at determining the hyperparameters. Following what stated in \autoref{sec:regr_statespace}, it is evident that a Wiener covariance function represents the best choice for regression of purely randomic signals. \Autoref{tab:3DOF_random} summarizes the selected initialization values of the necessary parameters for the algorithm. 
	\begin{table} [H]
		\centering
		\caption{ 3 \acsp{DOF} system, random load: estimators initialization values}
		\label{tab:3DOF_random}
		\renewcommand{\arraystretch}{2.0}
		\resizebox*{1\textwidth}{!}{
			\begin{threeparttable}			
				\begin{tabular}[H]{ccccccc}
					\toprule
					\begin{tabular}[c]{@{}c@{}} Initial state \\[-.4cm] mean\end{tabular}  & \begin{tabular}[c]{@{}c@{}} Initial error cov. \\[-.4cm] matrix\end{tabular}    & \begin{tabular}[c]{@{}c@{}}  Initial hyperparameters\\[-.4cm] Wiener\end{tabular}     & \begin{tabular}[c]{@{}c@{}}Initial hyperparameters\\[-.4cm] Matérn\end{tabular}    & \begin{tabular}[c]{@{}c@{}} Process noise \\[-.4cm] cov. matrix ($\mathbf{Q}$)\end{tabular} & \begin{tabular}[c]{@{}c@{}}  Measurement noise \\[-.4cm] cov. matrix ($\mathbf{R}$) \end{tabular}                                                                                                                                                                            
					\\ \midrule
					$\mathbf{\hat{x}}^a_{0|0} = \mathbf{0}$     & $\mathbf{\hat{P}}^x_{0|0} = \mathbf{0}$     & \begin{tabular}[c]{@{}c@{}} $\sigma^2 =1 \times 10^{-4}$  \end{tabular}                                                                                                                                                                                            & \begin{tabular}[c]{@{}c@{}} $\sigma^2 =5 $  \\[-.3cm] $l = 1 \times 10^{-2} $ \end{tabular} & \begin{tabular}[c]{@{}c@{}} $\mathbf{Q}_{\text{displ}} = 10^{-20} \times \mathbf{\mathbf{I}}_{\text{displ}}$   \\[-.3cm]     $\mathbf{Q}_{\text{vel}} = 10^{-10} \times \mathbf{\mathbf{I}}_{\text{vel}} $    \end{tabular}                                              & \begin{tabular}[c]{@{}c@{}} $\mathbf{R}_{\text{displ}} = 10^{-15} \times \mathbf{I}_{\text{displ}}$ \\[-.3cm] $\mathbf{R}_{\text{acc}} = 10^{-12} \times \mathbf{I}_{\text{acc}}$ \end{tabular} \\
					\bottomrule
				\end{tabular}
			\end{threeparttable}}
	\end{table}
\Multiref{fig:3DOF_random}{fig:3DOF_random_freq} respectively compare the actual input and $1^{st}$ mass displacement in time and frequency domains obtained via the proposed Wiener covariance function and the Matérn covariance function. The analysis of \multiref{fig:3DOF_random}{fig:3DOF_random_freq} proves that the use of the Matérn covariance function produces estimates featuring unrealistic frequency content, thus confirming that a Wiener covariance function allows to better capture the stochastic nature of the unknown input.
	\begin{figure} [!ht]
		\hspace{-9em}
		\begin{subfigure}{0.3\textwidth}
			\includestandalone{3DOF_random_input_freq}
		\end{subfigure}
		\hspace{9em}
		\begin{subfigure}{0.3\textwidth}
			\includestandalone{3DOF_random_resp_freq}
		\end{subfigure}
		\caption{3 \acsp{DOF} system: \acp{PSD} of the random input and the resulting $1^{st}$ \acs{DOF} displacement. ``Measured'' signals are shown by a solid black line, while those estimated (via filtering only) by making use of a Wiener and a Matérn covariance function ($\nu=1.5$) are respectively denoted via a dashed red and blue line.}
		\label{fig:3DOF_random_freq}
	\end{figure}\\
	 \Autoref{fig:3DOF_random_smooth} offers a time-domain comparison between the results obtained via filtering only and the ones achieved via sequential \ac{KF} and \ac{RTS} smoother. The input and response time histories estimated by combining filtering and smoothing show low frequency errors which are not generated when using Kalman filtering only.
	\\ \Autoref{tab:RMSE_3DOF_random} quantifies the estimation inaccuracy for the analyzed estimators by offering the \ac{NRMSE} values computed between the estimated and the ``measured'' signals for both the unknown force and the predicted responses. \Autoref{tab:RMSE_3DOF_random} confirms that adopting a Wiener covariance function within a \ac{GPLFM} solved with filtering only guarantees the highest performance for both input and response estimation in a random loading setting.
	\begin{table} [!ht]
		\centering
		\caption{3 \acsp{DOF} system, random load: \acs{NRMSE} values between ``measured'' and estimated signals (responses and force).}
		\label{tab:RMSE_3DOF_random}
		\resizebox*{1\textwidth}{!}{
			\begin{threeparttable}
				\begin{tabular}[H]{c|ccc|ccc|ccc|c|c}		
					& 	\multicolumn{3}{c|}{$\acs{NRMSE}_{displ} \times 10^{5}$} & \multicolumn{3}{c|}{$\acs{NRMSE}_{vel} \times 10^{5}$} & \multicolumn{3}{c|}{$\acs{NRMSE}_{acc} \times 10^{5}$} & $\acs{NRMSE} \times 10^{5}$ & \acs{NRMSE}\\
					& \acs{DOF}1 & \acs{DOF}2 & \acs{DOF}3 & \acs{DOF}1 & \acs{DOF}2 & \acs{DOF}3 & \acs{DOF}1 & \acs{DOF}2 & \acs{DOF}3 & mean resp. & force\\
					\hline
					Wiener (\acs{KF}) & 0.078&0.091&0.111&0.003&0.002&0.001&0.005&0.005&0.010&0.034&0.014\\
					Matérn (\acs{KF}) & 10178.163&11413.813&13798.689&5949.576&3661.975&1992.402&8319.623&7659.016&3143.413&734.630&0.459\\
					Wiener (\acs{KF}+\acs{RTS})  &32660.282&38181.965&46559.658&635.706&445.016&318.899&829.560&742.271&158.739&13392.455&13.077\\			
				\end{tabular}
			\end{threeparttable}
		}
	\end{table}
	\paragraph{Random multisine load}\mbox{} \\
	In this paragraph, the input-state predictions obtained for the 3 \acp{DOF} system subjected to a random multisine input at the $3^{rd}$ mass are summarized. A random multisine signal is constructed as a superposition of sines with random phase within a selected frequency range (0-20 Hz). The use of a quasiperiodic covariance function is hereby proposed for representing this type of load and performing the hyperparameters training by regression of the $3^{rd}$ mass acceleration. Indeed, adopting a quasiperiodic covariance function rather than a periodic one allows to simultaneously account for periodic and damping effects, which result from the random phase of the loading signal as well as the damped nature of the analyzed dynamical system. The initialization values for the algorithm under the different covariance assumptions are summarized in \autoref{tab:3DOF_multisine}. 
	\begin{table} [!ht]
		\centering
		\caption{3 \acsp{DOF} system, random multisine load: estimators initialization values}
		\label{tab:3DOF_multisine}
		\renewcommand{\arraystretch}{2.0}
		\resizebox*{1\textwidth}{!}{
			\begin{threeparttable}			
				\begin{tabular}[H]{ccccccc}
					\toprule
					\begin{tabular}[c]{@{}c@{}} Initial state \\[-.4cm] mean\end{tabular}  & \begin{tabular}[c]{@{}c@{}} Initial error cov. \\[-.4cm] matrix\end{tabular}    & \begin{tabular}[c]{@{}c@{}}  Initial hyperparameters\\[-.4cm] quasiperiodic\end{tabular}     & \begin{tabular}[c]{@{}c@{}} Initial hyperparameters\\[-.4cm] Matérn\end{tabular}    & \begin{tabular}[c]{@{}c@{}} Process noise \\[-.4cm] cov. matrix ($\mathbf{Q}$)\end{tabular} & \begin{tabular}[c]{@{}c@{}}  Measurement noise \\[-.4cm] cov. matrix ($\mathbf{R}$) \end{tabular}                                                                                                                                                                            
					\\ \midrule
					$\mathbf{\hat{x}}^a_{0|0} = \mathbf{0}$     & $\mathbf{\hat{P}}^x_{0|0} = \mathbf{0}$     & \begin{tabular}[c]{@{}c@{}} $\sigma^2 =2 \times 10^{-2}$ \\[-.3cm] $l = 3 \times 10^{-1} $ \\[-.3cm] $t_{period}=1$\\[-.3cm] $l_{matern}=1.3$ \end{tabular}                                                                                                                                                                                            & \begin{tabular}[c]{@{}c@{}} $\sigma^2 =5 $ \\[-.3cm] $l = 1 \times 10^{-2} $ \end{tabular} & \begin{tabular}[c]{@{}c@{}} $\mathbf{Q}_{\text{displ}} = 10^{-20} \times \mathbf{\mathbf{I}}_{\text{displ}}$   \\[-.3cm]     $\mathbf{Q}_{\text{vel}} = 10^{-10} \times \mathbf{\mathbf{I}}_{\text{vel}} $    \end{tabular}                                              & \begin{tabular}[c]{@{}c@{}} $\mathbf{R}_{\text{displ}} = 10^{-15} \times \mathbf{I}_{\text{displ}}$ \\[-.3cm] $\mathbf{R}_{\text{acc}} = 10^{-12} \times \mathbf{I}_{\text{acc}}$ \end{tabular} \\
					\bottomrule
				\end{tabular}
			\end{threeparttable}
		}
	\end{table}\\
\Multiref{fig:3DOF_multisine}{fig:3DOF_multisine_freq} compare the predictions obtained via the quasiperiodic and the Matérn covariance functions for the actual input and the $1^{st}$ mass displacement in time and frequency domains respectively. 
	\begin{figure} [!ht]
		\hspace{-9em}
		\begin{subfigure}{0.3\textwidth}
			\includestandalone{3DOF_multisine_input}
		\end{subfigure}
		\hspace{9em}
		\begin{subfigure}{0.3\textwidth}
			\includestandalone{3DOF_multisine_resp}
		\end{subfigure}
		\caption{3 \acsp{DOF} system: time and detailed time histories of the multisine input and the resulting $1^{st}$ \acs{DOF} displacement. ``Measured'' signals are shown by a solid black line, while those estimated (via filtering only) by making use of a quasiperiodic and a Matérn covariance function ($\nu=1.5$) are respectively denoted via a dashed red and blue line.}
		\label{fig:3DOF_multisine}
	\end{figure}
	\begin{figure} [!ht]
		\hspace{-9em}
		\begin{subfigure}{0.3\textwidth}
			\includestandalone{3DOF_multisine_input_freq}
		\end{subfigure}
		\hspace{9em}
		\begin{subfigure}{0.3\textwidth}
			\includestandalone{3DOF_multisine_resp_freq}
		\end{subfigure}
		\caption{3 \acsp{DOF} system: \acsp{PSD} of the multisine input and the resulting $1^{st}$ \acs{DOF} displacement. ``Measured'' signals are shown by a solid black line, while those estimated (via filtering only) by making use of a quasiperiodic and a Matérn covariance function ($\nu=1.5$) are respectively denoted via a dashed red and blue line.}
		\label{fig:3DOF_multisine_freq}
	\end{figure}
	\\ \Multiref{fig:3DOF_multisine}{fig:3DOF_multisine_freq} show that a quasiperiodic covariance function allows to better match the amplitude and the frequency content of the unknown input. The same conclusion can be drawn for the response estimate which, however, exhibits a slight mismatch between 15 and 20 Hz. This error can be explained by the use of only one acceleration observation, which reflects the input frequency content because of their direct feedthrough. Part of this content is then erroneously transferred to the response estimates of different type, i.e., displacement as the one in \multiref{fig:3DOF_multisine}{fig:3DOF_multisine_freq}. This issue is normally solved in practical applications by including different types of responses within the observations set. \\
	\Autoref{fig:3DOF_multisine_smooth} shows a time-domain comparison between the estimations obtained via filtering only and the ones achieved via sequential \ac{KF} and \ac{RTS} smoother. The input and response time histories estimated by combining filtering and smoothing exhibit a deviation error in the last two seconds of the analyzed time window. This effect only appears when smoothing is combined with filtering.
	\begin{figure} [!ht]
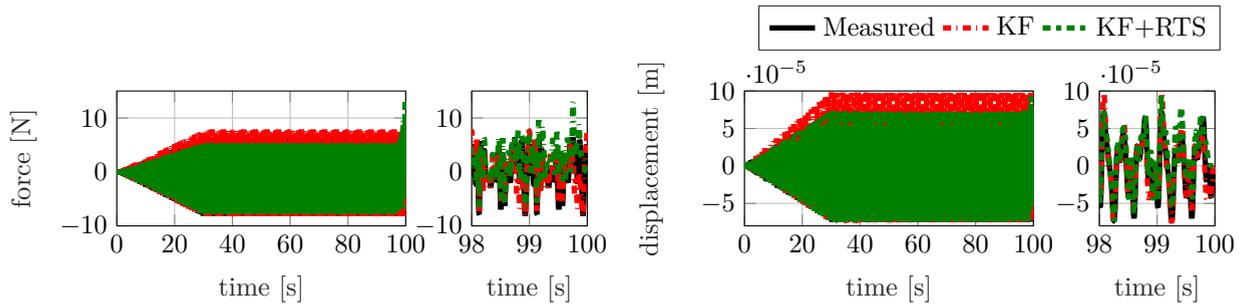

		\hspace{-9em}
		\begin{subfigure}{0.3\textwidth}
			\includestandalone{3DOF_multisine_input_smooth}
		\end{subfigure}
		\hspace{9em}
		\begin{subfigure}{0.3\textwidth}
			\includestandalone{3DOF_multisine_resp_smooth}
		\end{subfigure}
		\caption{3 \acsp{DOF} system: time and detailed time histories of the multisine input and the resulting $1^{st}$ \acs{DOF} displacement. ``Measured'' signals are shown by a solid black line, while those estimated by filtering only and a sequence of filtering and smoothing (quasiperiodic covariance function) are respectively denoted via a dashed red and green line.}
		\label{fig:3DOF_multisine_smooth}
	\end{figure}
	\\The overall highest prediction performance achieved under the assumption of a quasiperiodic covariance function (with filtering only) for a random multisine excitation is confirmed by \autoref{tab:RMSE_3DOF_multisine}, which reports on the \ac{NRMSE} values computed between the estimated and the ``measured'' signals for both the unknown force and the estimated responses.
	\begin{table} [H]
		\centering
		\caption{3 \acsp{DOF} system, random multisine load: \acs{NRMSE} values between ``measured'' and estimated signals (responses and force).}
		\label{tab:RMSE_3DOF_multisine}
		\resizebox*{1\textwidth}{!}{
			\begin{threeparttable}
				\begin{tabular}[H]{c|ccc|ccc|ccc|c|c}		
					& 	\multicolumn{3}{c|}{$\acs{NRMSE}_{displ}$} & \multicolumn{3}{c|}{$\acs{NRMSE}_{vel} $} & \multicolumn{3}{c|}{$\acs{NRMSE}_{acc} $} & $\acs{NRMSE} $ & \acs{NRMSE}\\
					& \acs{DOF}1 & \acs{DOF}2 & \acs{DOF}3 & \acs{DOF}1 & \acs{DOF}2 & \acs{DOF}3 & \acs{DOF}1 & \acs{DOF}2 & \acs{DOF}3 & mean resp. & force\\
					\hline
					Quasiperiodic (\acs{KF}) & 0.352&0.447&0.538&0.097&0.094&0.104&0.089&0.076&0.039&0.204&0.800\\
					Matérn (\acs{KF}) & 1.000&1.000&1.000&1.000&1.000&1.000&1.000&1.000&1.000&1.000&1.000\\
					Quasiperiodic (\acs{KF}+\acs{RTS})  &0.253&0.174&2.666&0.187&0.156&0.069&0.177&0.173&0.046&0.434&1.039\\
				\end{tabular}
			\end{threeparttable}
		}
	\end{table}
	
\begin{figure} [!ht]
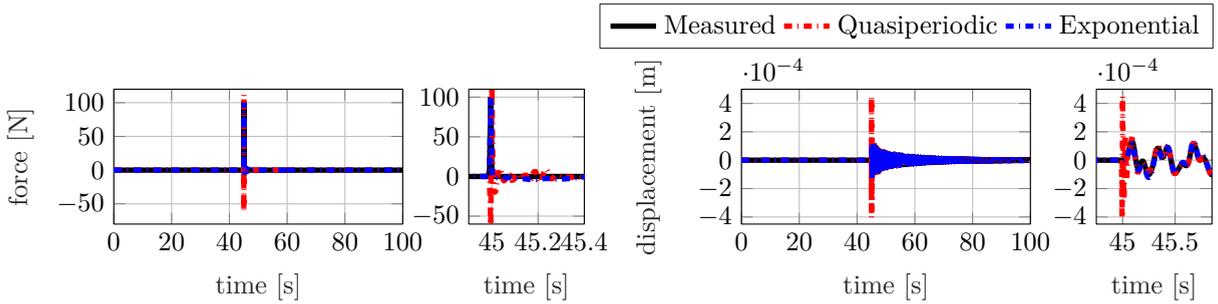

		\hspace{-10em}
		\begin{subfigure}{0.3\textwidth}
			\includestandalone{3DOF_impulse_input}
		\end{subfigure}
		\hspace{8em}
		\begin{subfigure}{0.3\textwidth}
			\includestandalone{3DOF_impulse_resp}
		\end{subfigure}
		\caption{3 \acsp{DOF} system: time and detailed time histories of the impulse input and the resulting $1^{st}$ \acs{DOF} displacement. ``Measured'' signals are shown by a solid black line, while those estimated (via filtering only) by making use of a quasiperiodic and an exponential covariance function are respectively denoted via a dashed red and blue line.}
		\label{fig:3DOF_impulse}
	\end{figure}

\begin{figure} [!htp]
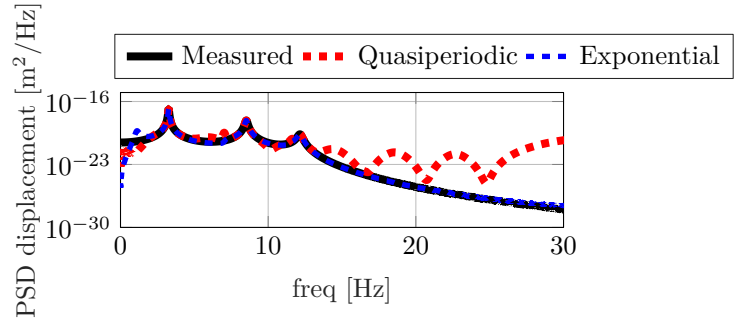

		\hspace{-9em}
		\begin{subfigure}{0.3\textwidth}
			\includestandalone{3DOF_impulse_resp_freq}
		\end{subfigure}
		\caption{3 \acsp{DOF} system: \acsp{PSD} of the $1^{st}$ \acs{DOF} displacement resulting from the impulse excitation. ``Measured'' signals are shown by a solid black line, while those estimated (via filtering only) by making use of a quasiperiodic and an exponential covariance function are respectively denoted via a dashed red and blue line.}
		\label{fig:3DOF_impulse_freq}
	\end{figure}

\begin{figure} [!ht]
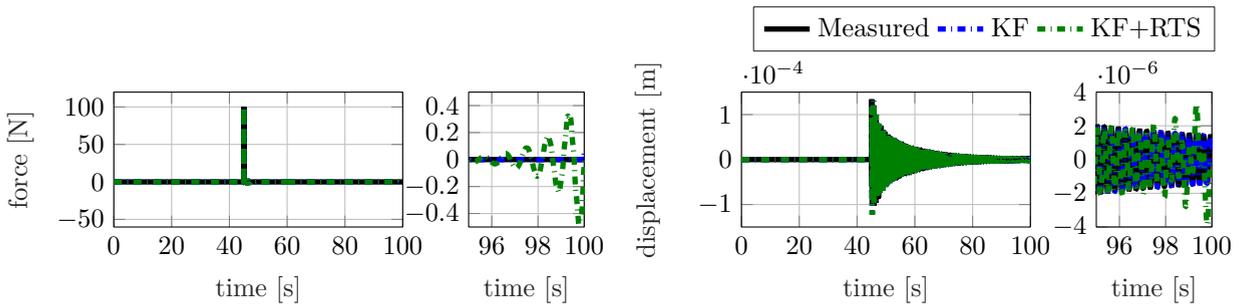

		\hspace{-9em}
		\begin{subfigure}{0.3\textwidth}
			\includestandalone{3DOF_impulse_input_smooth}
		\end{subfigure}
		\hspace{9em}
		\begin{subfigure}{0.3\textwidth}
			\includestandalone{3DOF_impulse_resp_smooth}
		\end{subfigure}
		\caption{3 \acsp{DOF} system: time and detailed time histories of the impulse input and the resulting $1^{st}$ \acs{DOF} displacement. ``Measured'' signals are shown by a solid black line, while those estimated by filtering only and a sequence of filtering and smoothing (exponential covariance function) are respectively denoted via a dashed blue and green line.}
		\label{fig:3DOF_impulse_smooth}
	\end{figure}

	\paragraph{Impulse load}\mbox{} \\
	This paragraph treats the input-state predictions obtained for the 3 \acp{DOF} system when an impulse load is applied at the $3^{rd}$ mass. A quasiperiodic covariance function is hereby compared with an exponential covariance function (Matérn with $\nu$=0.5) for regression of the $3^{rd}$ mass acceleration and subsequent construction of the unknown input model. Indeed, a quasiperiodic covariance function is optimal for regression of damped dynamical systems free responses since these are expected to exhibit an oscillatory response with damped frequency content linked to the system modal properties. The exponential covariance function, i.e., a Matérn covariance function with $\nu=0.5$, has been chosen as reference since this class of functions appear to be the most suitable for detecting large discontinuities in time due to the linked \acp{GP} being non-smooth, i.e., mean square non-differentiable. A summary of the initialization values selected for the algorithm parameters is reported in \autoref{tab:3DOF_impulse}. 
\vspace{-.5em}
	\begin{table} [!ht]
		\centering
		\caption{3 \acsp{DOF} system, impulse load: estimators initialization values}
		\label{tab:3DOF_impulse}
		\renewcommand{\arraystretch}{2.0}
		\resizebox*{1\textwidth}{!}{
			\begin{threeparttable}			
				\begin{tabular}[H]{ccccccc}
					\toprule
					\begin{tabular}[c]{@{}c@{}} Initial state \\[-.4cm] mean\end{tabular}  & \begin{tabular}[c]{@{}c@{}} Initial error cov. \\[-.4cm] matrix\end{tabular}    & \begin{tabular}[c]{@{}c@{}}  Initial hyperparameters\\[-.4cm] quasiperiodic\end{tabular}     & \begin{tabular}[c]{@{}c@{}} Initial hyperparameters\\[-.4cm] exponential \end{tabular}    & \begin{tabular}[c]{@{}c@{}} Process noise \\[-.4cm] cov. matrix ($\mathbf{Q}$)\end{tabular} & \begin{tabular}[c]{@{}c@{}}  Measurement noise \\[-.4cm] cov. matrix ($\mathbf{R}$) \end{tabular}                                                                                                                                                                            
					\\ \midrule
					$\mathbf{\hat{x}}^a_{0|0} = \mathbf{0}$     & $\mathbf{\hat{P}}^x_{0|0} = \mathbf{0}$     & \begin{tabular}[c]{@{}c@{}} $\sigma^2 =6 \times 10^{-1}$ \\[-.3cm] $l = 2.5 \times 10^{-1} $ \\[-.3cm] $t_{period}=0.3$\\[-.3cm] $l_{matern}=1$ \end{tabular}                                                                                                                                                                                            & \begin{tabular}[c]{@{}c@{}} $\sigma^2 =5 $ \\[-.3cm] $l = 1 \times 10^{-2} $ \end{tabular} & \begin{tabular}[c]{@{}c@{}} $\mathbf{Q}_{\text{displ}} = 10^{-20} \times \mathbf{\mathbf{I}}_{\text{displ}}$   \\[-.3cm]     $\mathbf{Q}_{\text{vel}} = 10^{-10} \times \mathbf{\mathbf{I}}_{\text{vel}} $    \end{tabular}                                              & \begin{tabular}[c]{@{}c@{}} $\mathbf{R}_{\text{displ}} = 10^{-15} \times \mathbf{I}_{\text{displ}}$ \\[-.3cm] $\mathbf{R}_{\text{acc}} = 10^{-12} \times \mathbf{I}_{\text{acc}}$ \end{tabular} \\
					\bottomrule
				\end{tabular}
			\end{threeparttable}
		}
	\end{table}\\
\Autoref{fig:3DOF_impulse} (left) compares the actual input time history against the predictions achieved using the \ac{GPLFM} approach with filtering only (both for a quasiperiodic and an exponential covariance function). \Autoref{fig:3DOF_impulse} (right) offers the time-domain estimation results for the $1^{st}$ \ac{DOF} displacement when a quasiperiodic or an exponential covariance function is adopted. Additionally, the corresponding \acp{PSD} are presented in \autoref{fig:3DOF_impulse_freq}.
\vspace{-.8em}
	\\From the analysis of \multiref{fig:3DOF_impulse}{fig:3DOF_impulse_freq} it can be concluded that the quasiperiodic covariance function generates larger estimation errors for both input and response estimations. This may be due to the quasiperiodic covariance function being centered on a specific frequency, e.g. the first natural frequency of the system, and only including its harmonics. This assumption excludes the possibility to include higher natural frequencies ($2^{nd}$ and $3^{rd}$ for this application) in the \ac{GP} representation and renders the hereby adopted quasiperiodic covariance function too strict for such application. A workaround for this limitation consists in constructing a covariance function as a product of quasiperiodic covariance functions centered at the 3 natural frequencies of the system. \\ \Autoref{fig:3DOF_impulse_smooth} shows a time-domain comparison between the estimations obtained via filtering only and the ones achieved via sequential \ac{KF} and \ac{RTS} smoother under the assumption of an exponential covariance function. The input and response time histories estimated by combining filtering and smoothing exhibit a deviation error in the last two seconds of the analyzed time window. These deviations are limited with respect to the ones experienced in the previously considered scenarios. This may be ascribed to the low signals amplitude at the end of the analyzed time frame for impulse response, which therefore generates lower accumulation errors.
	\\A quantification of the estimation errors obtained by the analyzed estimators is reported in \autoref{tab:RMSE_3DOF_impulse} by means of the \ac{NRMSE} values computed between the estimated and the ``measured'' signals for both the unknown force and the estimated responses. \Autoref{tab:RMSE_3DOF_impulse} confirms that the use of a quasiperiodic covariance function centered at a specific frequency within a \ac{GPLFM} generates large prediction inaccuracy for both the input and response estimations under an impulse excitation. 
	\begin{table} [!ht]
		\centering
		\caption{3 \acsp{DOF} system, impulse load: \acs{NRMSE} values between ``measured'' and estimated signals (responses and force).}
		\label{tab:RMSE_3DOF_impulse}
		\resizebox*{1\textwidth}{!}{
			\begin{threeparttable}
				\begin{tabular}[H]{c|ccc|ccc|ccc|c|c}		
					& 	\multicolumn{3}{c|}{$\acs{NRMSE}_{displ} \times 10^{2}$} & \multicolumn{3}{c|}{$\acs{NRMSE}_{vel} \times 10^{2}$} & \multicolumn{3}{c|}{$\acs{NRMSE}_{acc} \times 10^{2}$} & $\acs{NRMSE} \times 10^{2}$ & \acs{NRMSE}\\
					& \acs{DOF}1 & \acs{DOF}2 & \acs{DOF}3 & \acs{DOF}1 & \acs{DOF}2 & \acs{DOF}3 & \acs{DOF}1 & \acs{DOF}2 & \acs{DOF}3 & mean resp. & force\\
					\hline
					Quasiperiodic (\acs{KF}) & 37.570&44.491&54.658&7.738&5.337&2.895&10.131&10.102&9.753&20.297&1.792	\\
					Exponential (\acs{KF}) & 8.541&9.800&11.660&3.756&3.688&3.836&3.851&3.776&0.023&5.437&0.332 \\
					Exponential (\acs{KF}+\acs{RTS})  &8.605&9.876&11.751&3.768&3.708&3.864&3.854&3.780&0.074&5.475&0.334\\
				\end{tabular}
			\end{threeparttable}
		}
	\end{table}
	
\begin{figure} [!ht]
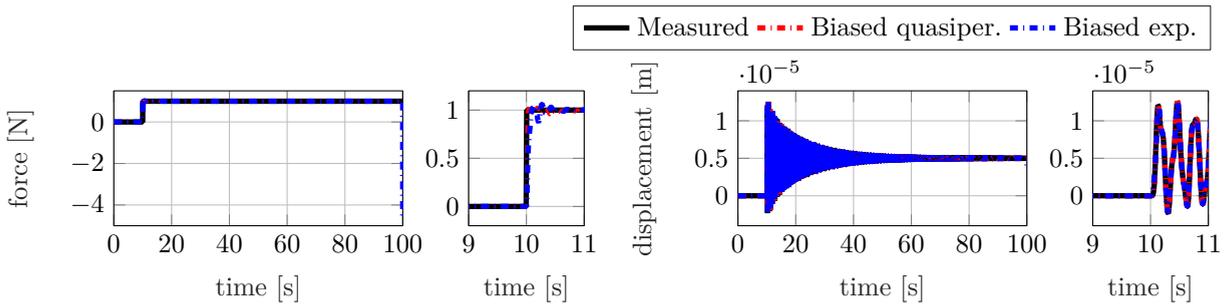

		\hspace{-11em}
		\begin{subfigure}{0.3\textwidth}
			\includestandalone{3DOF_step_input}
		\end{subfigure}
		\hspace{7em}
		\begin{subfigure}{0.3\textwidth}
			\includestandalone{3DOF_step_resp}
		\end{subfigure}
		\caption{3 \acsp{DOF} system: time and detailed time histories of the step input and the resulting $1^{st}$ \acs{DOF} displacement. ``Measured'' signals are shown by a solid black line, while those estimated (via filtering only) by making use of a biased quasiperiodic and a biased exponential covariance function are respectively denoted via a dashed red and blue line.}
		\label{fig:3DOF_step}
	\end{figure}

\begin{figure} [!ht]
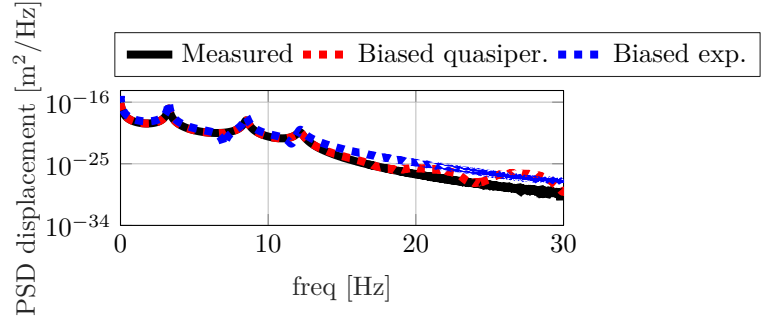

		\hspace{-9em}
		\begin{subfigure}{0.3\textwidth}
			\includestandalone{3DOF_step_resp_freq}
		\end{subfigure}
		\caption{3 \acsp{DOF} system: \acsp{PSD} of the $1^{st}$ \acs{DOF} displacement resulting from the step excitation. ``Measured'' signals are shown by a solid black line, while those estimated (via filtering only) by making use of a biased quasiperiodic and a biased exponential covariance function are respectively denoted via a dashed red and blue line.}
		\label{fig:3DOF_step_freq}
	\end{figure}

\begin{figure} [!ht]
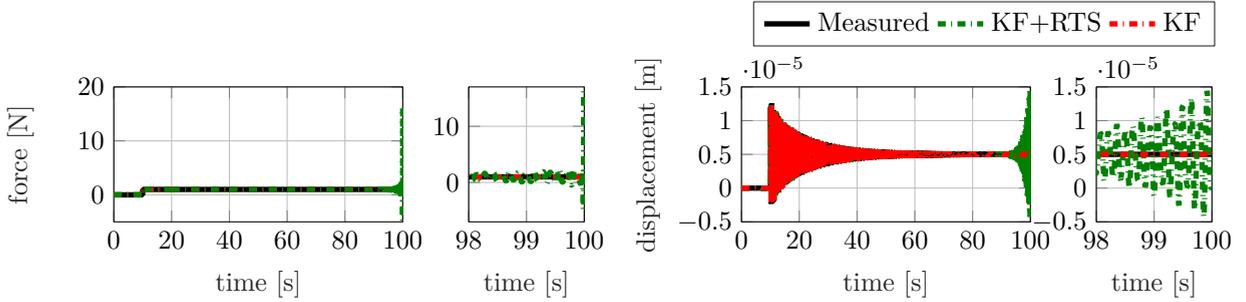

		\hspace{-9em}
		\begin{subfigure}{0.3\textwidth}
			\includestandalone{3DOF_step_input_smooth}
		\end{subfigure}
		\hspace{9em}
		\begin{subfigure}{0.3\textwidth}
			\includestandalone{3DOF_step_resp_smooth}
		\end{subfigure}
		\caption{3 \acsp{DOF} system: time and detailed time histories of the step input and the resulting $1^{st}$ \acs{DOF} displacement. ``Measured" signals are shown by a solid black line, while those estimated by filtering only and a sequence of filtering and smoothing (biased quasiperiodic covariance function) are respectively denoted via a dashed red and green line.}
		\label{fig:3DOF_step_smooth}
	\end{figure}

	\paragraph{Step load}\mbox{} \\
	In this paragraph, the input-state estimation results for the 3 \acp{DOF} system subjected to a step load are investigated. A quasiperiodic and an exponential covariance function have been selected for this problem following the same reasons standing for the impulse response. However, a step type of input features a static component which is normally difficult to identify via Bayesian estimators since these tools exploit a dynamical model of the system under study and a non-stationary model such as the random-walk for the unknown input. The \ac{GPLFM} approach can remedy to this problem if: i) displacement-level sensors are included in the observations set; ii) a constant offset is included in the covariance function used for regression of a displacement-level response. A biased quasiperiodic covariance function, as explained in \autoref{subsec:cov_functs}, embeds a static component in the prior when performing regression. This allows to fit the bias present in a displacement response caused by a step input, thus guaranteeing its representation within the resulting unknown input model employed in the input-state estimation. The same approach has been used to build a biased exponential covariance function as a summation of a constant and an exponential (Matérn with $\nu=0.5$) covariance function. The resulting covariance function has been adopted for estimation and results have been compared with the predictions obtained via the biased quasiperiodic covariance function. \Autoref{tab:3DOF_step} presents an overview of the initialization values selected for the algorithm necessary parameters. 
	\vspace{0.8em}
	\begin{table} [!ht]
		\centering
		\caption{3 \acsp{DOF} system, step load: estimators initialization values}
		\label{tab:3DOF_step}
		\renewcommand{\arraystretch}{2.0}
		\resizebox*{1\textwidth}{!}{
			\begin{threeparttable}			
				\begin{tabular}[H]{ccccccc}
					\toprule
					\begin{tabular}[c]{@{}c@{}} Initial state \\[-.4cm] mean\end{tabular}  & \begin{tabular}[c]{@{}c@{}} Initial error cov. \\[-.4cm] matrix\end{tabular}    & \begin{tabular}[c]{@{}c@{}}  Initial hyperparameters\\[-.4cm] biased quasiper.\end{tabular}     & \begin{tabular}[c]{@{}c@{}} Initial hyperparameters\\[-.4cm] biased exp.\end{tabular}    & \begin{tabular}[c]{@{}c@{}} Process noise \\[-.4cm] cov. matrix ($\mathbf{Q}$)\end{tabular} & \begin{tabular}[c]{@{}c@{}}  Measurement noise \\[-.4cm] cov. matrix ($\mathbf{R}$) \end{tabular}                                                                                                                                                                            
					\\ \midrule
					$\mathbf{\hat{x}}^a_{0|0} = \mathbf{0}$     & $\mathbf{\hat{P}}^x_{0|0} = \mathbf{0}$     & \begin{tabular}[c]{@{}c@{}} $\sigma^2 =2 \times 10^{-1}$ \\[-.3cm]$\sigma^2_{constant} =2 \times 10^{-1}$ \\[-.3cm] $l = 3 \times 10^{-1} $ \\[-.3cm] $t_{period}=0.3$ \\[-.3cm] $l_{matern}=1.3$ \end{tabular}                                                                                                                                                                                            & \begin{tabular}[c]{@{}c@{}} $\sigma^2 =2 \times 10^{-1}$ \\[-.3cm]$\sigma^2_{constant} =2 \times 10^{-1}$ \\[-.3cm] $l = 3 \times 10^{-1} $  \end{tabular} & \begin{tabular}[c]{@{}c@{}} $\mathbf{Q}_{\text{displ}} = 10^{-20} \times \mathbf{\mathbf{I}}_{\text{displ}}$   \\[-.3cm]     $\mathbf{Q}_{\text{vel}} = 10^{-10} \times \mathbf{\mathbf{I}}_{\text{vel}} $    \end{tabular}                                              & \begin{tabular}[c]{@{}c@{}} $\mathbf{R}_{\text{displ}} = 10^{-15} \times \mathbf{I}_{\text{displ}}$ \\[-.3cm] $\mathbf{R}_{\text{acc}} = 10^{-12} \times \mathbf{I}_{\text{acc}}$ \end{tabular} \\
					\bottomrule
				\end{tabular}
			\end{threeparttable}
		}
	\end{table}
\\ \Autoref{fig:3DOF_step} (left) proposes a time-domain comparison between the actual input and the estimations achieved using the \ac{GPLFM} approach with filtering only (both via the biased quasiperiodic covariance function and the biased exponential covariance function). 
\vspace{0.8em}
\\ The load time histories comparison shows that both the covariance functions allow to easily detect the step profile, but the biased exponential covariance function generates a large error at the last time instants of the analyzed time frame. Moreover, oscillations around the mean value are larger when adopting a biased exponential covariance function. \Autoref{fig:3DOF_step} (right) offers the actual $1^{st}$ \ac{DOF} displacement and the corresponding predictions achieved via the \ac{GPLFM} approach in time domain. The presented comparison shows that a good response prediction can be obtained via both the analyzed covariance functions. However, by analyzing the \acp{PSD} comparison in \autoref{fig:3DOF_step_freq}, it can be concluded that the response estimate produced by the biased quasiperiodic covariance function better matches the ``measured'' response frequency content. These results are in contrast to the impulse response case, where the exponential covariance function provides more accurate estimations. This can be due to a different effect introduced by the constant covariance function when used in combination with a quasiperiodic or an exponential covariance function. While in both cases it allows to capture the bias, it may influence differently the hyperparameters values of the ``dynamic'' (quasiperiodic or exponential) covariance function selected during the training phase. As a generic conclusion, it can be stated that both the proposed covariance functions can be adopted for the analyzed loading scenario as the results deviation is limited.
	\\ \Autoref{fig:3DOF_step_smooth} offers a time-domain comparison between the results obtained via filtering only and the ones achieved via sequential \ac{KF} and \ac{RTS} smoother in the step loading scenario. The input and response time histories estimated by combining filtering and smoothing show large deviations at the end of the analyzed time window. 
	\\ \Autoref{tab:RMSE_3DOF_step} quantifies the estimation inaccuracy for the analyzed estimators by offering the \ac{NRMSE} values computed between the estimated and the ``measured'' signals for both the unknown force and the predicted responses. \Autoref{tab:RMSE_3DOF_step} proves that adopting a biased quasiperiodic covariance function within a \ac{GPLFM} solved with filtering only allows for the optimal prediction performance for both input and responses under a step type of load.
	\begin{table} [!ht]
		\centering
		\caption{3 \acsp{DOF} system, step load: \acs{NRMSE} values between ``measured'' and estimated signals (responses and force).}
		\label{tab:RMSE_3DOF_step}
		\resizebox*{1\textwidth}{!}{
			\begin{threeparttable}
				\begin{tabular}[H]{c|ccc|ccc|ccc|c|c}		
					& 	\multicolumn{3}{c|}{$\acs{NRMSE}_{displ}$} & \multicolumn{3}{c|}{$\acs{NRMSE}_{vel}$} & \multicolumn{3}{c|}{$\acs{NRMSE}_{acc} $} & $\acs{NRMSE} $ & \acs{NRMSE}\\
					& \acs{DOF}1 & \acs{DOF}2 & \acs{DOF}3 & \acs{DOF}1 & \acs{DOF}2 & \acs{DOF}3 & \acs{DOF}1 & \acs{DOF}2 & \acs{DOF}3 & mean resp. & force\\
					\hline
					Biased quasiperiodic (\acs{KF}) & 0.016&0.017&0.005&0.173&0.251&0.180&0.287&0.428&0.513&0.208&0.055\\
					Biased exponential (\acs{KF}) & 	0.238&0.251&0.007&0.900&1.968&2.136&0.534&8.246&5.271&2.172&0.148 \\
					Biased quasiperiodic (\acs{KF}+\acs{RTS})  & 0.040&0.023&0.009&0.328&0.205&0.212&0.765&0.611&0.488&0.298&0.066\\

				\end{tabular}
			\end{threeparttable}
		}
	\end{table}

	\section{Case study: 3D-printed scaled titanium WT blade}
	\label{sec:3D_blade}
	The case study analyzed in this chapter concerns a 3D-printed scaled titanium WT blade \cite{brzhezinski2022dynamic}. The entire 3D-printed specimen,  manufactured by 3D Systems (Leuven, Belgium), comprises the scaled blade and a flange which was designed to allow the blade clamping. 
	\subsection{Measurement campaign}
	\label{subsec:3D_meas}
\begin{figure} [!htp]
		\hspace{-7em}
		\begin{subfigure}{1\textwidth}
			\includegraphics[width=0.35\textwidth]{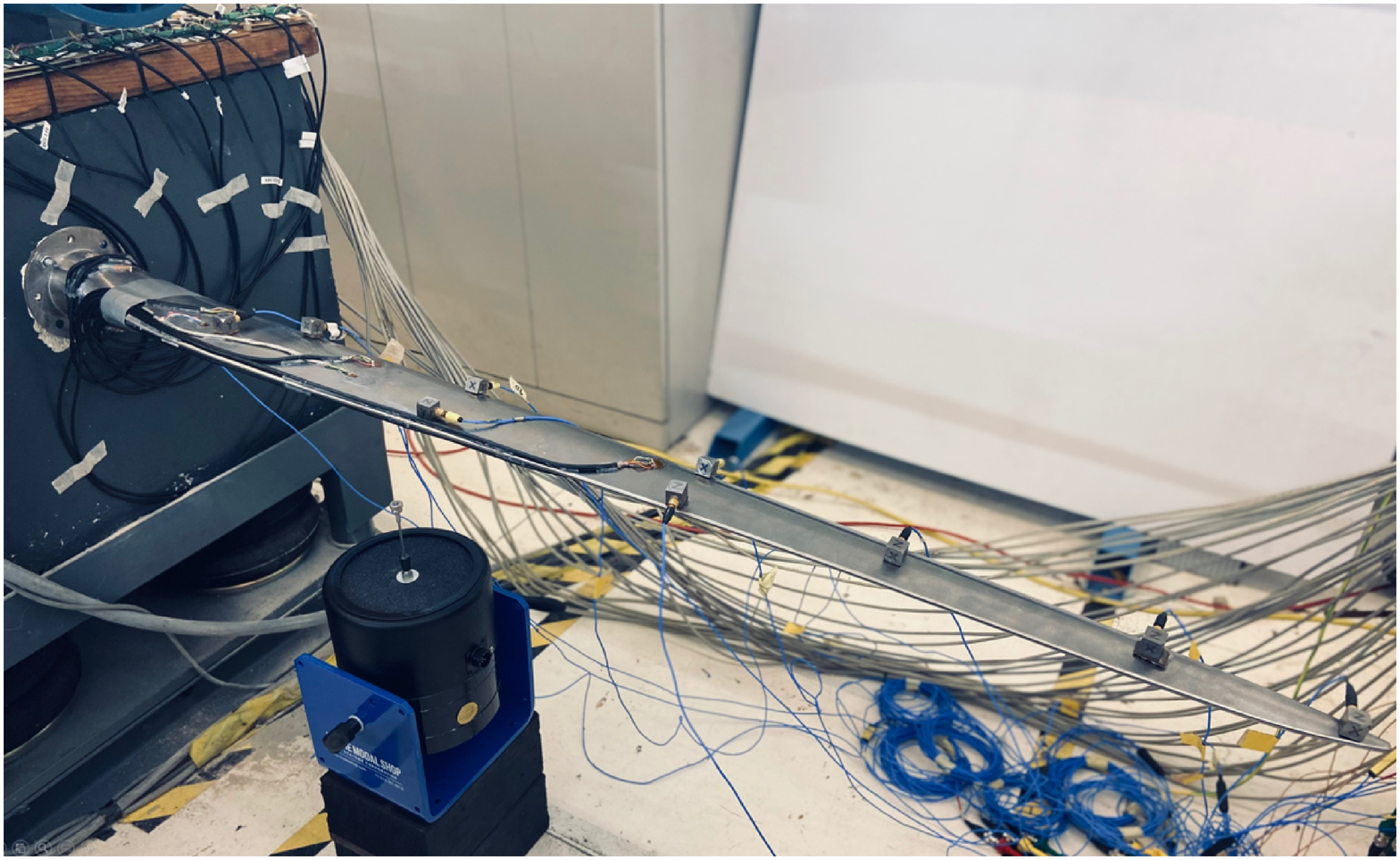}
		\end{subfigure}
		\hspace{-31.5em}
		\begin{subfigure}{0.46\textwidth}
			\includegraphics{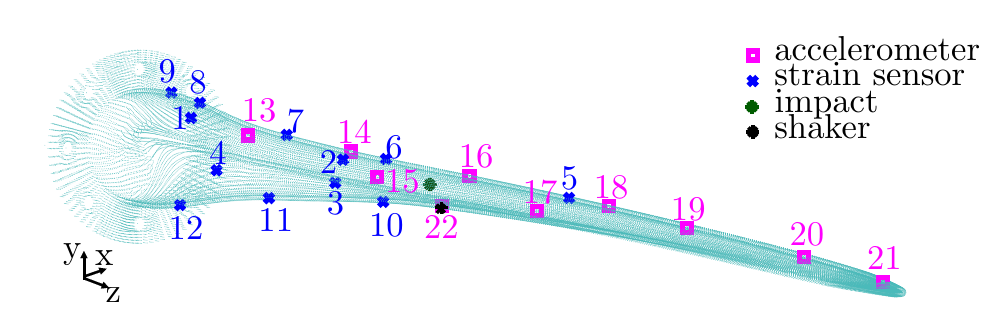}
		\end{subfigure}
		\caption{3D-printed scaled titanium \acs{WT} blade experimental setup (left) and sensing configuration (right): strain sensors (blue), accelerometers (magenta).}
		\label{fig:3D_setup}
	\end{figure}

	This section describes the setup adopted during the test campaign carried out at \ac{SISW} on the scaled blade, along with the series of tests performed with different types of excitation. \Autoref{fig:3D_setup} (left) shows the setup adopted for the measurements on the scaled \ac{WT} blade, during which the \ac{WT} blade was clamped to a concrete block. The entire setup includes 3 types of sensors: 4 uniaxial strain gauges, 8 rosettes and 10 triaxial accelerometers. \Autoref{fig:3D_setup} (right) shows the sensors locations. Strain sensors are arranged in sections along the blade on both the top and bottom surfaces, while accelerometers have been positioned only on the top surface. 

	During the test campaign, a Simcenter$^{TM}$ SCADAS system and Simcenter$^{TM}$ Testlab software have been used for data acquisition. Three types of tests have been performed according to the adopted excitation method: impact testing using a modal hammer, shaker testing and the so-called pull and release tests. \Autoref{fig:3D_setup} (right) shows the shaker and the impact locations. Data acquired during the impact test via the triaxial accelerometers has been used to determine the scaled \ac{WT} blade modal properties in the frequency range of interest 0-450 Hz, which have been extracted using Simcenter$^{TM}$ PolyMAX. During shaker testing, constant frequency sine tests and continuous random tests up to 450 Hz were carried out. Finally, pull and release tests were performed by applying an initial static load via a mass (known weight equal to 1.5 kg) at the blade tip. During this test the blade is then released by cutting the plastic tie used to hang the mass. This leads to free vibrations of the blade, which are recorded via the installed sensors. It is worth noting that during shaker and pull and release tests, the main component of the recorded acceleration and strain are respectively oriented along the vertical (Y) and axial (Z) directions. Therefore, the remainder of the section will only consider accelerations along Y and strains along Z.
	
	\subsection{Numerical model}
	\label{subsec:3D_FE}
\begin{figure} [!htp]
		\hspace{60em}
		\begin{subfigure}{1\textwidth}
			\includegraphics[width=.5\textwidth]{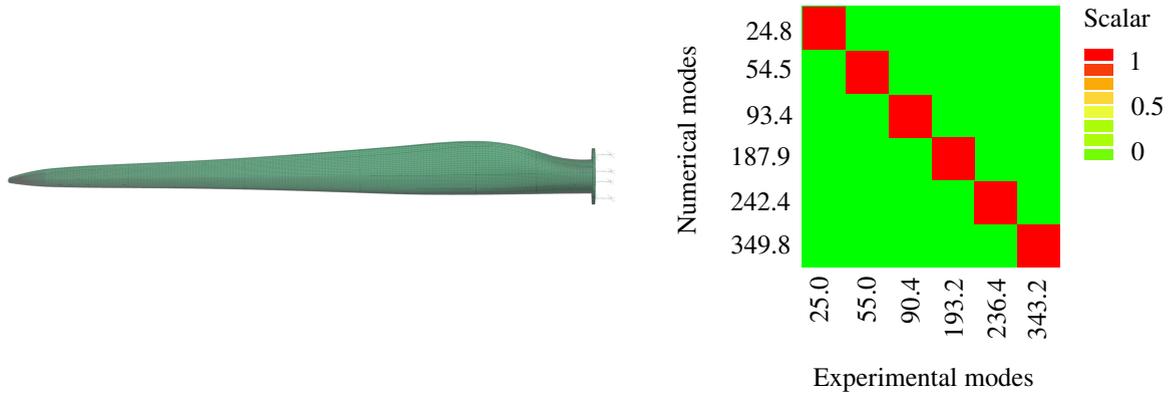}
			\vspace{7em}
		\end{subfigure}
		\hspace{-24em}
		\begin{subfigure}{0.48\textwidth}
			\input{MAC.TpX}
		\end{subfigure}
		\caption{3D-printed scaled titanium \acs{WT} blade \acs{FE} model (left). \acs{MAC} between numerical and experimental mode shapes from hammer test (right).}
		\label{fig:3D_blade_FE}
	\end{figure}

	The scaled \ac{WT} blade \ac{FE} model shown in \autoref{fig:3D_blade_FE} (left), has been developed in Simcenter$^{TM}$ 3D starting from the \ac{CAD} model used for 3D printing. The mesh is made up of around 65000 nodes and 33804 six-sided solid elements. The initial isotropic material (Ti6Al4V) properties have been defined according to the data sheet provided by the manufacturer. In order to reproduce the physical \acp{BC}, the flange has been divided into four parts, each connected to a spring via a \ac{RBE} connection. The free ends of the springs have been fixed. Additionally, concentrated masses have been added to the \ac{FE} model at the accelerometers locations to account for their presence and increase the correlation with experimental data. Numerical modes in the frequency range 0-450 Hz have been computed using NX$^{TM}$ Nastran SOL 103. NX$^{TM}$ Nastran SOL 200 has been adopted in Simcenter$^{TM}$ 3D for updating the \ac{FE} model using the experimental modal parameters as reference and a genetic algorithm for optimization. The following parameters have been set as design variables for the optimization process: springs stiffness, isotropic material Young's modulus, Poisson's ratio and density. \Autoref{fig:3D_blade_FE} (right) shows the final \ac{MAC} matrix between the reference experimental modes and the updated numerical modes. 
	\\A \ac{ROM} of the scaled blade has been built following the procedure outlined in \autoref{subsec:SSM} taking into account the first six modes (frequency range of interest: 0-450 Hz) and one residual attachment mode related to the unknown force to be estimated via the chosen Bayesian estimators. For shaker tests, the residual attachment mode has been computed by applying a unitary force at the shaker location pointed out in \autoref{fig:3D_setup} (right) along the positive Y direction. For pull and release tests, the unitary force for residual attachment mode computation has been applied at a tip node along the negative Y axis. 
	
	\subsection{Input-state estimation via \acp{GPLFM}}
	This section offers an experimental validation of the bespoke covariance function selection proposed in this work for input-state estimation via \acp{GPLFM}. By proposing a 3 \acp{DOF} example, \autoref{subsec:3DOFsexample} has proved that a proper covariance function selection plays an important role within the use of \acp{GPLFM} for estimating unknown loads and unmeasured structural responses in a Bayesian setting. In this section, three different loading conditions adopted during the 3D-printed scaled \ac{WT} blade tests are treated: pull and release, random and sinusoidal load. For each loading condition, the most suitable covariance function is proposed in contrast to a conventional Matérn covariance function for construction of the \ac{LFM} to be coupled with the scaled \ac{WT} blade combined deterministic-stochastic state-space model. It is worth noting that i) the filtering-only version is hereby adopted for the \ac{GPLFM} and ii) a single measurement is used for the training phase instead of the entire observations set. As mentioned in \autoref{subsec:LFMinputstate}, the most convenient choice for the time signal to be used for training consists in adopting a pre-recorded measurement collocated with the unknown input \cite{maes2015design}. Specifically, an acceleration signal is ideal for the random and sinusoidal load cases, while a displacement-level sensor, i.e., a strain signal, is preferred for the pull and release loading condition. 
	\subsubsection{Pull and release test}
\begin{figure} [!ht]
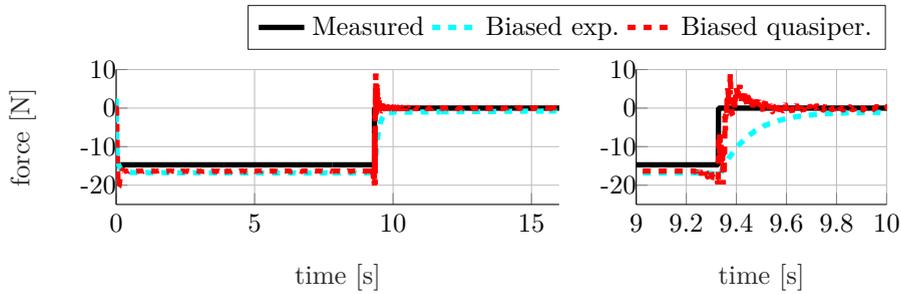

		\centering
		\includestandalone{TitaniumPR_input_cov}
		\caption{3D-printed scaled \acs{WT} blade pull and release test: input ``measured'' (black) and estimated (\acs{GPLFM} with biased exponential cov. - cyan, \acs{GPLFM} with biased quasiperiodic cov. - red) time histories (left) and detailed time histories (right)}
		\label{fig:TitaniumPR_input_cov}
	\end{figure}

\begin{figure} [!ht]
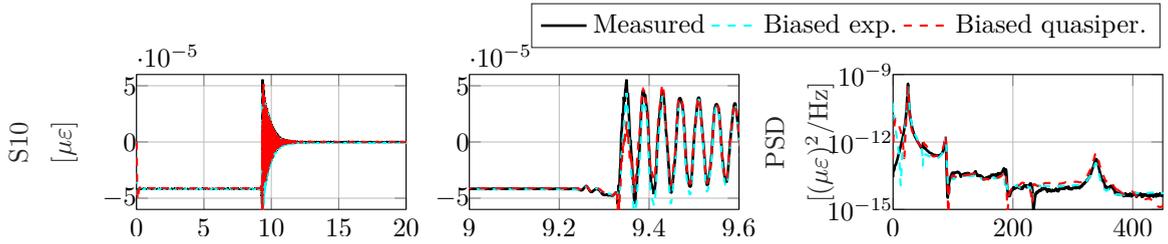

		\centering
		\includestandalone{TitaniumPR_response_sensor10_cov}
		\caption{3D-printed scaled \acs{WT} blade pull and release test: measured (black) and estimated (\acs{GPLFM} with biased exponential cov. - cyan, \acs{GPLFM} with biased quasiperiodic cov. - red) strain at location 10}
		\label{fig:TitaniumPR_sensor10_cov}
	\end{figure}

	This subsection proposes the use of a biased quasiperiodic covariance function for the \ac{GP} adopted to construct the \ac{LFM} for input-state estimation during the pull and release test performed on the 3D-printed scaled \ac{WT} blade. The bespoke covariance function selection is validated by comparing the estimation results against the ones achieved by making use of a biased exponential (equivalent to a Matérn function with $\nu=0.5$) covariance function. Specifically, sensor 1 has been hereby selected for training since no strain response collocated with the unknown load (applied at the tip) was available for this loading scenario. The hereby adopted estimators have been employed for input-state prediction using the mixed observations set (strain and acceleration) reported in \autoref{tab:initial_PR_cov}. The measurements acquired at the remaining locations have been used to validate the estimations provided by the investigated methods. The chosen values for the necessary initial conditions, the states process noise covariance and the measurement noise covariance matrices are also shown in \autoref{tab:initial_PR_cov}. Additionally, the analyzed \ac{GP} covariance functions initial hyperparameters are presented in \autoref{tab:initial_PR_cov}. 
	\begin{table} [!ht]
		\centering
		\caption{3D-printed scaled \acs{WT} blade pull and release test: \acsp{GPLFM} observations sets and initialization values for a biased exponential and a biased quasiperiodic covariance function }
		\label{tab:initial_PR_cov}
		\renewcommand{\arraystretch}{2.0}
		\resizebox*{1\textwidth}{!}{
			\begin{threeparttable}			
				\begin{tabular}[H]{ccccccc}
					\toprule
					Algorithm   & Observations                                                                                                                                     & \begin{tabular}[c]{@{}c@{}} Initial state \\[-.4cm] mean\end{tabular}                                                                                   & \begin{tabular}[c]{@{}c@{}} Initial error cov. \\[-.4cm] matrix\end{tabular}                                                                                                                                                                       &  Initial hyperparameters                                                                         & \begin{tabular}[c]{@{}c@{}} Process noise \\[-.4cm] cov. matrix ($\mathbf{Q}$)\end{tabular} & \begin{tabular}[c]{@{}c@{}}  Measurement noise \\[-.4cm] cov. matrix ($\mathbf{R}$) \end{tabular}                                                                                                                                                                             \\ \midrule
					Biased exponential & \textcolor{blue}{1}, \textcolor{blue}{3}, \textcolor{blue}{8}, \textcolor{purple}{15}, \textcolor{purple}{17}, \textcolor{purple}{21}             &  $\mathbf{\hat{x}}^a_{0|0} =\begin{bmatrix}
						\bm{\eta}\\\mathbf{0}
					\end{bmatrix}$                                                                                                                          & $\mathbf{\hat{P}}^x_{0|0} = 10^{-10} \times \mathbf{I}$                                                                                                                                                                                                & \begin{tabular}[c]{@{}c@{}}$\sigma^2 = 1.5$ \\[-.3cm]$\sigma^2_{constant} =1.1 \times 10^{-1}$ \\[-.3cm] $l = 1 \times 10^{-1} $ \end{tabular} &  \begin{tabular}[c]{@{}c@{}}$\mathbf{Q}_{displ} = 10^{-15} \times \mathbf{\mathbf{I}}$   \\[-.3cm]          $\mathbf{Q}_{vel} = 10^{-7} \times \mathbf{\mathbf{I}}$                                \end{tabular}            & \begin{tabular}[c]{@{}c@{}} $\mathbf{R}_{\text{strain}} = 10^{-14} \times \mathbf{I}$ \\[-.3cm] $\mathbf{R}_{\text{acc}} = 10^{-7} \times \mathbf{I}$ \end{tabular} \\ \hline
					
					Biased quasiperiodic & \textcolor{blue}{1}, \textcolor{blue}{3}, \textcolor{blue}{8}, \textcolor{purple}{15}, \textcolor{purple}{17}, \textcolor{purple}{21}             &  $\mathbf{\hat{x}}^a_{0|0} =\begin{bmatrix}
						\bm{\eta}\\\mathbf{0}
					\end{bmatrix}$                                                                                                                         & $\mathbf{\hat{P}}^x_{0|0} = 10^{-10} \times \mathbf{I}$                                                                                                                                                                                                & \begin{tabular}[c]{@{}c@{}} $\sigma^2 =2 \times 10^{-1}$ \\[-.3cm]$\sigma^2_{constant} =2 \times 10^{-1}$ \\[-.3cm] $l = 3 \times 10^{-1} $ \\[-.3cm] $t_{period}=4 \times 10^{-2}$ \\[-.3cm] $l_{matern}=10^{-1}$ \end{tabular} & \begin{tabular}[c]{@{}c@{}}$\mathbf{Q}_{displ} = 10^{-15} \times \mathbf{\mathbf{I}}$   \\[-.3cm]          $\mathbf{Q}_{vel} = 10^{-7} \times \mathbf{\mathbf{I}} $ \end{tabular} & \begin{tabular}[c]{@{}c@{}} $\mathbf{R}_{\text{strain}} = 10^{-14} \times \mathbf{I}$ \\[-.3cm] $\mathbf{R}_{\text{acc}} = 10^{-7} \times \mathbf{I}$ \end{tabular} \\
					\bottomrule
				\end{tabular}
			\end{threeparttable}
		}
	\end{table}
	\\ \Autoref{fig:TitaniumPR_input_cov} offers the predicted input signals provided by the \ac{GP}-based approach when the two investigated covariance functions are adopted. Predictions are compared against the ``measured'' force, i.e., the step-type of input reconstructed on the basis of the known weight of the applied mass. \Autoref{tab:TitaniumPR_cov_input_error} presents the \ac{SE} and the \ac{SD} inaccuracy indicators for the two analyzed cases. The \ac{SE} is formulated as the difference between the actual static input and the mean value of the estimated input profile. The \ac{SD}, instead, quantifies the oscillations that affect the estimated time history after the blade release. \Autoref{fig:TitaniumPR_input_cov} highlights that the \ac{GPLFM} with both the proposed covariance functions can easily detect the initial static load with limited inaccuracy. However, while the biased quasiperiodic covariance function allows for a correct detection of the release instant, the biased exponential covariance function produces a time history which does not follow the instantaneous release. Indeed, the estimated signal features a slow decay to zero which is reflected in both high \ac{SE} and \ac{SD}. It is therefore concluded that the use of a biased quasiperiodic covariance function to construct the \ac{GPLFM} outperforms the biased exponential covariance function on the input estimation task during pull and release tests on the scaled \ac{WT} blade.
	\begin{table}[htpb]
		\centering
		\caption {3D-printed scaled \acs{WT} blade pull and release test: input prediction errors for the \acs{GPLFM} with biased exponential and biased quasiperiodic covariance functions}
		\begin{tabular}{c|cc}
			& Biased exponential& Biased quasiperiodic \\
			\hline
			SE  &   $ 11.23 N$ & $ 1.06 N$\\    		
			\hline
			SD &    $ 1.91 N$ &   $0.20 N$ \\
		\end{tabular}
		\label{tab:TitaniumPR_cov_input_error}
	\end{table}
	\\ \Multiref{fig:TitaniumPR_sensor10_cov}{fig:TitaniumPR_sensor14_cov} show the time and frequency content of the responses estimated by the \ac{GPLFM} algorithm for strain gauge 10 and accelerometer 14 when the analyzed covariance functions are used. In both cases, the estimated signals match their measured counterparts with relatively high accuracy. 
	\begin{figure} [!ht]
		\centering
		\includestandalone{TitaniumPR_response_sensor14_cov}
		\caption{3D-printed scaled \acs{WT} blade pull and release test: measured (black) and estimated (\acs{GPLFM} with biased exponential cov. - cyan, \acs{GPLFM} with biased quasiperiodic cov. - red) acceleration at location 14}
		\label{fig:TitaniumPR_sensor14_cov}
	\end{figure}
	\\A global analysis of the prediction results is offered in \multiref{tab:TRAC_PR_sg_cov}{tab:TRAC_PR_acc_cov} via the \ac{TRAC} \cite{VETTORI2023109654} values for the entire set of estimated responses (respectively strain and acceleration). The \ac{FRAC} \cite{VETTORI2023109654} values for the strain and acceleration responses are instead offered in \multiref{tab:FRAC_PR_sg_cov}{tab:FRAC_PR_acc_cov} respectively. The analysis of the \ac{TRAC} and \ac{FRAC} indicators shows that, while a similar result is achieved in terms of strain response estimation, the accuracy of acceleration response prediction is substantially higher when adopting a biased quasiperiodic covariance function. This is also demonstrated by the frequency content provided in \autoref{fig:TitaniumPR_sensor14_cov}, where a low frequency component erroneously affects the signal estimated by the biased exponential covariance function. To the contrary, the biased quasiperiodic covariance function produces a \ac{PSD} which matches the measured signal in the entire reported frequency range (including low frequency).
	\begin{table} [!ht]
		\centering
		\caption{3D-printed scaled \acs{WT} blade pull and release test: \acs{TRAC} values between measured and estimated (\acs{GPLFM} with biased exponential and biased quasiperiodic covariance functions) strain responses. Observations are underlined. }
		\label{tab:TRAC_PR_sg_cov}
		\resizebox*{1\textwidth}{!}{
			\begin{threeparttable}
				\renewcommand{\arraystretch}{1.7}
				\setlength{\tabcolsep}{7pt}
				\begin{tabular}[H]{c|cccccccccccc}				
					Estimator & \textcolor{blue}{1} & \textcolor{blue}{2} & \textcolor{blue}{3} & \textcolor{blue}{4} & \textcolor{blue}{5} & \textcolor{blue}{6} & \textcolor{blue}{7} & \textcolor{blue}{8} & \textcolor{blue}{9} & \textcolor{blue}{10} & \textcolor{blue}{11} & \textcolor{blue}{12} \\
					\hline
					Biased exponential & \underline{0.997}	&0.997	&\underline{0.998}&0.998	&0.996	&0.998	&0.998	&\underline{0.998}	&0.995	&0.997&	0.998	&0.998
					\\
					Biased quasiperiodic  &		\underline{0.994}&	0.995	&\underline{0.995}&	0.994&	0.995&	0.995&	0.995	&\underline{0.993}	&0.989	&0.995	&0.995&	0.995
					\\
				\end{tabular}
			\end{threeparttable}
		}
	\end{table}
	\begin{table} [!ht]
		\centering
		\caption{3D-printed scaled \acs{WT} blade pull and release test: \acs{TRAC} values between measured and estimated (\acs{GPLFM} with biased exponential and biased quasiperiodic covariance functions) acceleration responses. Observations are underlined. }
		\label{tab:TRAC_PR_acc_cov}
		\resizebox*{0.81\textwidth}{!}{
			\begin{threeparttable}
				\renewcommand{\arraystretch}{1.7}
				\setlength{\tabcolsep}{7pt}
				\begin{tabular}[H]{c|ccccccccccc}				
					Estimator & \textcolor{magenta}{13} & \textcolor{magenta}{14} & \textcolor{magenta}{15} & \textcolor{magenta}{16} & \textcolor{magenta}{17} & \textcolor{magenta}{18} & \textcolor{magenta}{19} & \textcolor{magenta}{20} & \textcolor{magenta}{21} \\
					\hline
					Biased exponential	& 0.282	&0.459	&\underline{0.997}	&0.961	&\underline{1.000}	&0.971	&0.408	&0.729	&\underline{1.000}  \\
					Biased quasiperiodic & 			0.762	&0.866&	\underline{1.000}&	0.961&	\underline{1.000}&	0.973&	0.904&	0.931&\underline{	1.000}\\
				\end{tabular}
			\end{threeparttable}
		}
	\end{table}
	\begin{table} [!ht]
		\centering
		\caption{3D-printed scaled \acs{WT} blade pull and release test: \acs{FRAC} values between measured and estimated (\acs{GPLFM} with biased exponential and biased quasiperiodic covariance functions) strain responses. Observations are underlined. }
		\label{tab:FRAC_PR_sg_cov}
		\resizebox*{1\textwidth}{!}{
			\begin{threeparttable}
				\renewcommand{\arraystretch}{1.7}
				\setlength{\tabcolsep}{7pt}
				\begin{tabular}[H]{c|ccccccccccccc}
					
					Estimator & \textcolor{blue}{1} & \textcolor{blue}{2} & \textcolor{blue}{3} & \textcolor{blue}{4} & \textcolor{blue}{5} & \textcolor{blue}{6} & \textcolor{blue}{7} & \textcolor{blue}{8} & \textcolor{blue}{9} & \textcolor{blue}{10} & \textcolor{blue}{11} & \textcolor{blue}{12} \\
					\hline
					Biased exponential   &\underline{0.992}&	0.988	&\underline{0.996}	&0.999&	0.919&	0.993	&0.998	&\underline{0.999}	&0.999	&0.990	&0.997&	0.998\\				
					Biased quasiperiodic &			\underline{0.982}&	0.991	&\underline{0.994}	&0.997	&0.972&	0.993	&0.996&	\underline{0.997}&	0.997	&0.992&	0.995	&0.996
					\\             
				\end{tabular}			
			\end{threeparttable}
		}
	\end{table}
	\begin{table} [!ht]
		\centering
		\caption{3D-printed scaled \acs{WT} blade pull and release test: \acs{FRAC} values between measured and estimated (\acs{GPLFM} with biased exponential and biased quasiperiodic covariance functions) acceleration responses. Observations are underlined. }
		\label{tab:FRAC_PR_acc_cov}
		\resizebox*{0.81\textwidth}{!}{
			\begin{threeparttable}
				\renewcommand{\arraystretch}{1.7}
				\setlength{\tabcolsep}{7pt}
				\begin{tabular}[H]{c|ccccccccccc}					
					Estimator & \textcolor{magenta}{13} & \textcolor{magenta}{14} & \textcolor{magenta}{15} & \textcolor{magenta}{16} & \textcolor{magenta}{17} & \textcolor{magenta}{18} & \textcolor{magenta}{19} & \textcolor{magenta}{20} & \textcolor{magenta}{21} \\
					\hline
					Biased exponential &0.445&	0.604	&\underline{1.000}&	0.991	&\underline{1.000}&	0.984	&0.481	&0.954	&\underline{1.000}\\				
					Biased quasiperiodic &				0.837	&0.864	&\underline{1.000}&	0.992	&\underline{1.000}	&0.981&	0.994	&0.985&\underline{	1.000}\\
				\end{tabular}			
			\end{threeparttable}
		}
	\end{table}

	\subsubsection{Random test}
\begin{figure} [!ht]
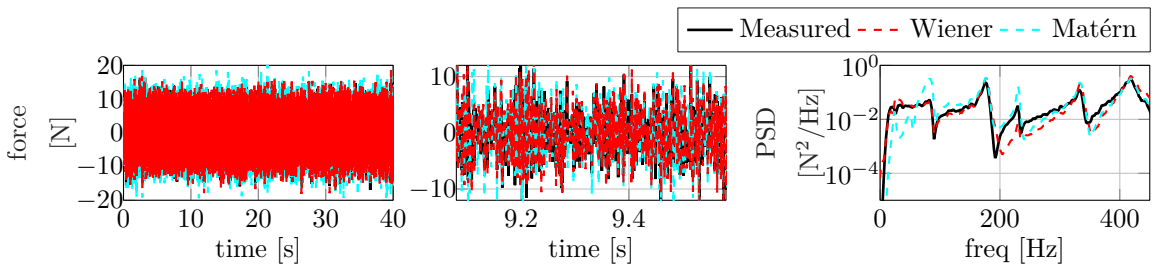

		\centering
		\includestandalone{TitaniumRnd_input_cov}
		\caption{3D-printed scaled \acs{WT} blade random test: input measured (black) and estimated (\acs{GPLFM} with Matérn cov. - cyan, \acs{GPLFM} with Wiener cov. - red) force signals}
		\label{fig:TitaniumRnd_input_cov}
	\end{figure}

	This subsection proposes the use of a Wiener covariance function for the \ac{GP} adopted within the \ac{LFM} constructed for input-state estimation during random tests performed on the 3D-printed scaled \ac{WT} blade. The bespoke covariance function selection is validated by comparing the estimation results against the ones achieved by making use of a conventional Matérn covariance function ($\nu=1.5$) for the mixed observations set (strain and acceleration) reported in \autoref{tab:initial_random_cov}. The chosen values for the necessary initial conditions and the process and measurement noise covariance matrices are also summarized in \autoref{tab:initial_random_cov}. Additionally, the analyzed \ac{GP} covariance functions initial hyperparameters are presented in \autoref{tab:initial_random_cov}. The hyperparameters have been tuned during the training phase by maximizing the log marginal likelihood of the acceleration collocated with the unknown input, i.e., at the shaker location indicated in \autoref{fig:3D_setup} (right). 
	\begin{table} [!ht]
		\centering
		\caption{3D-printed scaled \acs{WT} blade random test: \acsp{GPLFM} observations sets and initialization values for a Matérn ($\nu$=1.5) and a Wiener covariance function }
		\label{tab:initial_random_cov}
		\renewcommand{\arraystretch}{2.0}
		\resizebox*{1\textwidth}{!}{
			\begin{threeparttable}				
				\begin{tabular}[H]{c|cccccc}
					Estimator   & Observations                                                                                                                                                                            & \begin{tabular}[c]{@{}c@{}} Initial state \\[-.4cm] mean\end{tabular} & \begin{tabular}[c]{@{}c@{}} Initial error cov. \\[-.4cm] matrix\end{tabular} & Initial hyperparameters                                                                  & \begin{tabular}[c]{@{}c@{}} Process noise \\[-.4cm]cov. matrix ($\mathbf{Q}$)\end{tabular} & \begin{tabular}[c]{@{}c@{}}  Measurement noise \\[-.4cm]cov. matrix ($\mathbf{R}$) \end{tabular}                                                                                                                                                                             \\ \hline
					Matérn   & \textcolor{blue}{1}, \textcolor{blue}{3}, \textcolor{blue}{8},\textcolor{blue}{12}, \textcolor{magenta}{15}, \textcolor{magenta}{18}, \textcolor{magenta}{21}, \textcolor{magenta}{22}                            & $\mathbf{\hat{x}}^a_{0|0} = \mathbf{0}$                                       & $\mathbf{\hat{P}}^a_{0|0} = 10^{-10} \times \mathbf{I}$                          & \begin{tabular}[c]{@{}c@{}} $\sigma^2 =3 \times 10^{3}$ \\ [.3cm] $l=2 \times 10^{-4}$ \end{tabular}                                                                                                            & \begin{tabular}[c]{@{}c@{}}$\mathbf{Q}_{displ} = 10^{-20} \times \mathbf{\mathbf{I}}$   \\[-.3cm]      $\mathbf{Q}_{vel} = 10^{-10}  \times \mathbf{\mathbf{I}} $ \end{tabular}       & \begin{tabular}[c]{@{}c@{}} $\mathbf{R}_{\text{strain}} = 10^{-14} \times \mathbf{\mathbf{I}}$ \\[-.3cm] $\mathbf{R}_{\text{acc}} = 10^{-7} \times \mathbf{\mathbf{I}}$ \end{tabular}      \\ \hline
					
					Wiener & \textcolor{blue}{1}, \textcolor{blue}{3}, \textcolor{blue}{8}, \textcolor{blue}{12}, \textcolor{magenta}{15}, \textcolor{magenta}{18}, \textcolor{magenta}{21}, \textcolor{magenta}{22} & $\mathbf{\hat{x}}^a_{0|0} = \mathbf{0}$                                       & $\mathbf{\hat{P}}^x_{0|0} = 10^{-10} \times \mathbf{I}$                          &  $\sigma^2 =3 \times 10^{3}$  &\begin{tabular}[c]{@{}c@{}}$\mathbf{Q}_{displ} = 10^{-20} \times \mathbf{\mathbf{I}}$   \\[-.3cm]          $\mathbf{Q}_{vel} = 10^{-10}  $ \end{tabular}   & \begin{tabular}[c]{@{}c@{}} $\mathbf{R}_{\text{strain}} = 10^{-14} \times \mathbf{\mathbf{I}}  $ \\[-.3cm] $\mathbf{R}_{\text{acc}} = 10^{-7} \times \mathbf{\mathbf{I}} $ \end{tabular} \\
				\end{tabular}
			\end{threeparttable}
		}
	\end{table}
	\\ \Autoref{fig:TitaniumRnd_input_cov} offers a comparison in both time and frequency domains between the actual force, recorded by a force cell during the test, and the input predictions obtained by adopting the \ac{GPLFM} in combination with a Wiener and a Matérn covariance function ($\nu=1.5$). A quantification of the input estimations accuracy in time and frequency domains is respectively reported in \multiref{tab:TRAC_random_acc_cov}{tab:FRAC_random_acc_cov} (last column) by means of the \ac{TRAC} and \ac{FRAC} estimators. From the results reported in \multiref{tab:TRAC_random_acc_cov}{tab:FRAC_random_acc_cov}, it can be concluded that a Wiener covariance function allows for higher input estimation accuracy. \Autoref{fig:TitaniumRnd_input_cov} indeed highlights that the Matérn covariance function generates a distorted frequency content between 0 and 100 Hz.
	\\ \Multiref{fig:TitaniumRnd_sg_cov}{fig:TitaniumRnd_acc_cov} present the comparison between measured and estimated signals respectively for a strain and an accelerometer that are not included in the observations set. These results demonstrate that a good prediction accuracy can be achieved by the \ac{GPLFM} for both strain and acceleration predictions when each of the two analyzed covariance functions is adopted. However, the Matérn covariance function seems to produce a mismatch between 0 and 90 Hz in both the strain and the acceleration response. An overall information regarding the response estimation precision is reported in \multirefs{tab:TRAC_random_sg_cov}{tab:TRAC_random_acc_cov}{tab:FRAC_random_sg_cov}{tab:FRAC_random_acc_cov} by means of the \ac{TRAC} and \ac{FRAC} indicators. The presented indicators show higher values (especially for strains) when the Wiener covariance function is adopted, confirming the conclusion drawn from  \multiref{fig:TitaniumRnd_sg_cov}{fig:TitaniumRnd_acc_cov}. 
	\begin{figure} [!htp]
		\centering
		\includestandalone{TitaniumRnd_sensor6_cov}
		\caption{3D-printed scaled \acs{WT} blade random test: measured (black) and estimated (\acs{GPLFM} with Matérn cov. - cyan, \acs{GPLFM} with Wiener cov. - red) strain at location 6}
		\label{fig:TitaniumRnd_sg_cov}
	\end{figure}
	\begin{figure} [!htp]
		\centering
		\includestandalone{TitaniumRnd_sensor19_cov}			
		\caption{3D-printed scaled \acs{WT} blade random test: measured (black) and estimated (\acs{GPLFM} with Matérn cov. - cyan, \acs{GPLFM} with Wiener cov. - red) acceleration at location 19}
		\label{fig:TitaniumRnd_acc_cov}
	\end{figure}
	\begin{table} [!htp]
		\centering
		\caption{3D-printed scaled \acs{WT} blade random test: \acs{TRAC} values between measured and estimated (\acs{GPLFM} with Matérn and Wiener covariance functions) strain responses. Observations are underlined. }
		\label{tab:TRAC_random_sg_cov}
		\resizebox*{1\textwidth}{!}{
			\begin{threeparttable}
				\renewcommand{\arraystretch}{1.7}
				\setlength{\tabcolsep}{7pt}
				\begin{tabular}[H]{c|cccccccccccc}		
					Estimator & \textcolor{blue}{1} & \textcolor{blue}{2} & \textcolor{blue}{3} & \textcolor{blue}{4} & \textcolor{blue}{5} & \textcolor{blue}{6} & \textcolor{blue}{7} & \textcolor{blue}{8} & \textcolor{blue}{9} & \textcolor{blue}{10} & \textcolor{blue}{11} & \textcolor{blue}{12} \\
					\hline
					Matérn &\underline{0.595}	&0.584&	\underline{0.513}	&0.510	&0.743	&0.601&	0.652&	\underline{0.654}&	0.693&	0.581&	0.630&	\underline{0.607}
					\\ 
					Wiener  &	\underline{0.938}&	0.848&\underline{	0.825}&	0.889	&0.828	&0.903&	0.889&\underline{	0.959	}&0.951	&0.853&	0.943&\underline{	0.942}
					\\
				\end{tabular}
			\end{threeparttable}
		}
	\end{table}
	\begin{table} [!ht]
		\centering
		\caption{3D-printed scaled \acs{WT} blade random test: \acs{TRAC} values between measured and estimated (\acs{GPLFM} with Matérn and Wiener covariance functions) force and acceleration responses. Observations are underlined.}
		\label{tab:TRAC_random_acc_cov}
		\resizebox*{0.93\textwidth}{!}{
			\begin{threeparttable}
				\renewcommand{\arraystretch}{1.7}
				\setlength{\tabcolsep}{7pt}
				\begin{tabular}[H]{c|ccccccccccc}				
					Estimator & \textcolor{magenta}{13} & \textcolor{magenta}{14} & \textcolor{magenta}{15} & \textcolor{magenta}{16} & \textcolor{magenta}{17} & \textcolor{magenta}{18} & \textcolor{magenta}{19} & \textcolor{magenta}{20} & \textcolor{magenta}{21} & \textcolor{magenta}{22} & \textbf{force} \\
					\hline
					Matérn &0.914&	0.494	&\underline{1.000}	&0.305	&0.095	&\underline{1.000}	&0.882	&0.558	&\underline{1.000}	&\underline{0.977	}&0.642\\
					Wiener  &				0.913&	0.532&	\underline{0.988}&	0.329	&0.109	&\underline{0.964}	&0.960&	0.628	&\underline{0.989}	&\underline{0.999	}&0.863\\
				\end{tabular}
			\end{threeparttable}
		}
	\end{table}
	\begin{table} [!ht]
		\centering
		\caption{3D-printed scaled \acs{WT} blade random test: \acs{FRAC} values between measured and estimated (\acs{GPLFM} with Matérn and Wiener covariance functions) strain responses. Observations are underlined.}
		\label{tab:FRAC_random_sg_cov}
		\resizebox*{1\textwidth}{!}{
			\begin{threeparttable}
				\renewcommand{\arraystretch}{1.7}
				\setlength{\tabcolsep}{7pt}
				\begin{tabular}[H]{c|ccccccccccccc}				
					Estimator & \textcolor{blue}{1} & \textcolor{blue}{2} & \textcolor{blue}{3} & \textcolor{blue}{4} & \textcolor{blue}{5} & \textcolor{blue}{6} & \textcolor{blue}{7} & \textcolor{blue}{8} & \textcolor{blue}{9} & \textcolor{blue}{10} & \textcolor{blue}{11} & \textcolor{blue}{12} \\
					\hline
					Matérn           & \underline{0.730}	&0.673&	\underline{0.937}&	0.943&	0.907&	0.912&	0.855&	\underline{0.851}&	0.899&	0.794	&0.756	&\underline{0.878		}						\\
					Wiener            &   				\underline{0.961}&	0.890	&\underline{0.931}	&0.974	&0.786	&0.936&	0.9867&	\underline{0.968}&	0.977&	0.920&	0.993	&\underline{0.997}				\\             
				\end{tabular}
				
			\end{threeparttable}
		}
	\end{table}
	\vspace{-.5em}
	\begin{table} [!ht]
		\centering
		\caption{3D-printed scaled \acs{WT} blade random test: \acs{FRAC} values between measured and estimated (\acs{GPLFM} with Matérn and Wiener covariance functions) force and acceleration responses. Observations are underlined.}
		\label{tab:FRAC_random_acc_cov}
		\resizebox*{0.93\textwidth}{!}{
			\begin{threeparttable}
				\renewcommand{\arraystretch}{1.7}
				\setlength{\tabcolsep}{7pt}
				\begin{tabular}[H]{c|ccccccccccc}					
					Estimator & \textcolor{magenta}{13} & \textcolor{magenta}{14} & \textcolor{magenta}{15} & \textcolor{magenta}{16} & \textcolor{magenta}{17} & \textcolor{magenta}{18} & \textcolor{magenta}{19} & \textcolor{magenta}{20} & \textcolor{magenta}{21} & \textcolor{magenta}{22} & \textbf{force} \\
					\hline
					Matérn             & 0.981&	0.239&	\underline{1.000}&	0.107	&0.702&	\underline{1.000}&	0.979&	0.842&	\underline{1.000}&\underline{	0.943}&	0.781\\
					Wiener         &  			 0.969&	0.280	&\underline{0.973}&	0.109&	0.594&	\underline{0.966}&	0.997&	0.831&	\underline{1.000}&	\underline{0.997}	&0.937	\\		 				
				\end{tabular}			
			\end{threeparttable}
		}
	\end{table}
	\vspace{-.5em}
	\subsubsection{Sine test}
\begin{figure} [!ht]
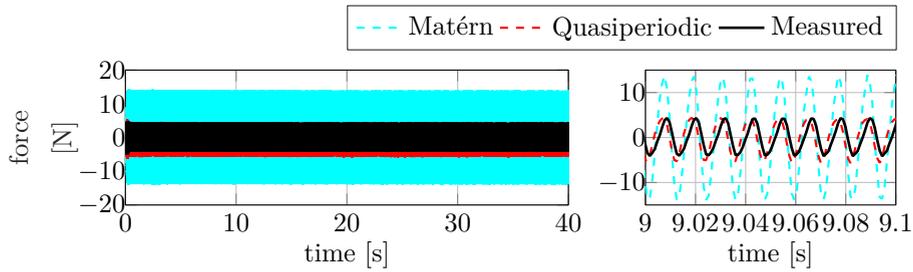

		\centering
		\includestandalone{TitaniumSine_input_cov}
		\caption{3D-printed scaled \acs{WT} blade sine test: input measured (black) and estimated (\acs{GPLFM} with Matérn cov. - cyan, \acs{GPLFM} with quasiperiodic cov. - red) force signals}
		\label{fig:TitaniumSine_input_cov}
	\end{figure}

\begin{figure} [!ht]
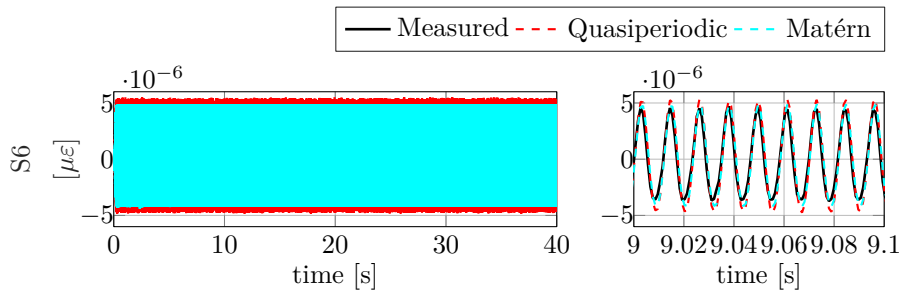

		\centering
		\includestandalone{TitaniumSine_response_sensors6_cov}
		\caption{3D-printed scaled \acs{WT} blade sine test: measured (black) and estimated (\acs{GPLFM} with Matérn cov. - cyan, \acs{GPLFM} with quasiperiodic cov. - red) strain at location 6}
		\label{fig:TitaniumSine_sg_cov}
	\end{figure}

	The use of a quasiperiodic covariance function for constructing a \ac{GPLFM} is hereby explored for input-state estimation during sinusoidal tests ($f_{sine}=86 Hz$) performed on the 3D-printed scaled \ac{WT} blade. As anticipated in \autoref{subsec:LFMinputstate}, experimentally recorded data are often affected by external sources of disturbance. For this reason, even when a sinusoidal signal is selected to drive the shaker, the actual load acting on the structure and recorded by the load cell may deviate from the pure sine assumption. To take this effect into account when constructing the \ac{GPLFM}, a quasiperiodic covariance function is preferred to a periodic covariance function. The bespoke covariance function selection is hereby adopted for prediction during sine tests using the mixed observations set (strain and acceleration) reported in \autoref{tab:initial_sine_cov}. The estimation results are compared against the ones achieved by making use of the same measurements set in combination with a conventional Matérn covariance function ($\nu=1.5$). The chosen values for the necessary initial conditions and the process and measurement noise covariance matrices are also shown, along with the \ac{GP} covariance functions initial hyperparameters. The initial period of the quasiperiodic covariance function has been set such that $t_{period} = 1/f_{sine}$. The hyperparameters have been selected by maximizing the log marginal likelihood of the acceleration collocated with the unknown input, i.e., at the shaker location indicated in \autoref{fig:3D_setup} (right). 
	\vspace{-.7em}
	\begin{table} [!ht]
		\centering
		\caption{3D-printed scaled \acs{WT} blade sine test: \acsp{GPLFM} observations sets and initialization values for a Matérn ($\nu$=1.5) and a quasiperiodic covariance function }
		\label{tab:initial_sine_cov}
		\renewcommand{\arraystretch}{2.0}
		\resizebox*{1\textwidth}{!}{
			\begin{threeparttable}				
				\begin{tabular}[H]{c|cccccc}
					Estimator   & Observations                                                                                                                                                                            & \begin{tabular}[c]{@{}c@{}} Initial state \\[-.4cm] mean\end{tabular} & \begin{tabular}[c]{@{}c@{}} Initial error cov. \\[-.4cm] matrix\end{tabular} & \begin{tabular}[c]{@{}c@{}}  Initial hyperparameters                                                   \end{tabular}               & \begin{tabular}[c]{@{}c@{}} Process noise \\[-.4cm]cov. matrix ($\mathbf{Q}$)\end{tabular} & \begin{tabular}[c]{@{}c@{}}  Measurement noise \\[-.4cm]cov. matrix ($\mathbf{R}$) \end{tabular}                                                                                                                                                                             \\ \hline
					Matérn   & \textcolor{blue}{1}, \textcolor{blue}{3}, \textcolor{blue}{8},\textcolor{blue}{12}, \textcolor{magenta}{15}, \textcolor{magenta}{18}, \textcolor{magenta}{21}, \textcolor{magenta}{22}                            & $\mathbf{\hat{x}}^a_{0|0} = \mathbf{0}$                                       & $\mathbf{\hat{P}}^a_{0|0} = 10^{-10} \times \mathbf{I}$                          & \begin{tabular}[c]{@{}c@{}} $\sigma^2 =3 \times 10^{3}$ \\ [-.3cm] $l=2 \times 10^{-4}$ \end{tabular}                                                                                                            & \begin{tabular}[c]{@{}c@{}}$\mathbf{Q}_{displ} = 10^{-20} \times \mathbf{\mathbf{I}}$   \\[-.3cm]      $\mathbf{Q}_{vel} = 10^{-10}  \times \mathbf{\mathbf{I}} $ \end{tabular}       & \begin{tabular}[c]{@{}c@{}} $\mathbf{R}_{\text{strain}} = 10^{-14} \times \mathbf{\mathbf{I}}$ \\[-.3cm] $\mathbf{R}_{\text{acc}} = 10^{-7} \times \mathbf{\mathbf{I}}$ \end{tabular}      \\ \hline
					
					Quasiperiodic & \textcolor{blue}{1}, \textcolor{blue}{3}, \textcolor{blue}{8}, \textcolor{blue}{12}, \textcolor{magenta}{15}, \textcolor{magenta}{18}, \textcolor{magenta}{21}, \textcolor{magenta}{22} & $\mathbf{\hat{x}}^a_{0|0} = \mathbf{0}$                                       & $\mathbf{\hat{P}}^x_{0|0} = 10^{-10} \times \mathbf{I}$                          &  \begin{tabular}[c]{@{}c@{}}$\sigma^2 =5 \times 10^{-1}$ \\[-.3cm] $l=5 \times 10^{-1}$\\[-.3cm] $t_{period}=1.2\times10^{-2}$\\[-.3cm] $l_{matern}=8\times10^{-1}$\end{tabular} &\begin{tabular}[c]{@{}c@{}}$\mathbf{Q}_{displ} = 10^{-20} \times \mathbf{\mathbf{I}}$   \\[-.3cm]          $\mathbf{Q}_{vel} = 10^{-10}  $ \end{tabular}   & \begin{tabular}[c]{@{}c@{}} $\mathbf{R}_{\text{strain}} = 10^{-14} \times \mathbf{\mathbf{I}}  $ \\[-.3cm] $\mathbf{R}_{\text{acc}} = 10^{-7} \times \mathbf{\mathbf{I}} $ \end{tabular} \\
				\end{tabular}
			\end{threeparttable}
		}
	\end{table}
\vspace{-.5em}
	\\ \Autoref{fig:TitaniumSine_input_cov} offers a time domain comparison between the actual force and the input predictions obtained via the \ac{GPLFM} with a quasiperiodic and a Matérn covariance function. During the tests, the force applied by the shaker has been recorded by a force cell placed between the shaker and the structure. A quantification of the input estimations accuracy is reported in \autoref{tab:TRAC_acc_sine_cov} (last column) by means of the \ac{TRAC} estimator. From the results reported in \autoref{fig:TitaniumSine_input_cov}, it can be concluded that while the \ac{GPLFM} with a Matérn covariance function overestimates the load amplitude, a more accurate estimation is obtained via the \ac{GPLFM} with a quasiperiodic covariance function. This result is confirmed by the \ac{TRAC} values offered in \autoref{tab:TRAC_acc_sine_cov}.
	\\ \Multiref{fig:TitaniumSine_sg_cov}{fig:TitaniumSine_acc_cov} present the comparison between measured and estimated signals respectively for a strain and an accelerometer that are not included in the observations set. These results demonstrate that a good and comparable prediction accuracy can be achieved by the \ac{GPLFM} with a quasiperiodic or a Matérn covariance function for both strain and acceleration predictions. An overall information regarding the response estimation precision is reported in \multiref{tab:TRAC_sg_sine_cov}{tab:TRAC_acc_sine_cov} by means of the \ac{TRAC} values. The presented indicators show comparable values for both strain and acceleration responses when the \ac{GPLFM} is adopted with one of the two analyzed covariance functions.  
	\begin{figure} [!ht]
		\centering
		\includestandalone{TitaniumSine_response_sensors19_cov}			
		\caption{3D-printed scaled \acs{WT} blade sine test: measured (black) and estimated (\acs{GPLFM} with Matérn cov. - cyan, \acs{GPLFM} with quasiperiodic cov. - red) acceleration at location 19}
		\label{fig:TitaniumSine_acc_cov}
	\end{figure}
	\begin{table} [!ht]
		\centering
		\caption{3D-printed scaled \acs{WT} blade sine test: \acs{TRAC} values between measured and estimated (\acs{GPLFM} with Matérn and quasiperiodic covariance functions) strain responses. Observations are underlined. }
		\label{tab:TRAC_sg_sine_cov}
		\resizebox*{1\textwidth}{!}{
			\begin{threeparttable}
				\renewcommand{\arraystretch}{1.7}
				\setlength{\tabcolsep}{7pt}
				\begin{tabular}[H]{c|cccccccccccc}			
					Estimator & \textcolor{blue}{1} & \textcolor{blue}{2} & \textcolor{blue}{3} & \textcolor{blue}{4} & \textcolor{blue}{5} & \textcolor{blue}{6} & \textcolor{blue}{7} & \textcolor{blue}{8} & \textcolor{blue}{9} & \textcolor{blue}{10} & \textcolor{blue}{11} & \textcolor{blue}{12} \\
					\hline
					Matérn &			\underline{	0.994}&	0.988&	\underline{0.221}&	0.999	&1.000	&0.946	&0.997&\underline{	0.994}&	0.992&	0.987&	0.995&	\underline{0.992}
					\\ 
					Quasiperiodic  &				\underline{	0.981}	&0.985&	\underline{0.685}	&0.889	&0.997&	0.968&	0.988	&\underline{0.994}	&0.985	&0.996	&1.000&	\underline{0.999}
					\\
				\end{tabular}
			\end{threeparttable}
		}
	\end{table}
	\begin{table} [!ht]
		\centering
		\caption{3D-printed scaled \acs{WT} blade sine test: \acs{TRAC} values between measured and estimated (\acs{GPLFM} with Matérn and quasiperiodic covariance functions) force and acceleration responses. Observations are underlined.}
		\label{tab:TRAC_acc_sine_cov}
		\resizebox*{0.93\textwidth}{!}{
			\begin{threeparttable}
				\renewcommand{\arraystretch}{1.7}
				\setlength{\tabcolsep}{7pt}
				\begin{tabular}[H]{c|ccccccccccc}				
					Estimator & \textcolor{magenta}{13} & \textcolor{magenta}{14} & \textcolor{magenta}{15} & \textcolor{magenta}{16} & \textcolor{magenta}{17} & \textcolor{magenta}{18} & \textcolor{magenta}{19} & \textcolor{magenta}{20} & \textcolor{magenta}{21} & \textcolor{magenta}{22} & \textbf{force} \\
					\hline
					Matérn &				0.986&	0.976	&\underline{1.000}	&0.923&0.998	&\underline{1.000}	&0.997	&0.998	&\underline{1.000}	&\underline{1.000}&	0.767\\
					Quasiperiodic &							0.989&	0.986&\underline{	1.000}	&0.947	&0.998	&\underline{0.999}	&0.995	&0.999	&\underline{1.000}	&\underline{1.000}	&0.812\\
				\end{tabular}
			\end{threeparttable}
		}
	\end{table}
	
	The content offered in this section provides an experimental validation of the concepts elaborated in \autoref{subsec:LFMinputstate} and proved numerically in \autoref{subsec:3DOFsexample}. The bespoke covariance function selection for \acp{GPLFM} construction in the framework of joint input-state estimation is hereby evaluated during tests on the 3D-printed scaled blade under different loading conditions. The produced results prove that a proper a priori selection of the covariance function allows to considerably enhance the achievable load estimation accuracy.
	
	\section{Conclusions}
	\label{sec:concl}
	This paper elaborates on the necessary unknown input modeling, referred to as latent force modeling, for construction of Bayesian input-state estimators. Specifically, the use of structured \acp{LFM} is hereby proposed as a more comprehensive and flexible alternative to the commonly adopted random-walk transition model. A structured \ac{LFM} can be constructed by exploiting the \ac{SDE} representation for \ac{GP} regression, which exhibits different features according to the chosen \ac{GP} kernel. This work offers the \ac{SDE} formulation for non-conventional covariance functions in the structural dynamics context. The analysis has been conducted through an analogy with harmonic oscillators, which has highlighted the tendency of certain classes of kernels towards a better regression performance in specific loading conditions. In particular, periodic or quasiperiodic kernels can be used to model expected harmonic components respectively in absence or in presence of damping. Constant covariance functions can be exploited to model biases in data, which may occur under static loading conditions. Finally, the Wiener covariance function provides a \ac{GPLFM} formulation equivalent to the random-walk model, thus representing an optimal solution for estimation of randomic signals. The selection of bespoke \ac{GP} covariance functions for estimation of standard excitation signals in vibration measurements is thus proposed in this work. The study is validated via a simulated 3 \acp{DOF} example and an experimental case study concerning a 3D-printed scaled \ac{WT} blade, where regression is performed on a single observation instead of adopting the entire set of measured responses. Additionally, the suggested framework foresees the exclusion of the \ac{RTS} smoother step from the \ac{GP}-based algorithm for input-state estimation to allow for real-time inference. Both the simulated and the experimental applications show that a proper selection of the covariance function adopted for \ac{GP} regression allows for a more accurate input estimation, which features a good accuracy even when a single response measurement is used for training the covariance function hyperparameters. Moreover, the simulated example also demonstrates that the use of smoothing can introduce accumulation errors in the input and response predictions.
	
	\section*{Acknowledgements}
	The authors gratefully acknowledge the European Commission for its support of the Marie Sklodowska Curie program through the ITN DyVirt project (GA 764547). 
	
	%
	
	
	\bibliographystyle{unsrt}
	
	\bibliography{cas-refs}

	%
	%
	%
	%
\end{document}

%% file: acro.tex
\begin{acronym}
	\acro{A-AKF}{Adaptive-noise Augmented Kalman Filter}
	\acro{AKF}{Augmented Kalman Filter}
	\acro{AR}{Autoregressive}
	\acro{BARC}{Box Assembly with Removable Component}
	\acro{BC}{Boundary Condition}
	\acro{BCC}{Boundary Condition Challenge}
	\acro{BDM}{Bayesian Dynamic Model}
	\acro{BM}{Brownian Motion}
	\acro{CAD}{Computer Aided Design}
	\acro{CMS}{Component Mode Synthesis}
	\acro{CMS-ME}{Component Mode Synthesis - Modal Expansion}
	\acro{DKF}{Dual Kalman Filter}
	\acro{DOF}{Degrees of Freedom}
	\acro{DT}{Digital Twin}
	\acro{EDM}{Electrical Discharge Machining}
	\acro{EoM}{Equation of Motion}
	\acro{EMA}{Experimental Modal Analysis}
	\acro{FE}{Finite Element}
	\acro{FT}{Fourier Transform}
	\acro{FRAC}{Frequency Response Assurance Criterion}
	\acro{GDF}{Gillijns De Moor Filter}
	\acro{GP}{Gaussian Process}
	\acro{GPLFM}{Gaussian Process Latent Force Model}
	\acro{GFRP}{Glass Fiber Reinforced Plastic}
	\acro{KF}{Kalman Filter}
	\acro{KS}{Kalman Smoother}
	\acro{LFM}{Latent Force Model}
	\acro{LTI}{Linear Time Invariant}
	\acro{LTV}{Linear Time Variant}
	\acro{ML}{Machine Learning}
	\acro{MAP}{Maximum a Posteriori}
	\acro{MCMC}{Markov chain Monte Carlo}
	\acro{MAC}{Modal Assurance Criterion}
	\acro{ME}{Modal Expansion}
	\acro{MOR}{Model Order Reduction}
	\acro{NRMSE}{Normalized Root Mean Square Error}
	\acro{OMA}{Operational Modal Analysis}
	\acro{OSP}{Optimal Sensor Placement}
	\acro{PBH}{Popov-Belevitch-Hautus}
	\acro{PDE}{Partial Differential Equation}
	\acro{PSD}{Power Spectral Densitie}
	\acro{PE}{Predictive Engineering}
	\acro{PDF}{Probability Density Function}
	\acro{POD}{Proper Orthogonal Decomposition}
	\acro{QoI}{Quantities of Interest}
	\acro{RIRA}{Residual Inertia-Relief Attachment}
	\acro{RB}{Reduced Basis}
	\acro{ROM}{Reduced Order Model}
	\acro{RTS}{Rauch-Tung-Striebel}
	\acro{RW}{Random Walk}
	\acro{RBE}{Rigid Body Element}
	\acro{RMS}{Root Mean Square}
	\acro{RMSE}{Root Mean Square Error}
	\acro{SD}{Standard Deviation}
	\acro{SDE}{Stochastic Differential Equation}
	\acro{SE}{Static Error}
	\acro{SHM}{Structural Health Monitoring}
	\acro{SISW}{Siemens Industry Software}
	\acro{SSM}{State-Space Model}
	\acro{STP}{Student-t Process}
	\acro{STPLFM}{Student-t Process Latent Force Model}
	\acro{TF}{Transfer Function}
	\acro{TRAC}{Time Response Assurance Criterion}
	\acro{UKF}{Unscented Kalman Filter}
	\acro{VS}{Virtual Sensing}
	\acro{WT}{Wind Turbine}
\end{acronym}